\documentclass[%
 pre, twocolumn,
 amsmath,amssymb,
 aps,
]{revtex4-2}
\DeclareMathOperator{\Tr}{Tr}
\DeclareMathOperator{\Cov}{Cov}
\usepackage{hyperref}
\usepackage{mathptmx}
\usepackage{supertabular}
\usepackage{longtable}
\usepackage{graphicx}% Include figure files
\usepackage{dcolumn}% Align table columns on decimal point
\usepackage{bm}% bold math
\usepackage{xcolor}
\usepackage{cancel}
\usepackage{soul}

\newcommand{\Rev}[1]{\textcolor{black}{#1}}
\newcommand{\RevTwo}[1]{\textcolor{black}{#1}}
\newcommand{\Corr}[1]{\textcolor{black}{#1}}

\begin{document}

\preprint{APS/123-QED}

\title{Broken detailed balance and entropy production in directed networks}% Force line breaks with \\

\author{Ramón Nartallo-Kaluarachchi\textsuperscript{1,2,*},
Malbor Asllani\textsuperscript{3,§},
Gustavo Deco\textsuperscript{4,5,6},
Morten L. Kringelbach\textsuperscript{2,7,8},
Alain Goriely\textsuperscript{1,†} and 
Renaud Lambiotte\textsuperscript{1,9,‡}}
\affiliation{%
1. Mathematical Institute, University of Oxford, Woodstock Road, Oxford, OX2 6GG, United Kingdom
\\
2. Centre for Eudaimonia and Human Flourishing, University of Oxford, 7 Stoke Pl, Oxford, OX3 9BX, United Kingdom
\\
3. Department of Mathematics, Florida State University, Tallahassee, FL 32306, United States of America
\\
4. Centre for Brain and Cognition, Computational Neuroscience Group, Universitat Pompeu Fabra, 08018
Barcelona, Spain\\
5. Department of Information and
Communication Technologies, Universitat Pompeu
Fabra, 08018 Barcelona, Spain\\
6. Institucio Catalana de la Recerca i Estudis Avancats (ICREA), 08010 Barcelona, Spain\\
7. Center
for Music in the Brain, Aarhus University, \& The
Royal Academy of Music, Aarhus/Aalborg, Denmark\\
8. Department of Psychiatry, University of Oxford, Oxford, OX3 7JX United Kingdom\\
9. The Turing Institute, British Library, 96 Euston Rd, London, NW1 2DB, United Kingdom\\
{\normalfont{\textsuperscript{*}\{nartallokalu\}\textsuperscript{†}\{goriely\}\textsuperscript{‡}\{lambiotte\}@maths.ox.ac.uk}, \textsuperscript{§}\{masllani\}@fsu.edu}
}

\date{\today}% It is always \today, today,

\begin{abstract}
\noindent The structure of a complex network plays a crucial role in determining its dynamical properties. In this work, we show that the \Rev{degree to which a} network is \Rev{directed and hierarchically organised is closely associated with the degree to which its dynamics break detailed balance and produce entropy}. We consider a range of dynamical processes and show how different directed network features \Rev{affect their entropy production rate}. \Rev{We begin with an analytical treatment of a 2-node network followed by numerical simulations of synthetic networks using the preferential attachment and Erdös-Renyi algorithms. Next, we analyse a collection of 97 empirical networks to determine the effect of complex real-world topologies.} Finally, we present a simple method for inferring broken detailed balance and directed network structure from multivariate time-series and apply our method to identify non-equilibrium dynamics and hierarchical organisation in both human neuroimaging and financial time-series. Overall, our results shed light on the consequences of directed network structure \Rev{on non-equilibrium dynamics} and \Rev{highlight} the importance and ubiquity of hierarchical organisation and non-equilibrium dynamics in real-world systems.
\end{abstract}

\maketitle

%\tableofcontents

\section{Introduction}
\label{sec: intro}
\noindent The abstraction of large complex systems as networks of interconnected elements has been instrumental in the modelling of systems in ecology \cite{dunne2002ecology}, economics \cite{jackson2010socialeconomic}, sociology \cite{wasserman2012socialnetwork}, bio-medicine \cite{barabasi2011networkmedicine}, neuroscience \cite{bullmore2009brainnetwork,basset2017networkneuro} and beyond \cite{newman2018networks}. In particular, dynamical processes evolving on networks have become prototypical models of real-world systems in a range of diverse fields \cite{newman2003structurefunction,boccaletti2006complexdynamis,barrat2008dynamical}. Reconciling the relationship between the structure of interactions and the emergent dynamical phenomena of such systems remains a outstanding challenge.
Many physical, chemical and biological systems operate far from thermodynamic equilibrium \cite{Prigogine1968Thermodynamics}. These non-equilibrium systems consume energy and dissipate heat to their surroundings, producing entropy. In particular, energy consumption and entropy production represent a key mechanism by which living systems are able to stave off thermodynamic  equilibrium and so-called `heat death' \cite{Schrodinger1944whatislife}. Equilibrium systems are often characterised by symmetric interactions between identical elements that in-turn yield time-reversible dynamics.\begin{figure}
    \centering
    \includegraphics[width = 0.5\textwidth]{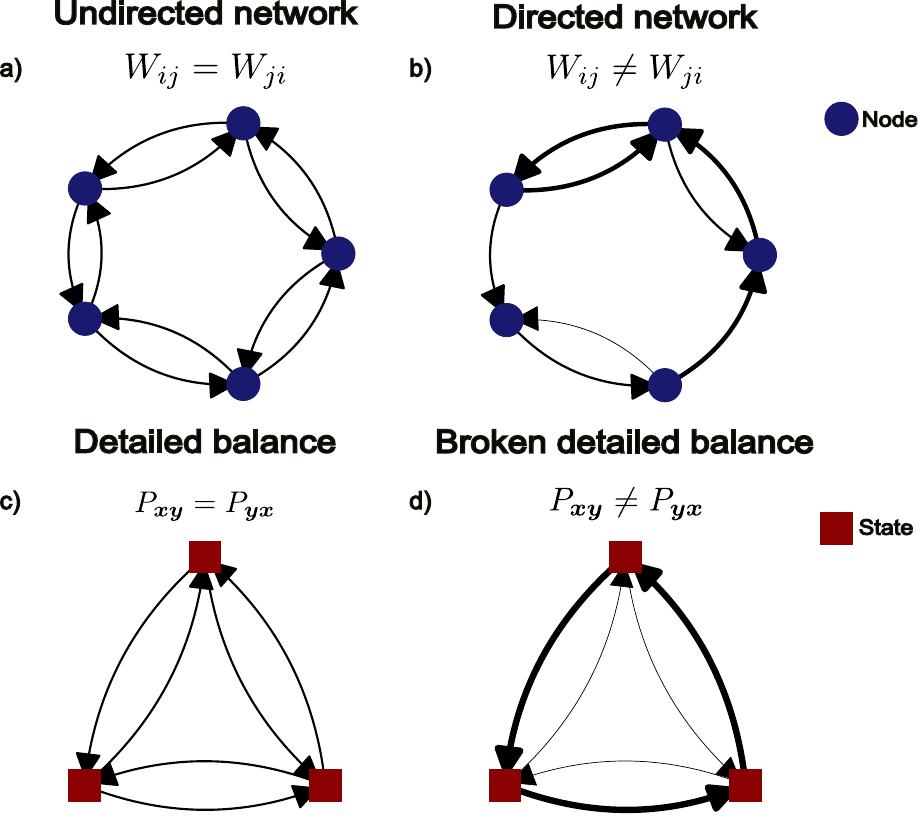}
    \caption{\textbf{Asymmetry in complex systems} Symmetric nodal interactions \Rev{usually} lead to symmetric transition rates between system states whilst asymmetric interactions \Rev{usually} lead to broken detailed balance. \textbf{Top:} $a)$ An undirected network with $W_{ij}=W_{ji}$. $b)$ A directed network with $W_{ij}\neq W_{ji}$. Blue circles represents nodes or elements of the system. \textbf{Bottom:} $c)$ A system in detailed balance with $P_{\bm{xy}}=P_{\bm{yx}}$. $d)$ A system violating detailed balance with $P_{\bm{xy}}\neq P_{\bm{yx}}$. Red squares represent distinct, discrete system states. The thickness of connections represents the weight/\RevTwo{joint transition probability}, respectively.}
    \label{fig: schematicbdb}
\end{figure} Similarly, non-reciprocal interactions between elements cause violations of the so-called `detailed balance condition' given, equivalently, by the equalities $P_{\bm{y}\bm{x}} = P_{\bm{x}\bm{y}}$ and $\pi_{\bm{y}}\cdot P_{\bm{x}|\bm{y}}  = \pi_{\bm{x}}\cdot P_{\bm{y}|\bm{x}}$,
and illustrated in Figure \ref{fig: schematicbdb}. Here $P_{\bm{x}\bm{y}}$ is the joint transition probability from state $\bm{x}$ to $\bm{y}$, $P_{\bm{x}|\bm{y}}$ is the transition probability from $\bm{y}$ to $\bm{x}$ conditional on the system being in state $\bm{y}$, $\pi_{\bm{y}}$ is the steady-state probability of being in state $\bm{y}$ \Rev{and $P_{\bm{x}\bm{y}}=P_{\bm{y}|\bm{x}}\pi_{\bm{x}}$}. Violation of detailed balance leads to non-equilibrium steady states and irreversibility in the system's dynamics. The degree to which a system diverges from thermodynamic equilibrium can be quantified through the rate at which it produces entropy \cite{seifert2012thermodynamics}. The entropy production rate (EPR) quantifies the distance of the system from equilibrium and the irreversibility of its dynamics by measuring the divergence between the probability of observing system trajectories and their time-reversals \cite{roldan2010dissipation}. In this work, we make progress on this front by demonstrating a novel and important link between the structure of directed networks and the \Rev{non-equilibrium nature of} dynamical processes evolving on them.\\\\
Broken detailed balance and non-equilibrium dynamics have been observed in a range of microscopic \cite{Gnesotto2018brokendetailedbalance,brangwynne2008cytoplasmicdiffusion,yin1999nonequilibriumRNA,Huang2003ecoli,mehta2012cellularcomputation,stuhrman2012cytoskeleton} and mesoscopic \cite{battle2016brokendetailedbalance} processes at the molecular and cellular level in living systems. At the macroscopic scale, temporal irreversibility has been observed in evolutionary dynamics \cite{England2013selfreplication} and large-scale neural dynamics \cite{lynn2021detailedbalance,deco2023tenet,bolton2023AoT}. However, despite many advancements in modern non-equilibrium statistical physics \cite{seifert2012thermodynamics}, results in stochastic thermodynamics have been limited to the study of small systems, with large, complex systems only attracting attention very recently \cite{aguilera2023sherrington,herpich2018collective,Herphic2020manybody,sune2019clock}. When we abstract systems as networks of nodes and edges, symmetric interactions correspond to undirected networks, i.e., when a pair of edges exist between two nodes and they have the same weight in both directions. Conversely, asymmetric interactions correspond to directed networks, where the existence and strength of edges can vary in each direction. Previous attempts to reconcile network science and non-equilibrium thermodynamics have focused on network representations of \Rev{Markov chains} \cite{schnakenberg1976networktheory}, where nodes represent mesoscopic states or thermodynamic quantities \cite{oster1971networkthermodynamics} whilst edges represent transition rates. However, little is known about the role of network structure in real-space, where nodes and edges represent elements of the system and their interaction strengths, with the exception of chemical reaction networks \cite{rao2016crn,polettini2015crn,DalCengio2023geometrycrn} \Rev{or linear network dynamics \cite{vaidya2021sociallearning}}. Despite this, in systems such as the human brain, pairwise interactions have been shown to be the dominant contribution to the EPR, highlighting the importance of network structure in a system's thermodynamics \cite{lynn2023decomposing}. Understanding the role of network topology remains an important unsolved problem in the \Rev{non-equilibrium dynamics} of complex systems \cite{papo2024braincomplexnetwork}.\\\\
As anticipated, symmetry-breaking in the organisation of complex systems is the structural feature that drives network dynamical processes out of equilibrium. Dynamical processes on directed networks differ drastically from their undirected counterparts \cite{krakauer2023brokensymmetry, johnson2020digraphs} including in their phase-transitions \cite{fruchart2021nonreciprocal}, synchronisation properties \cite{muolo2020sychnronisation, muolo_persistence_2024}, topological resilience \cite{asllani2018topologicalresilience, nicoletti_resilience_2019} and pattern formation \cite{asllani2014patterns,muolo2019patterns}. Recent studies on extensive datasets of real-world networks, have shown a ubiquity of strong directedness and clear signs of hierarchical organisation \cite{johnson2014trophiccoherence,asllani2018nonnormal,johnson2020digraphs,mackay2020directed,obrien2021hierarchical}. As a result, the dynamics of directed networks are more indicative of the dynamics of real-world complex systems. In particular, this strong directedness results in a marked non-normality of the operators defined on such networks \cite{trefethen2001nonnormal,asllani2018nonnormal}. Consequently, non-normality can \Rev{result in} the underlying networked systems \Rev{diverging from equilibria} \cite{asllani2018nonnormal}. Motivated by the ubiquity of directed structures in the real world and their significance in non-linear dynamics, this paper aims to further strengthen this link \Rev{using the formalism of stochastic thermodynamics}.\\\\
In this work, we bring new insight into the role of network structure in the emergence of broken detailed balance and irreversibility by demonstrating \Rev{the close association between the} directedness of the interaction network \Rev{and the divergence of} a dynamical process from thermodynamic equilibrium. We first define a range of measures of directedness in networks and a scheme to smoothly parameterise the directedness of a network. Secondly, we introduce three network dynamical processes and calculate their EPR, namely the discrete- and continuous-time random walks (RW) \cite{Masuda2017randomwalks,lambiotte2023continuoustimerandomwalk}, Ornstein-Uhlenbeck (OU) \cite{Godreche2018OU} and Ising dynamics \cite{Nishimori2001statphys}. \Rev{Next, we consider a solvable 2-node system and show analytically, for the OU, and numerically, for the Ising, that the EPR is a function of the directedness. Two further solvable models, the OU process on both circulant and 2-node interpolated networks, are considered in Appendices \ref{sec: app: 2node} \& \ref{sec: app: circulant}. Thirdly, we consider hierarchically asymmetric networks generated using preferential attachment and show that increased directedness drives an increase in the EPR for all processes across network size.} Fourthly, using the Erdös-Rényi (ER) \cite{erdos1959random} as a null model, we decouple different measures of directedness and show that locally evolving processes, like the RW, produce more entropy when the system becomes more `locally directed', whereas the EPR of globally coupled processes, like the Ising and OU dynamics, is dictated by the `global directedness' of the underlying network. Subsequently, we consider structural data in the form of 97 real-world directed networks from  a range of fields including
biology, social interactions, ecology, transport, and language \cite{obrien2021hierarchical}. By considering the processes evolving on real-world topologies, we are able to further confirm the link between the directedness measures and the EPR \Rev{of the synthetic dynamics but on complex, real-world topologies}. Finally, we describe a simple, but powerful, method for extracting directed networks and the EPR from multivariate time-series (MVTS) using a linear auto-regression technique \cite{Shumway2017timeseries}. We apply this method to MVTS from human neuroimaging and the stock-market to reveal the hierarchical organisation of brain-regions and stock price interactions as well as the non-equilibrium nature of their dynamics. In particular we confirm the increased EPR in task-based brain states, as previously found \cite{lynn2021detailedbalance,deco2022insideout,deco2023tenet,deco2023violations}, but additionally \Rev{link this observation to an associated} reorganisation of the hierarchy of brain regions. Moreover, we show that consumer goods/services emerge as leader nodes in the hierarchy of stocks that governs market movements. Overall, this paper draws an important link between the directed structure of a complex system and \Rev{the degree to which it violates the detailed balance condition and time-reversal symmetry}. This work puts to the forefront the importance of considering asymmetries in interactions when studying the dynamics of complex networks, a consideration often overlooked in areas such as neuroscience \cite{kale2018directed,friston2021parcels}, and presents a new perspective for analysing both non-equilibrium systems and directed networks in both theory and empirical data.
\section{Measures of directedness in networks}
\label{sec: measures}
\noindent As anticipated, non-reciprocal interactions in complex systems disrupt detailed balance, leading to an increase in the EPR. This section delineates four established metrics to quantify the overall directedness of networks and to examine their influence on EPR within both random graphs and empirical datasets. These systems are represented as directed networks with non-negative weighted adjacency matrices, $\bm{W}=(W_{ij})$, where $W_{ij} \geq 0$ signifies the strength of the directed link $i \rightarrow j$, including self-edges $W_{ii} \geq 0$ and characterized by inherent asymmetry $\bm{W} \neq \bm{W}^\top$ \Corr{\footnote{\Corr{The definition of $W_{ij}$ to represent the link from $i$ to $j$ is one of two conventions. We choose this convention to be consistent with the definition of trophic coherence \cite{mackay2020directed}. The opposite, is the more typical convention for network dynamics \cite{newman2018networks}.}}}. By exploring four distinct metrics for network directedness, we seek to elucidate its effect on EPR, highlighting the emergence of hierarchical structuring within networks - an outcome deeply influenced by non-reciprocal interactions \cite{corominas2013hierarchy,obrien2021hierarchical}.
\subsection{Irreciprocity}
\noindent As a measure of the directedness of a network, we first consider the (ir)reciprocity \footnote{We consider the (ir)reciprocity of weighted networks as defined in Ref. \cite{squartini2013weightedreciprocity}. Alternative definitions of reciprocity for unweighted graphs are given in Ref. \cite{garlaschelli2004reciprocity,newman2018networks}.}\cite{garlaschelli2004reciprocity,squartini2013weightedreciprocity}. The reciprocity quantifies how reciprocated pairwise relationships are on average across the network by comparing each connection $i\rightarrow j$ with $j\rightarrow i$. We first define the reciprocated portion of the pairwise relationship between two nodes,
\begin{align}
    \overleftrightarrow{W}_{ij} &= \min(W_{ij},W_{ji})=\overleftrightarrow{W}_{ji},
\end{align}
and the unreciprocated portion in each direction,
\begin{align}
    \overleftarrow{W}_{ij} &= W_{ji}-  \overleftrightarrow{W}_{ij}
    = \overrightarrow{W}_{ji}.
\end{align}
The reciprocity is then quantified by,
\begin{align}
    r(\bm{W})=\frac{\sum_{i,j\neq i}\overleftrightarrow{W}_{ij}}{\sum_{i,j\neq i}{W}_{ij}},
\end{align}
which is in the range $[0,1]$ with $r=0$ corresponding to a perfectly unreciprocated network, where edges can only run in a single direction, and $r=1$ corresponding to an undirected network. We quantify the overall directedness of the network as $1-r$, which we define to be the \textit{irreciprocity}.
Whilst the irreciprocity gives a measure of directedness for the network as a whole, we describe it as a `local' measure, meaning it averages over each pairwise relationship in turn without analysing the structure of the network `globally'. As a result, this measure fails to distinguish between `loop-like' structures and motifs that cause the network to globally follow a single direction. Directedness and non-zero irreciprocity are equivalent.
\subsection{Trophic directedness}
\noindent The question of whether a network globally follows a direction in its structure is another important notion of directedness. This idea is intimately linked to the idea of a hierarchy within the system where nodes can be organized into levels indicating their position in the top-down organisation of the network. First being put forward in the field of ecology \cite{johnson2014trophiccoherence}, in the context of food webs, \textit{trophic (in)coherence} is a measure quantifying how neatly a network can be organized into so-called trophic levels \cite{mackay2020directed}.\\\\
Inspired by the Helmholtz-Hodge decomposition \cite{Jiang2011Hodge}, the trophic incoherence of a network is given by,
\begin{align}
\label{eq: cost}F_0(\bm{W},\bm{h})=\min_{\bm{h}}\frac{\sum_{i,j} W_{ij}(h_j-h_i-1)^2}{\sum_{i,j} W_{ij}},
\end{align}
where $\bm{h}=(h_1,...,h_N)$ is the vector with entries corresponding to the trophic levels for each of the $N$ nodes, that minimizes the cost function \footnote{Equivalently, the `SpringRank' formulation considers directed springs between nodes and aims to find a ranking that minimizes the total energy of these springs \cite{debacco2018springrank}}. \Rev{The edge-weights, when used to define a flow on the edges, can be decomposed into a gradient and conservative part \cite{Jiang2011Hodge, kichikawa2019community}. The trophic levels represent a function defined on the nodes whose gradient best reflects the edge structure of the network. This is defined in such a way that a low trophic level corresponds to nodes with a large outgoing flow which have a place at the `top' of the hierarchy.} The trophic levels are found as solution of the linear system
\begin{align}
\label{eq:trophicsystem}
    \bm{\Lambda} \bm{h} = \bm{v},
\end{align}
where $v_i = \sum_j W_{ji} - W_{ij}$ and $\bm{\Lambda} = \text{diag}(\bm{u}) - \bm{W} - \bm{W}^{\top}$ defines the (symmetric) weighted graph Laplacian with $u_i = \sum_j W_{ji} + W_{ij}$ \cite{mackay2020directed}.\\\\
 The nodes of any network can be partitioned into weakly connected components which are disjoint sets of nodes where node $i$ belongs to a component if there is a node $j$ in the component with $\max(W_{ij},W_{ji})>0$. Furthermore, the number of weakly connected components corresponds to the nullity of the Laplacian. Consequently, the system (\ref{eq:trophicsystem}) has a non-unique solution corresponding to the nullity of the Laplacian. By enforcing that the lowest trophic level in each (weakly) connected component is equal to 0, one can obtain a unique solution to the equation (\ref{eq:trophicsystem}) and calculate $F_0$. The trophic incoherence, $F_0$, is restricted to the range $[0,1]$ with $F_0=1$ corresponding to completely non-hierarchical (including undirected) networks and $F_0=0$ corresponding to networks that can be perfectly organized into trophic levels. We will consider \textit{trophic directedness}, $\sqrt{1-F_0}$, as a measure of directedness in a network.
Trophic directedness assumes the direction of the flow to be bottom-up meaning a low trophic level corresponds to the `top' of the hierarchy if we were to consider a more intuitive top-down visualisation. By convention, throughout this paper, we display the trophic levels of a network by plotting nodes with the lower trophic levels at the top and inverting the \(y\)-axis.

\subsection{Non-normality}
\noindent A third notion of directedness is \textit{network non-normality} \cite{asllani2018nonnormal}.
A matrix $\bm{W}$ is said to be normal if it satisfies,
\begin{align}
    \label{eq: normal}
    \bm{W}\bm{W}^{\top}=\bm{W}^{\top}\bm{W}.
\end{align}
It is, therefore, non-normal if $\bm{W}\bm{W}^{\top}\neq\bm{W}^{\top}\bm{W}$ \cite{trefethen2001nonnormal}. Symmetric matrices are normal whilst non-normal matrices are necessarily asymmetric. Correspondingly, undirected networks are normal whilst non-normal networks are directed. The significance of developing a metric to quantify non-normality stems from the prevalence of non-normal matrices across a broad spectrum of applications, encompassing both linear systems and networks \cite{trefethen2001nonnormal}.\\\\
The eigenvectors of a non-normal matrix do not form an orthonormal basis or in other words such matrices are not diagonalisable by unitary matrices. As a result, the autonomous system of linear differential equations,
    $\dot{\bm{x}}(t)  = \bm{W}\bm{x}(t)$
where $\bm{W}$ is non-normal, can undergo transient growth such that small perturbations can excite the system away, temporarily, from asymptotically stable equilibria. This feature has lead to the investigation of the role and uses of non-normal interactions in linear systems \cite{asllani2018nonnormal}, neuronal \cite{hennequin2012nonnormal} and communication networks \cite{baggio2020communication} as
well as in pattern formation \cite{muolo2019patterns}, synchronisation \cite{muolo2020sychnronisation}, resilience to perturbation \cite{asllani2018topologicalresilience} and network control of instabilities \cite{lindmark2021control,duan2022instability}.\\\\
A range of measures can be used to quantify the non-normality of a matrix \cite{asllani2018nonnormal}. Of particular interest are those derived from its spectrum, $\sigma(\bm{W})$, which governs its behavior as a linear system, and its pseudo-spectrum, $\sigma_{\epsilon}(\bm{W}) = \{\sigma(\bm{W}+\bm{E}):||\bm{E}||<\epsilon\}$, which governs its response to perturbations \cite{trefethen2001nonnormal}. The transient behavior of non-normal linear systems cannot be explained by the traditional spectral abscissa, $\alpha(\bm{W})= \max \Re (\sigma(\bm{W}))$, which determines the asymptotic dynamics. Instead, measures such as the numerical abscissa, $\omega(\bm{W})=\max \sigma(\frac{1}{2}(\bm{W}+\bm{W}^{\top}))$, capture the transient short-term growth of non-normal systems whilst pseudo-spectral measures such as the $\epsilon$-pseudo-spectral abscissa, $\alpha_{\epsilon}(\bm{W})= \max \Re (\sigma_{\epsilon}(\bm{W}))$, and the Kreiss constant, $\mathcal{K}(\bm{W}) = \max_{\epsilon>0} \frac{\alpha_{\epsilon}(\bm{W})}{\epsilon}$, capture their sensitivity to perturbation \cite{trefethen2001nonnormal}. We quantify the degree to which a network breaks the normality condition using a common spectral measure, the \textit{Henrici index},
\begin{align}
\label{eq: henrici}
    d_{H}(\bm{W}) & = \sqrt{||\bm{W}||_F^2-\sum_{i=1}^N |\lambda_i|^2},
\end{align}
where $||\bm{W}||_{F} = \sqrt{\sum_i\sum_jW_{ij}^2}$ is the Frobenius norm and $\{\lambda_i\}$ is the set of eigenvalues \cite{trefethen2001nonnormal}. The Henrici index is 0 when a matrix is normal and positive otherwise. To facilitate the comparison of various networks in terms of non-normality, irrespective of their size, the \textit{normalized Henrici index} has been introduced \cite{asllani2018nonnormal},
\begin{align}
    \label{eq: normalisedhenrici}
     \hat{d}_{H}(\bm{W}) & = \frac{d_H(\bm{W})}{||\bm{W}||_F},
\end{align}
which has values between 0 and 1. We opt for the Henrici index as it captures the spectral properties of $\bm{W}$, without directly assuming linear dynamics. In order to understand the nature of non-normality as a measure of hierarchical asymmetry, we consider an unweighted adjacency matrix $\bm{A}$. The entry $(\bm{A}^{\top}\bm{A})_{ij}$ represents the number of common sources connecting into nodes $i$ and $j$ whilst $(\bm{A}\bm{A}^{\top})_{ij}$ represents the number of common targets from nodes $i$ and $j$ \cite{mackay2020directed}. Therefore, in a normal unweighted network, for every given pair of nodes, the number of common sources and targets will coincide whereas in a non-normal unweighted network they will not coincide. In the case of weighted edges, non-normality captures the hierarchical asymmetry and the quantity $|\bm{WW}^{\top}-\bm{W}^{\top}\bm{W}|$ is maximised is the case of a directed acyclic graph (DAG) where, after node relabelling, $\bm{W}$ is upper triangular yielding a clear net direction and nodal hierarchy \cite{asllani2018nonnormal}.\\\\
Nevertheless, trophic directedness and non-normality are not equivalent and there are a range of networks which are trophically undirected yet non-normal and vice-versa (see Appendix \ref{sec: app: trophicdirectednessnn}). Previous work has shown that trophic directedness and non-normality are closely, but non-linearly, correlated \cite{mackay2020directed,johnson2020digraphs}. Furthermore, non-normality does not capture all directed networks, as the set of asymmetric matrices contains, but is much larger than, the set of non-normal matrices. There is no complete characterisation of normal, asymmetric matrices but examples include circulant, block-circulant with circulant blocks, and skew-symmetric matrices. In Appendix \ref{sec: app: trophicdirectednessnn}, we revisit the notion of trophic flatness (\(F_0=1\)) extending the concept to show its inequivalence to normality in weighted networks. We further contribute a new example of a trophically flat, non-normal network, uniquely without using a self-loop, thus broadening the discourse beyond previous studies.

\subsection{Parameterising directedness in networks}
\begin{figure}[h]
\centering
\includegraphics[width=\linewidth]{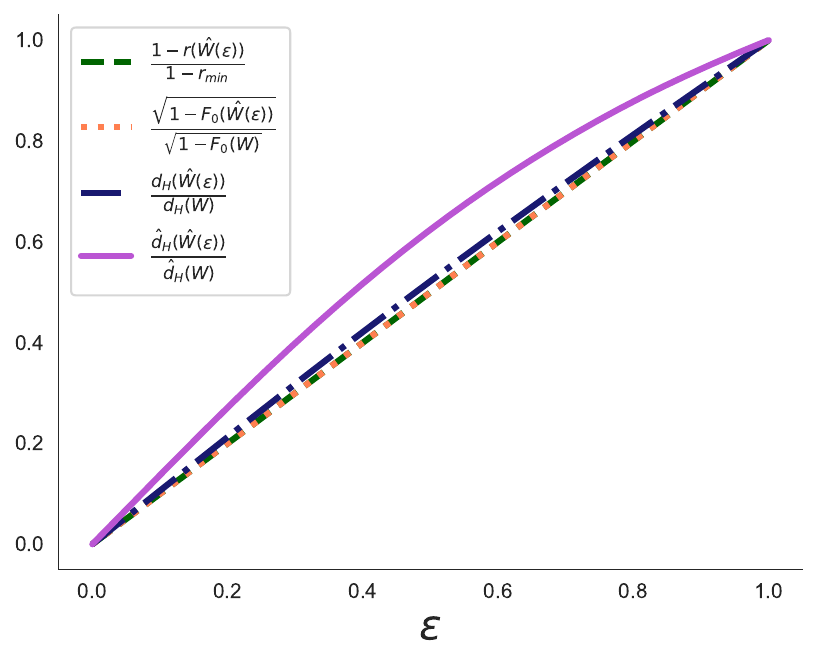}
\caption{\textbf{Parameterising directedness}: Beginning with a hierarchical, non-normal network $\bm{W}$, we can measure the directedness of the parameterised network as a function of $\epsilon$, the interpolation parameter. Increasing $\epsilon$ increases the directedness almost linearly for each of the four measures. Here each measure is normalised by its maximum value which occurs at $\epsilon=1$.}
\label{fig: interpolation}
\end{figure}

\noindent To investigate the dynamical effects of directedness on network dynamics, it would be preferable to continuously vary the level of asymmetry in a network. To this aim, we first generate highly hierarchical, non-normal networks using a preferential attachment (PA) scheme with weak reciprocal links (see Appendix \ref{sec: app: PA} for details) \cite{asllani2018nonnormal}. We then linearly interpolate between this strongly directed network and its undirected Hermitian,
\begin{align}
\label{eq: interpolation1}
    \bm{\hat{W}}(\epsilon) & = (1-\epsilon)\bm{\Tilde{W}} + \epsilon\bm{W},
\end{align}
for $\epsilon\in [0,1]$. Here, the Hermitian network is given by $\bm{\Tilde{W}} = \frac{1}{2}\left( \bm{W} + \bm{W}^{\top}\right).$ By increasing the parameter $\epsilon$, we are able to continuously increase the directedness of the network up to some maximal value. For a highly non-normal, hierarchical $\bm{W}$, Figure \ref{fig: interpolation} shows the measures of directedness for the interpolated networks as a function of $\epsilon$.\\
 % avoid blank space here
\newline For the irreciprocity and trophic directedness, we can show that the interpolation parameterises the measure exactly linearly (for a proof see Appendix \ref{sec: app: exactinterpolation}) i.e.,
\begin{align}
    1-r(\bm{\hat{W}}(\epsilon)) &= \epsilon (1-r(\bm{W})),\\
    \sqrt{1-F_0(\bm{\hat{W}}(\epsilon)} &= \epsilon \sqrt{1-F_0(\bm{W})}.
\end{align}
The eigenvalues of the matrix $\bm{\hat{W}}(\epsilon)$ are not calculable from $\bm{\Tilde{W}}$ and $\bm{W}$ so the same cannot be done for the Henrici indices. However, the numerical experiments, shown for just one network in Figure \ref{fig: interpolation}, consistently show an increase with $\epsilon$.
Furthermore, we note that any weakly connected component in $\bm{W}$ becomes strongly connected in the interpolated network $\bm{\hat{W}}(\epsilon)$ for $\epsilon<1$. Using the PA scheme detailed in Appendix \ref{sec: app: PA}, the network is strongly connected, by construction.

\section{Entropy production rate of dynamics on directed networks}
\label{sec: dynamics}
\noindent The investigation of the effect of directedness in breaking detailed balance and increasing the EPR in networked systems, will be focused on three prototypical network dynamical processes: the random walk in discrete and continuous time \cite{Masuda2017randomwalks,lambiotte2023continuoustimerandomwalk}, the Ornstein-Uhlenbeck \cite{scwarze2021motifs} and the Ising model \cite{Nishimori2001statphys}. Each of the dynamics shares three important traits. First, they represent a broad spectrum of stochastic dynamics, with both continuous and symbolic variables as well as continuous and discrete time, that are well-studied and have found applications in many disparate fields. Second, under certain conditions, all the processes converge to equilibrium steady states on undirected networks and non-equilibrium, entropy-producing steady states on directed networks. Third, for each system, we can either explicitly calculate or numerically estimate the EPR with minimal sampling from the dynamics, alleviating the need for computational approaches.
\subsection{Random walks on directed networks}
\noindent We first consider the dynamics of non-interacting walkers randomly exploring the graph \cite{Masuda2017randomwalks,lambiotte2023continuoustimerandomwalk}. \Rev{Random walks (RW) can evolve in either discrete time-steps or continuous time where the waiting times have a given probability distribution. Here, we consider two walks, the discrete-time random walk (DTRW) \cite{Masuda2017randomwalks} and the continuous-time random walk (CTRW) with waiting-times given by the Poisson distribution \cite{lambiotte2023continuoustimerandomwalk}, both taking place on a weighted network with weight matrix $\bm{W}$. Both processes yield a Markov chain where the state of the random walker, $x(t)$, at any time $t$ is the node on which the walker is standing. For both the DT and CTRW, at each step, the walker moves from node $i$ to node $j$ according to the transition probability},
\begin{align}
    T_{ij} & = \frac{W_{ij}}{\sum_jW_{ij}}.
\end{align}
 Clearly, a walker can only move from one node to its neighbours and the probability that the walker moves to a particular node is proportional to the weight of the edge connecting their current position and the destination node. \Rev{The Perron-Frobenius theorem dictates an irreducible, aperiodic Markov chain will converge to its unique stationary distribution \cite{Haggstrom2020MarkovChain}. As a result, on undirected networks the RW converges to an equilibrium steady state, with the exception of bipartite graphs in discrete-time which are periodic}. On directed networks, the graph must be strongly connected in order for the existence of a single steady state.\\\\
 \Rev{To derive the steady-state distribution for the DTRW}, we assume a population of random walkers that is originally distributed on the network with density $\bm{\pi}(0)$, then after $t$ steps, the distribution on the network is given by,
\begin{align}
    \bm{\pi}(t)&=\bm{\pi}(0)\bm{T}^t,
\end{align}
where $\bm{T}=(T_{ij})$. Therefore the steady-state satisfies
\begin{align}
    \bm{\pi}=\bm{\pi}\bm{T},
\end{align}
namely, it is the Perron left eigenvector of the transition matrix $\bm{T}$ \cite{Haggstrom2020MarkovChain}. Consequently, the joint transition probability is given by,
\begin{align}
    P_{ij} &= T_{ij}\pi_i,
\end{align}
where $\pi_i$ is the steady-state probability of the walker being at node $i$.\\\\
%\Rev{For the CTRW, we assume that the waiting times are distributed independently of the walker's position. As a result the probability that the walker is on node $i$ after time $t$ is given by,
%\begin{align}
%\label{eq: CTRWdynamics}
%    \pi_i(t) = \sum_{n=0}^{\infty}\pi_i[n]\cdot p(n,t).
%\end{align}
%Here $\pi_i[n]$ is the probability of the walker arriving at node $i$ after $n$ jumps and $p(n,t)$ is the probability of the walker taking exactly $n$ jumps in time $t$, making use of the fact that these probabilities are independent. Assuming that the waiting times are Poisson distributed, and following Ref. \cite{lambiotte2023continuoustimerandomwalk}, we obtain that the probability density evolves according to,
\RevTwo{For the CTRW, assuming that the waiting times are Poisson distributed and following Ref. \cite{lambiotte2023continuoustimerandomwalk}, we obtain that the probability density evolves according to}
\begin{align}
    \frac{d \bm{\pi}}{dt}&= \bm{\pi}(t)\bm{L},
\end{align}
\RevTwo{where $\bm{\pi}(t)= (\pi_1(t),...,\pi_N(t))$, $\pi_i(t)$ is the probability that the walker is on each node $i$ at time $t$ and $\bm{L}=\bm{T}-\bm{I}$ is the random-walk Laplacian}. Consequently, the stationary distribution is given by $\bm{\pi}\bm{L}=0$, thus it coincides with the stationary distribution of the DTRW, $\bm{\pi} = \bm{\pi T}$.\\\\
The EPR for Markovian dynamics at stationarity is given by the Schnakenberg formula \cite{schnakenberg1976networktheory},
\begin{align}
    \Phi &= \frac{1}{2}\sum_{i,j}(T_{ji}\pi_{j} - T_{ij}\pi_{i})\log\left(\frac{T_{ji}\pi_{j}}{T_{ij}\pi_{i}}\right),
\end{align}
which can, alternatively, be written as the Kullback-Leibler (KL) divergence between the forward and backward joint transition rates between all pairs of states,
\begin{align}
\label{eq: entropyprodmarkov2}
\Phi & =\sum_{i,j}P_{ij}\log \frac{P_{ij}}{P_{ji}}.
\end{align}
However, we note that not all states are achievable from other states. \Rev{Given the existence of an edge $i\rightarrow j$ with no reciprocal link, $P_{ij}>P_{ji}=0$, then (\ref{eq: entropyprodmarkov2}) will diverge to infinity. Therefore on a graph containing at least one entirely unreciprocated edge, the EPR of the RW at stationary is, in fact, infinite. Whilst there have been some recent attempts to extend the thermodynamic theory of Markov chains to systems with unidirectional transitions \cite{busiello2020undirectional,Mazano2024Thermocomputation}, we instead quantify the degree to which the detailed balance condition is violated using Jensen-Shannon (JS) divergence \cite{nielsen2019jsd} instead of the KL-divergence. Whilst we are no longer quantifying the exact time-derivative of the entropy, the JS-divergence allows us to quantify the difference between forward and backward transition probabilities without the quantity diverging to infinity. We define our JS-divergence EPR to be,}
\begin{align}
\label{eq: JSDRW}
    \Phi & =\sum_{i\rightarrow j}P_{ij}\log \frac{P_{ij}}{\Tilde{P}_{ij}} + P_{ji}\log \frac{P_{ji}}{\Tilde{P}_{ij}},
\end{align}
where the sum is over all directed edges and $\bm{\Tilde{P}} = \frac{1}{2}(\bm{P}+\bm{P^{\top}})$ is the Hermitian, or averaged, distribution. We can directly verify that this quantity does not diverge as $\Tilde{P}_{ij}$ does not vanish on connected nodes. \Rev{As the EPR formula and the stationary distribution coincide for the DT and CTRW, we treat these models as equivalent in the subsequent analysis. However we note that in the special case of bipartite graphs, the CTRW remains ergodic and converges to the stationary distribution, whilst the DTRW does not converge unless the walkers are initialised according to the stationary distribution as the Markov chain is periodic \cite{Haggstrom2020MarkovChain}}. Importantly, unlike in the other models, directedness is not a sufficient condition to guarantee that the steady state is non-equilibrium and entropy producing, as will be shown later on.
\subsection{The network-based multivariate Ornstein-Uhlenbeck process}
\noindent The Ornstein-Uhlenbeck (OU) process in one dimension is a linear stochastic dynamical system modelling a particle in Brownian motion \cite{uhlenbeck1930brownian}. It can be extended to the multivariate
case which models a number of interacting particles \cite{Godreche2018OU}. In addition, it can be re-cast as a network dynamical system \cite{scwarze2021motifs}, with interactions being constrained to the weighted edges of a network, as will be considered here.\\\\
Consider a system of $N$ interacting particles, the multivariate OU process is given by the Langevin system,
\begin{align}
    \frac{d\bm{x}}{dt}&=-\bm{B}\bm{x}(t) + \bm{\xi}(t),
\end{align}
where $\bm{x}(t)\in \mathbb{R}^N$ is the time-dependent state vector, $\bm{B} \in \mathbb{R}^{N \times N}$ is the \textit{friction matrix}, and $\bm{\xi}(t)\in \mathbb{R}^N$ is additive white noise with covariance given by,
\begin{align}
\langle \bm{\xi}(t)\bm{\xi}^{\top}(t')\rangle = 2\bm{D} \delta(t-t'),
\end{align}
where $\bm{D} \in \mathbb{R}^{N \times N}$ is the \textit{noise covariance matrix} which is, by definition, symmetric.\newline\\
\noindent Given a weighted network, $\bm{W}$, we can constrain the interactions such that they occur along the edges of the network. In addition, we assume the additive noise to be applied independently to each node in the network. Under these assumptions, the OU is given by the Langevin system,
\begin{align}
\frac{d\bm{x}}{dt}&= \Theta(\gamma \Corr{\bm{W}^{\top}}-\bm{I})\bm{x}(t) + \bm{\nu}(t),
\end{align}
where $\bm{I}$ is the identity matrix, $\Theta$ is the reversion rate and $\gamma$ is the coupling parameter \cite{scwarze2021motifs}. As the thermal fluctuations are assumed to be uncorrelated in time and between nodes, the additive noise satisfies,
\begin{align}
\langle \bm{\nu}(t)\bm{\nu}^{\top}(t')\rangle = 2\sigma \bm{I} \delta(t-t'),
\end{align}
where $\sigma$ is the noise intensity. We relate the networked system to the generalised OU process with the following relations,
\begin{align}
\bm{B}&:= \Theta(\bm{I}-\gamma \Corr{\bm{W}^{\top}}),\\
\bm{D}&:= \sigma \bm{I}.
\end{align}
Returning to the generalised multivariate case, we will derive the EPR rate of the OU in a steady state (see Appendix \ref{sec: app: EPROU} and Ref. \cite{Godreche2018OU} for further details). If each eigenvalue of the friction matrix, $\bm{B}$, has positive real part, then, in the absense of noise, the system relaxes exponentially fast to $\bm{x}=0$. Therefore, in the presence of noise, the process will relax to a steady state with Gaussian fluctuations. Generally, this is a non-equilibrium steady state with the EPR being dependent on the matrices $\bm{B}$ and $\bm{D}$. The steady state is Gaussian with mean $\bm{x} = \bm{0}$ and steady state covariance given by,
\begin{align}
\bm{S} &=  \lim_{t\rightarrow \infty}\langle \bm{x}(t)\bm{x}^{\top}(t)\rangle.
\end{align}
It can be shown that $\bm{S}$ satisfies the following Sylvester equation \cite{zabczyk2020mathematicalcontrol} (see Appendix \ref{sec: app: EPROU}),
\begin{align}
\label{eq: sylvester1}
\bm{B}\bm{S} + \bm{S}\bm{B}^{\top}&=2\bm{D}.
\end{align}
The condition for the steady state to be reversible and in equilibrium is known to be \cite{lax1960fluctuationsnonequilibrium},
\begin{align}
\label{eq: reversiblecondition}
    \bm{B}\bm{D} = \bm{D}\bm{B}^{\top},
\end{align}
where the covariance is given by,
\begin{align}
    \bm{S}&=\bm{B}^{-1}\bm{D}.
\end{align}
In the case of the networked system, the reversibility condition becomes,
\begin{align}
    \bm{W}&=\bm{W}^{\top},
\end{align}
which corresponds to the interaction network being undirected. When condition (\ref{eq: reversiblecondition}) is not satisfied, the EPR, $\Phi$, can be written in the form,
\begin{align}
\Phi &=-\Tr (\bm{D}^{-1}\bm{B}\bm{Q}).
\end{align}
where $\bm{Q}=\bm{BS}-\bm{D}$. Therefore, given a matrix pair $\bm{B},\bm{D}$ - or correspondingly, a network, $\bm{W}$, and noise intensity, $\sigma$ - one has only to numerically solve the Sylvester equation (\ref{eq: sylvester1}) in order to calculate the EPR $\Phi$. We note that a range of numerical techniques and linear algebra packages exist for the accurate and efficient solution of equations of this type \cite{simoncini2016matrix}.
\subsection{The Ising model}
\noindent The final stochastic dynamical system we consider is the ubiquitous \textit{Ising spin-glass model} \cite{Nishimori2001statphys}. The Ising model with $N$ nodes, is made up of $N$ discrete spins that can take values in $\{+1,-1\}$, `up' and `down' spins respectively. We consider the system, in the absence of external fields, evolving in discrete time with either sequential or parallel spin updates. Given the state of the system, a spin configuration, $\bm{x}(t) = (x_{1}(t),...,x_N(t))$, the spins at time $t+1$ are updated as a discrete time Markov chain,
\begin{align}
\label{eq: conditionalising1}
    P(\bm{x}(t+1)|\bm{x}(t)) &= \prod_i \frac{\exp\left(x_i(t+1) \sum_j\Corr{W_{ji}}x_j(t)/T\right)}{2\cosh\left(\sum_j\Corr{W_{ji}}x_j(t)/T\right)},
\end{align}
where $T$ is the thermodynamic temperature and $\bm{W} = (W_{ij})$ are the pairwise coupling strengths defined by a weighted network. In the absence of external fields, the Ising model has reversible dynamics and is in equilibrium when the coupling strengths are symmetric i.e. $W_{ij}=W_{ji}$, corresponding to an undirected network, and irreversible dynamics when the coupling strengths are asymmetric, corresponding to a directed network \cite{huang2014asymmetric}. The joint transition probability between two spin configurations, or states, $\bm{y},\bm{z}$, is given by,
\begin{align}
    P_{\bm{y}\bm{z}} = P(\bm{x}(t+1)=\bm{z}, \bm{x}(t)=\bm{y}).
\end{align}
We recall the EPR for Markovian dynamics, given by,
\begin{align}
\label{eq: isingentropy}
    \Phi &= \sum_{\bm{y},\bm{z}}P_{\bm{y}\bm{z}}\log \frac{P_{\bm{y}\bm{z}}}{P_{\bm{z}\bm{y}}}.
\end{align}
Note that $\bm{z,y}$ are spin configurations, not nodes as in the RW. We can factorise the joint transition probabilities as follows,
\begin{align}
    P_{\bm{y}\bm{z}} = P_{\bm{z|y}}\cdot \pi_{\bm{y}},
\end{align}
where $P_{\bm{z|y}} = P(\bm{x}(t+1)=\bm{z}|\bm{x}(t)=\bm{y})$ is the conditional transition probability and $\pi_{\bm{y}}$ is the steady state probability of being in state $\bm{y}$. In order to estimate the EPR, we can estimate the steady state probabilities $\pi_{\bm{y}}$ using numerical sampling and use the conditional transition probability given by equation (\ref{eq: conditionalising1}) to calculate $\Phi$ using (\ref{eq: isingentropy}). However, we note that there are $2^N$ possible configurations for a system with $N$ spins. Therefore, as the system gets large, estimating the steady state probability becomes computationally challenging. For this reason, our analysis of the Ising model bisects into the case of small networks $(N\leq 10)$, where we can employ this approach, and the case of large networks $(N>10)$, where we require a mean-field approximation \cite{aguilera2021meanfield}.
\subsubsection{Small networks of Ising spins}
\noindent As mentioned above, for small networks of spins ($N\leq10$) we estimate the steady state probabilities by sampling from the model using Glauber dynamics \cite{Newman1999MonteCarlo}, a sequential spin update rule given by, 
\begin{align}
\label{eq: glauber}
    P[x_{i}(t+1)=1|\bm{x}(t)] = [1 + \text{exp}(-\frac{2}{T}\sum_j\Corr{W_{ji}}x_j(t))]^{-1}.
\end{align}
We note that Glauber dynamics simulate the Ising model with sequential, not parallel updates as described in equation (\ref{eq: conditionalising1}), but that these converge to the same steady state dynamics over time. After a sufficient burn-in period, we count the number of occurrences of each of the $2^N$ states and divide that by the total number of time-steps to given an estimate of the steady state probabilities, which we then use to estimate the EPR.

\subsubsection{Large networks of Ising spins}
\noindent The combinatorial explosion of state space with the number of spins, means that estimating the steady state probabilities is computationally challenging for large systems. Alternative approaches include coarse-graining the state space by clustering together states and estimating transition probabilities \cite{Esposito2012coarsegraining,roldan2010dissipation,lynn2021detailedbalance}, or using mean-field approximations \cite{aguilera2021meanfield}. Whilst using a coarse-grained state-space has been shown to capture the effect on the EPR of changing the temperature of the Ising model \cite{lynn2021detailedbalance}, it proved to be an inaccurate when varying the directedness of the network (see Appendix \ref{sec: app: coarsegraining}) and so we hereby opt for a mean-field approach.\\\\
In order to approximate the EPR of the system, we focus on two key statistical properties of the system,
\begin{widetext}
\begin{align}
    m_{i}(t)&=\sum_{\bm{x}(t)}x_{i}(t)P(\bm{x}(t)),\\
    D_{il}(t)&=\sum_{\bm{x}(t),\bm{x}(t-1)}\left[ x_{i}(t)x_{l}(t-1)P(\bm{x}(t),\bm{x}(t-1))-m_{i}(t)m_{l}(t-1)\right].
\end{align}
\end{widetext}
where $\bm{m}(t)=(m_1(t),...,m_N(t))$ is the average activation rate of the system
and $\bm{D}(t)$ is the delayed correlation matrix. These are sufficient statistics to define a particular Ising model \cite{aguilera2021meanfield}.\\\\
Given a particular network, $\bm{W}$, and the time-delayed correlations, $\bm{D}(t)$, we can calculate the time-dependent EPR, given by,
\begin{align}
\label{eq: isingentropy2}
    \Phi(t) &=\sum_{i,j} (\Corr{W_{ji}-W_{ij}})D_{ij,t}.
\end{align}
\begin{figure*}
\centering
\includegraphics[width=0.9\linewidth]{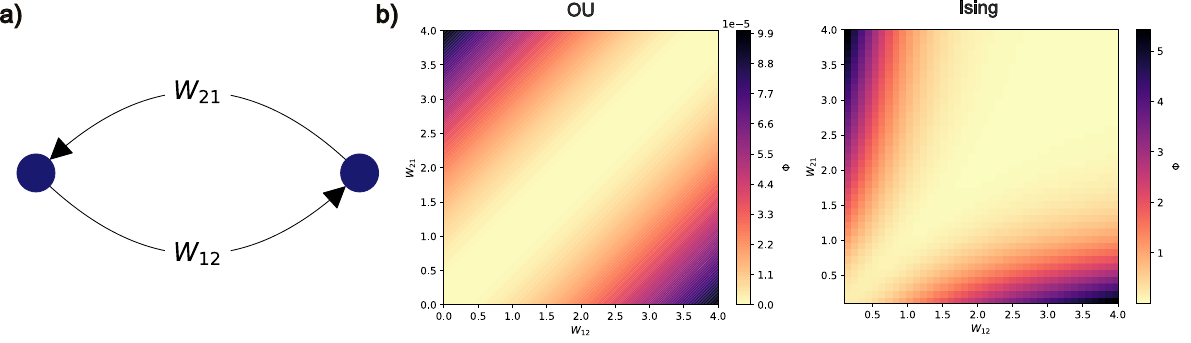}
\caption{\textbf{Entropy production in the 2-node network}: $a)$ We first consider a minimal directed network with only 2 nodes, where we can explore the entire space of networks by varying the interaction strengths. $b)$ \textbf{Left:} The exact EPR of the OU as given in equation (\ref{eq: entropyOUP2}) for different 2-node networks. The positive diagonal represents undirected networks which do not produce entropy. Further from this line, the EPR is higher. \textbf{Right:} The approximated EPR of the Ising model also increases as the network becomes more directed, but with a non-linear relationship.}
\label{fig: 2nodes}
\end{figure*}\\
We note that $\Phi(t)=0$ if the network is undirected i.e. $W_{ij}=W_{ji}$. We will estimate the average activation rate and delayed correlations using the so-called \textbf{classical naive mean field} (NMF) given by,
\begin{align}
    m_{i}(t)&\approx \tanh \sum_j \Corr{W_{ji}}m_{j}(t-1),\\
    D_{il}(t)&\approx \Corr{W_{li}}(1-m^2_{i}(t))(1-m^2_{l}(t-1)).
\end{align}
For a derivation see Appendix \ref{sec: app: nMF}. Beginning with all spins set to 1, for a given network, we can use the NMF to approximate the time-delayed correlations and use equation (\ref{eq: isingentropy2}) to estimate the time-dependent EPR. \Rev{As the NMF approximates the Ising model at stationarity, the EPR converges to a time-independent value $\Phi$}.
\section{Entropy production in synthetic hierarchical networks}
\label{sec: entropydirected}
\subsection{An exactly treatable case: 2-node networks}
\noindent Before considering large directed networks, we first consider a simple 2-node network with asymmetric coupling, as shown in Figure \ref{fig: 2nodes}.\\\\The 2-node directed network is defined by the weight matrix,
\begin{align}
    \bm{W} = \begin{bmatrix}
0 & W_{12}\\ 
W_{21} & 0
\end{bmatrix}. 
\end{align}
\noindent This network has only two free parameters and so we can explore the space of directed networks and see how the directedness and EPR change. Firstly, the irreciprocity of this network is given by,
\begin{align}
    1-r& = \frac{|W_{12}-W_{21}|}{W_{12}+W_{21}}.
\end{align}
The trophic levels, $\bm{h}$, are given by the solution to the equation,
\begin{align}
    \begin{bmatrix}W_{12}+W_{21} & 0\\ 
0 & W_{12}+W_{21}
\end{bmatrix}\begin{bmatrix}
h_1\\ 
h_2
\end{bmatrix}=\begin{bmatrix}
W_{21}-W_{12}\\ 
W_{12}-W_{21}
\end{bmatrix},
\end{align}
which has general solution,
\begin{align}
    h_1-h_2=\frac{W_{21}-W_{12}}{W_{21}+W_{12}},
\end{align}
and trophic directedness,
\begin{align}
    \sqrt{1-F_0} &= \frac{|W_{12}-W_{21}|}{W_{12}+W_{21}},
\end{align}
which coincides with the irreciprocity. The Henrici and normalised Henrici indices are given by,
\begin{align}
    d_H & = |W_{12}-W_{21}|,\\
    \hat{d}_H&=\frac{|W_{12}-W_{21}|}{\sqrt{W_{12}^2+W_{21}^2}}.
\end{align}
\noindent Each of the measures somehow captures the asymmetry between pair of weights, thereby quantifying the directedness, but with unique nuances. The irreciprocity and trophic directedness are normalised by the $l_1$ norm of the matrix whilst the normalised Henrici index uses the $l_2$ norm and the Henrici index is unnormalised. Next we consider the dynamics on this network, beginning with the RW. The DTRW does not converge to a steady state on the 2-node network unless the walkers are originally distributed with the steady state distribution. This is due to the fact that the network is bipartite meaning at each time-step a random walker can only move to the other node. The transition matrix is given by,
\begin{align}
    \bm{T} = \begin{bmatrix}
0 & 1\\ 
1 & 0
\end{bmatrix}.
\end{align}
Therefore, $\bm{\pi}=\left ( \frac{1}{2},\frac{1}{2}\right)$ is the Perron left eigenvector and the steady-state. Yet, despite the directedness of the network, this steady state clearly preserves detailed balance as $P_{12}=P_{21}=\frac{1}{2}$. \RevTwo{Moreover, whilst this 2-node example is directed but normal, normality is not a necessary condition for the random walk to obey detailed balance as shown by the counter-example in Appendix \ref{sec: app: trophicdirectednessnn}}. Next we consider the OU process. We note that the eigenvalues of the associated friction matrix,
\begin{align}
    \bm{B}=\Theta(\bm{I}-\gamma \Corr{\bm{W}^{\top}}),
\end{align}
are given by the pair,
\begin{align}
    \lambda_{\pm}&=\Theta \pm \Theta \gamma \sqrt{W_{12}W_{21}}.
\end{align}
Therefore, to guarantee decay to a steady state, we enforce the condition,
\begin{align}
    \gamma<\frac{1}{\sqrt{W_{21}W_{12}}}.
\end{align}
\begin{figure*}[t]
    \centering
\includegraphics[width=\linewidth]{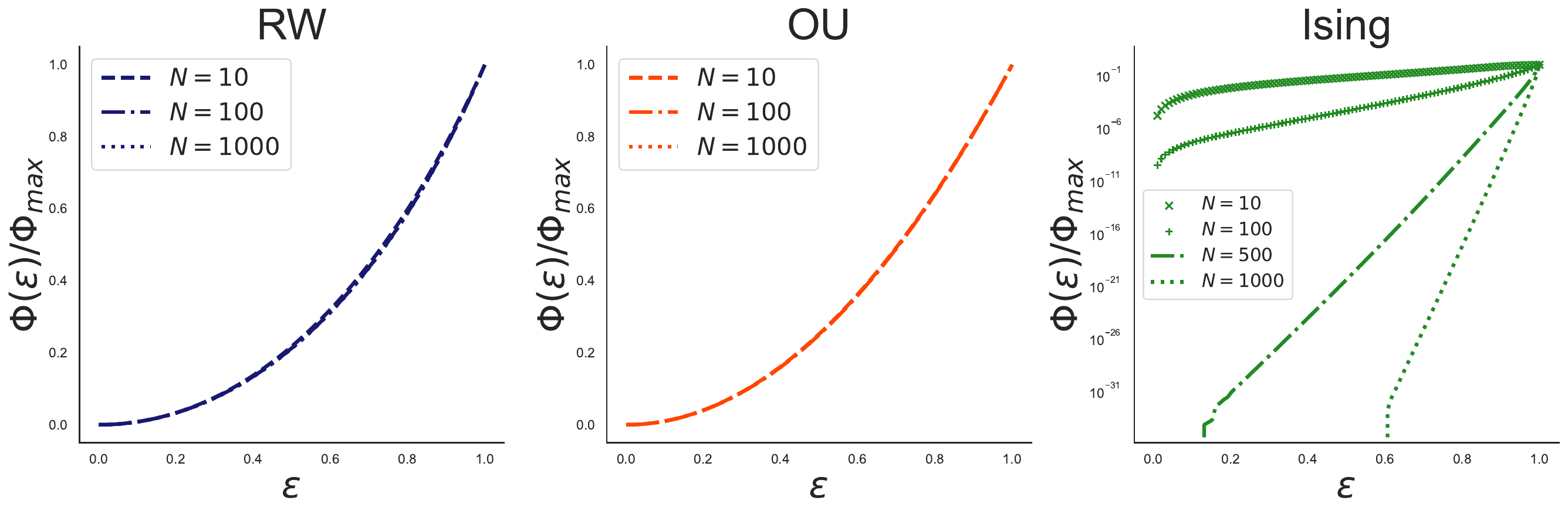}
    \caption{\textbf{\Corr{Entropy production rate in interpolated synthetic networks}}:
\Rev{Normalised EPR, $\Phi(\epsilon)/\Phi_{\text{max}}$, in the RW, OU and Ising systems as a function of the directedness parameter $\epsilon$. \textbf{Column 1} The first panel show the results for the RW. Across network sizes, $\Phi(\epsilon)/\Phi_{\text{max}}$ scales quadratically in $\epsilon$.}
\textbf{Column 2:} The second column shows the results for the OU process. The normalised EPR, $\Phi(\epsilon)/\Phi_{\text{max}}$, increases quadratically in $\epsilon$, as found analytically in the 2-node network. 
\textbf{Column 3:} The final column shows the result for the NMF Ising approximation. The normalised EPR, $\Phi(\epsilon)/\Phi_{\text{max}}$, increases faster than quadratically and approaches an exponential increase as $N\rightarrow \infty$. Note that the $y-$axis is logarithmic from 0 to 1 for this pane.}
\label{fig: interpolation entropy}
\end{figure*}\\
\noindent Handily, the Sylvester equation (\ref{eq: sylvester1}) is explicitly solvable for systems of size $N=2$ and so the EPR has a closed form solution \cite{Godreche2018OU}. For the 2-node network, the EPR becomes,
\begin{align}
\label{eq: entropyOUP2}
    \Phi &=\frac{\Theta \gamma^2}{2}(W_{12}-W_{21})^2,
\end{align}
a function of the difference,  $|W_{12}-W_{21}|$, between the weights of the pair of edges. Importantly, this means the EPR increases quadratically with the (unnormalised) Henrici index, a relationship that we will show to hold consistently across experiments for the OU. Furthermore, the EPR is not normalised with respect to the total strength of the network. Trivially, when the connection is symmetrical, $W_{12}=W_{21}$, the system preserves detailed balance.\\\\
Finally, we can consider the Ising dynamics on this network. Whilst we do not have the exact solution, there are only four possible spin configurations in this system, therefore we can accurately estimate the steady state probabilities and consequently the EPR. The EPR, $\Phi$, for both the OU and Ising systems is plotted as a function of the network weights $(W_{12},W_{21})$ as shown in Figure \ref{fig: 2nodes}. The positive diagonal $W_{12}=W_{21}$ represents the subset of undirected networks, which preserve detailed balance. In both cases, as the asymmetry of the network increases the EPR increases. The EPR in the OU is a function purely of the distance between the two weights, whereas for the Ising model, the relationship appears to be non-linear.
\subsection{Entropy production in parameterised synthetic networks}
\noindent Equipped with a range of stochastic network dynamics where we can estimate the EPR, as well as a mechanism to systematically vary the directedness of the network, we can investigate the effect of directedness on the degree of non-equilibrium. Considering networks of sizes $N=$ 10, 100, 500 and 1000, we generate hierarchical networks using preferential attachment (see Appendix \ref{sec: app: PA}) and then interpolate between these networks and their Hermitian as described in Section \ref{sec: measures}. Recall the interpolation is given by,
\begin{align}
    \bm{\hat{W}}(\epsilon) & = (1-\epsilon)\bm{\Tilde{W}} + \epsilon\bm{W},
\end{align}
for $\epsilon\in [0,1]$, with $\epsilon=0$ corresponding to an undirected network and $\epsilon=1$ being maximally directed. Recall further, that Figure \ref{fig: interpolation} showed that all four measures of asymmetry scaled linearly or almost linearly with the parameter $\epsilon$. Therefore we can consider $\epsilon$ to be a normalised (relative) measure of directedness. Figure \ref{fig: interpolation entropy} shows the results of varying $\epsilon$ for all three dynamics and for different network sizes. Clearly, in each case, the directedness $\epsilon$ causes a non-linear increase in the relative EPR that remains consistent as the network size increases. Furthermore, there are clear similarities between dynamics. \Rev{Firstly, we consider the two leftmost panels, corresponding to the RW and OU dynamics. We see that the normalised EPR $\Phi(\epsilon)/\Phi_{\text{max}}$ increases quadratically in $\epsilon$. In fact, this quadratic increase in $\epsilon$ can be proven to be exact in the special cases of the OU process on 2-node and circulant networks (see Appendices \ref{sec: app: 2node} and \ref{sec: app: circulant})}. \begin{figure*}[t]
    \centering
    \includegraphics[width=\linewidth]{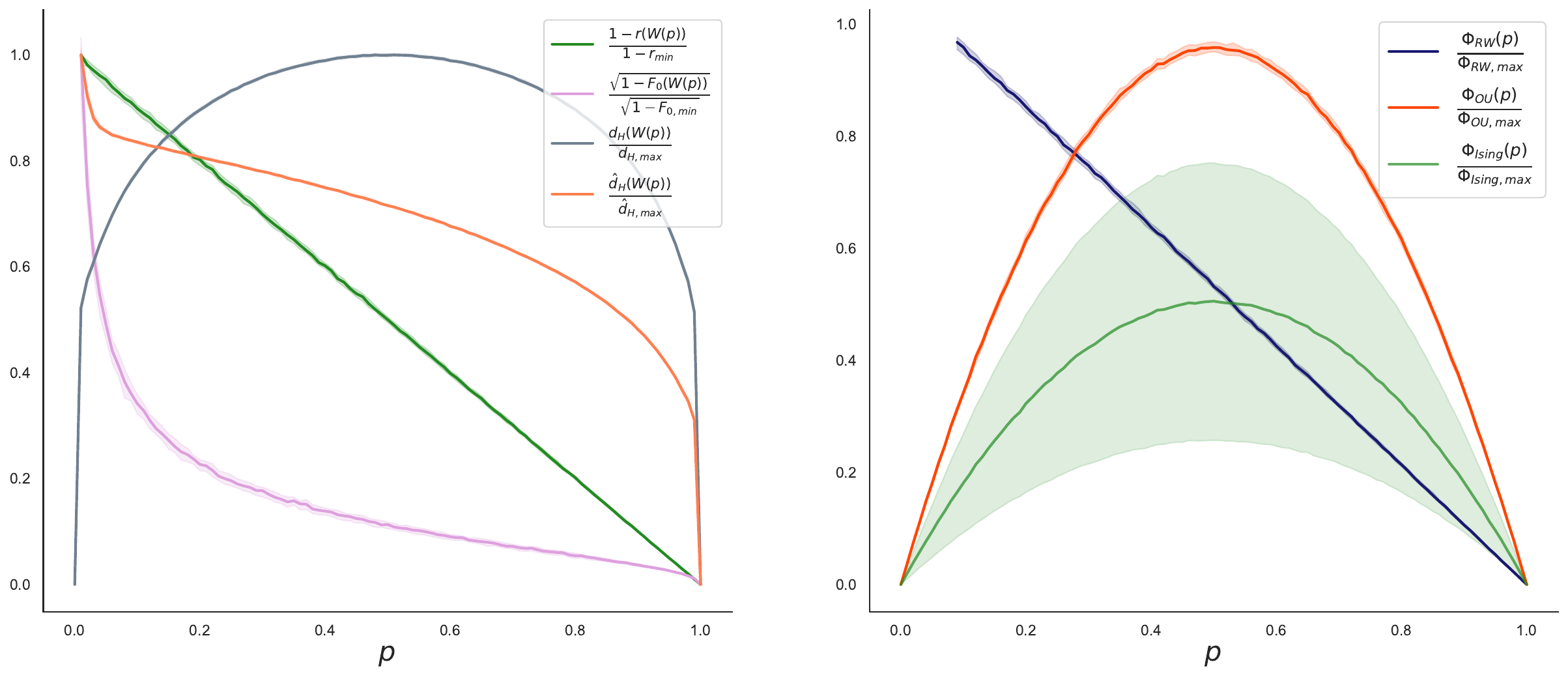}
    \caption{\textbf{Directedness and entropy production rate in Erdős–Rényi graphs}: 
\textbf{Left:} This plot shows the mean, with standard deviation shading, of normalised directedness measures as a function of the ER parameter $p$ for $N=100$ over 100 samples. We can see that the irreciprocity, trophic directedness and normalised Henrici index decrease as a function of $p$, with different decay rates. On the other hand, the Henrici index has a parabolic shape and peaks at $p=0.5$. \textbf{Right:} This plot shows the mean, with standard deviation shading, of the EPR, normalised by the maximum of the EPR $\Phi$, for the three systems as a function of the ER parameter $p$ for $N=100$ over 100 samples. The RW curve follows the irreciprocity whilst both the OU and Ising curves peak at $p=0.5$, closely following the Henrici index of the network. The Ising curve shows very large variance at each value of $p$ suggesting that small structural differences lead to large differences in $\Phi$ and that the model is potentially close to the critical temperature.}
\label{fig: ER entropy}
\end{figure*}
In the final panel, we show the results for the NMF Ising model. The EPR also increases with $\epsilon$, but this increase is faster than quadratic (note the $y-$axis is logarithmic) and approaches an exponential increase as $N\rightarrow\infty$. Across both dynamics and system size, \Rev{an increase in the directedness of the network is associated with a larger violation of detailed balance and an increase in the EPR of the system}.
\section{Entropy production and directedness in Erdős–Rényi graphs}
\label{sec: er}
\noindent In the previous section, we used the interpolation to vary the directedness of a given network. We were able to show that the EPR scales, non-linearly, with the relative directedness of the network for all the considered dynamics. However, as all the directedness measures increased (almost) linearly in $\epsilon$, it is unclear which `type' of directedness is \Rev{most closely associated with} the violation of detailed balance and the EPR. In an effort to decouple the different measures of directedness, we consider a different approach to generating networks. Instead of specifically generating hierarchical, directed networks and interpolating, as before, we now utilise directed Erdős–Rényi graphs as a null network model \cite{erdos1959random}. A directed Erdős–Rényi (ER) graph, $G(N,p)$, is a randomly sampled, unweighted network with $N$ nodes. Each directed pair of nodes $i \rightarrow j$ is connected with independent probability $p\in [0,1]$. The ER generating process does not assume or enforce a hierarchical structure, however hierarchical directedness can emerge spontaneously. For each value of $p$, we randomly sample ER graphs and measure their directedness with each measure. In addition, we measure the EPR of each of the three dynamics on these networks.\\
\\
Figure \ref{fig: ER entropy} shows the behaviour of the directedness measures as a function of the ER parameter $p$, normalised by the maximum value across all trials, (left pane) as well as the behaviour of the EPR for each of the three systems as a function of the ER parameter $p$ (right pane), normalised by the maximum of the median value, $\hat{\Phi}$. For these calculations, we consider $N=100$ and sample 100 graphs at each value of $p$. The figure shows the mean, with standard deviation shading, over the 100 samples. Firstly, we note the different profiles of the directedness measures. The unique behaviour of each measure shows we have been able to decouple the different notions of directedness, allowing us to identify which one is closely correlated with the EPR. The irreciprocity, trophic directedness and normalised Henrici index all decay as a function of $p$ but with different decay rates. On the other hand, the Henrici index has a parabolic shape, peaking at $p=0.5$. We can compare this to the different profiles of each of the dynamics in the right pane. \Rev{The RW curve decays linearly in $p$ mirroring the curve of the irreciprocity}. We note that $\Phi_{\text{RW}}$ is only defined when the network is strongly connected and so is only calculated for samples generated with $p>0.1$ where we checked that the sampled graph was strongly connected. Both the OU and Ising curves follow the parabolic shape of the Henrici index, peaking at $p=0.5$. \Rev{The Ising curve show far greater variance at each value of $p$ suggesting that the system is more susceptible to small structural changes in the network or that the model is operating close to the critical temperature, where the EPR is maximised \cite{martynec2020entropycriticality,aguilera2023sherrington}}.\\\\
\Rev{The RW is a process that only sees the network locally. Therefore the dynamics, and by extension the EPR, depend only on the directedness of individual edges in or out of the walker's momentary position. As a result, the EPR is strongly correlated with the irreciprocity, which strictly measures differences between the forward and backward connection strength. Furthermore, the RW dynamics are invariant to scaling of the network weights i.e. the dynamics of the process evolving on $\bm{W}$ is the same as that of $\alpha \bm{W}$ for $\alpha>0$. Therefore, the fact that the irreciprocity is a normalised measure also explains the strong correlation.} On the other hand, the OU and Ising dynamics, are complex interacting systems where, at any given time-point, the entire network structure is influencing the dynamics. Furthermore, they are not invariant to scaling of the network. As a result, the unnormalised Henrici's index, measuring non-normality, appears to be most strongly correlated with the EPR of the systems, indicating that the global hierarchy, as well as the magnitude of the weights, of the network are responsible for driving the interacting dynamics from equilibrium.
\section{Entropy production and directedness in real world networks}
\label{sec: realworld}
\noindent In this section, we consider a dataset of 97 real-world directed networks from ecology, sociology, biology, language, transport and economics \cite{obrien2021hierarchical}. For a full description of the dataset see Appendix \ref{sec: app: networkdata}. We measure the directedness of these networks with the four measures and then consider the three dynamical systems evolving on these networks and measure the EPR. Furthermore, we differentiate between networks from different domains, and plot the correlations between the directedness measures and the EPR. For real-world networks, the irreciprocity, trophic directedness and non-normality are all strongly positively correlated (see Appendix \ref{sec: app: trophicdirectednessnn} and Refs. \cite{mackay2020directed,johnson2020digraphs,asllani2018nonnormal}). In addition, most directedness is trophically or hierarchically organised, indicating a lack of `loop-like' structures, that are asymmetric but not hierarchical, in real-world networks \cite{bianconi2008localstructuredirected,johnson2017loops}. Next, by considering the three dynamical systems of interest evolving on the real world networks, we measure the EPR and plot the correlations with the different measures of directedness.\\
\begin{figure*}
    \centering
\includegraphics[width=\linewidth]{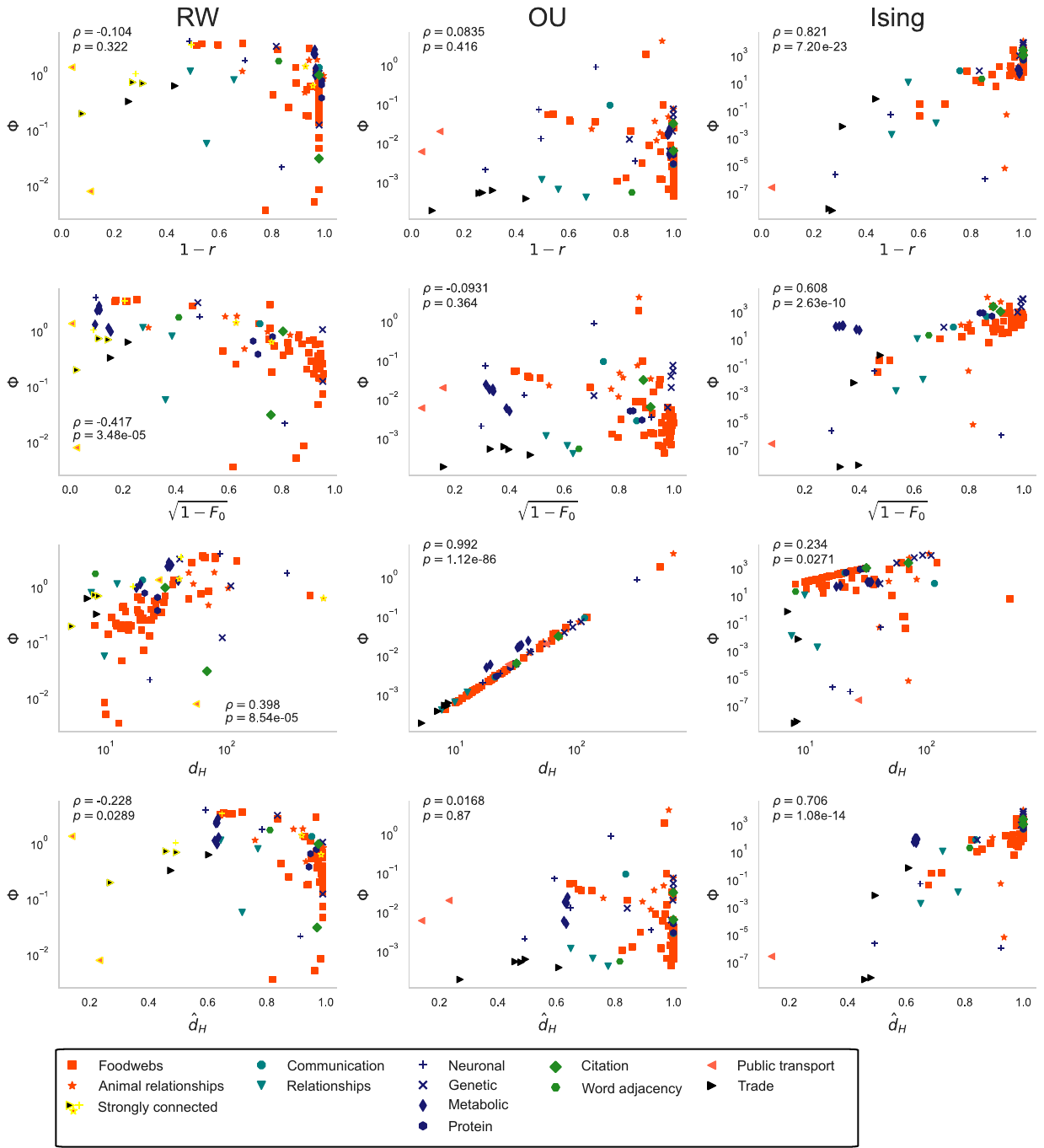}
    \caption{\textbf{\Corr{Entropy production rate and directedness in real-world networks}}:
We plot the EPR of three dynamical systems against four directedness measures on real world networks from twelve fields. The first column shows the EPR of the RW. \Rev{These dynamics are restricted to 95 weakly connected networks. We use a perturbation to render the networks strongly connected and guarantee a stationary distribution. The networks which are already strongly connected, and thus remain unperturbed, are denoted by a yellow border. The second column shows the EPR of the OU and the third the EPR of the Ising model which consider all 97 networks. The RW and OU models are significantly positively correlated with the unnormalised Henrici index whilst the Ising model is significantly positively correlated with all four measures. The OU model is uncorrelated with the other measures indicating that normalised measures do not accurately predict the EPR when comparing across network sizes, weight distributions and substructures. The RW model is weakly negatively correlated with normalised measures due to the spread in the EPR for the almost perfectly directed networks, that is not captured by the normalised measures.} Ecological networks are in orange, social in teal, biological in blue, language in green, transport in red and economic in black. The Pearson correlation coefficient is denoted by $\rho$ and its significance by $p$, \RevTwo{using a two-sided $t-$test assuming uncorrelated samples}.}
\label{fig: RW entropy}
\end{figure*}
\\
\Rev{The RW dynamics are only guaranteed to converge to a steady-state on strongly connected networks. This dataset contains only 9 strongly connected networks and many of the 95 weakly connected networks are almost DAGs meaning the largest strongly connected component contains a small fraction of the nodes. In order to analyse as many real-world networks as possible for the RW dynamics, we add a small reciprocal edge to the weakly connected networks to render them strongly connected. Given a weakly connected network with weight matrix $\bm{W}$, we define the perturbed weight matrix $\bm{W}_{\delta}$, by,
\begin{align}
    W_{\delta,ij} = \left\{\begin{matrix}
W_{ij} & \text{if } W_{ij}\neq0\\
\delta & \text{if } W_{ij}=0 \text{ and } W_{ji}\neq 0\\
0 & \text{otherwise}
\end{matrix}\right..
\end{align}
In this section, we consider $\delta = 0.01$. Such a perturbation could result in changes in the directedness of the networks, but Figure \ref{fig: RW directedness} in Appendix \ref{sec: app: trophicdirectednessnn} shows that this effect is minimal. For the networks which are already strongly connected, we do not apply a perturbation. Such unperturbed networks are highlighted by a yellow border in columns 1 \& 2 of Figure \ref{fig: RW entropy}. The results presented in Figure \ref{fig: RW entropy} show the EPR of each system against each of the four directedness measures. For the RW and the OU model there is a significant positive correlation between the unnormalised Henrici index the EPR and the OU model has poor correlation with the other measures. This indicates that the size, weight distribution and substructure of the network play a role in determining the EPR. The unnormalised Henrici index is able differentiate between highly directed networks of different sizes and weight distributions whilst the normalised measures group the many DAG-like networks, which all have almost perfect irreciprocity, trophic directedness and normalised Henrici index, together. Nevertheless, these networks display a range of different EPR values. This results in a negative correlation between the RW and the normalised measures. On the other hand the Henrici index is able to spread these networks over a larger range of value resulting in strong, significant correlations. In previous experiments, using the interpolation or the ER null model, network substructure, size and weight distribution was not varied with the complex heterogeneity seen in real-world network data from diverse fields. The almost perfect correlation between the EPR of the OU model and the Henrici index further supports our analytical findings in the 2-node solvable model and the numerical simulations for synthetic networks in Sections \ref{sec: entropydirected} \& \ref{sec: er}. Finally the rightmost column shows the results for the Ising model. The Ising model is strongly positively correlated with the normalised measures of directedness and significantly with all measures. Whilst the previous section suggested that the Henrici index should be the most strongly correlated to the EPR in the Ising model, the high variance in the EPR also indicated that the Ising model could be susceptible to small changes in network structure, or proximity to criticality, which could explain noisier relationships in the real-world networks.\\\\
The analysis on real-world networks indicates a link between directedness and broken detailed balance in dynamical processes. However, unlike in the previous experiments, this result highlights the role of overall network size and weight-distribution on the EPR through the failure of unnormalised measures of directedness to predict the EPR of the dynamics, in particular for the RW. On the other hand, the results show that the Henrici index is strongly correlated with the EPR, particularly for the OU model, indicating the effectiveness of unnormalised directedness measures in capturing the asymmetries that result in broken detailed balance and increased EPR. Nevertheless, the analysis considers synthetic models on the real-world topologies, where the models do not necessarily reflect the dynamics associated with that particular system. Furthermore, structural network data represents a complex system at a single chosen resolution and is often noisy, containing missing and spurious links, which can influence the results of simulations \cite{Newman2018noisy}. As a consequence, the results in this section aim to examine the role of complex, real-world topologies on the link between directedness and EPR but do not allow for conclusions to be drawn on the true dynamics of the systems the networks represent. \RevTwo{Furthermore, real-world networks contain many other topological features, for example modularity or small-worldness, whose contribution to the EPR is unknown. To investigate the significance of directedness in particular, we perform null-modelling on a subset of empirical networks using the (directed) configuration model \cite{newman2001randomgraphs} in Appendix \ref{sec: app: configuration}}. In order to reach more concrete conclusions about the role of directed network structure on non-equilibrium in real-world dynamics, one must consider empirical dynamic data, in the form of multivariate time-series, alongside structural network data.} 
\section{Entropy production and directed networks from multivariate time-series}
\label{sec: timeseries}
\noindent In the previous section, we considered a plethora of real-world structural network data from a diverse range of fields and \Rev{assumed synthetic dynamics evolving on the networks. However, such synthetic models are not necessarily representative of the dynamics associated with those particular systems}. Furthermore, in many complex systems, such (directed) structural network data is either unavailable or of less interest than empirical spatiotemporal data of node activities in the form of a multivariate time-series (MVTS). Extracting information about the structural or dynamical organisation of a system from the complex patterns in a MVTS is a ubiquitous problem in complex systems science. \Rev{In addition, the problem of identifying non-equilibrium, irreversible dynamics in experimental data is a challenging one \cite{seifert2019inference}. Nevertheless, a range of computational methods exist to quantify broken detailed balance in time-varying data \cite{Seif2021machinelearning, lynn2021detailedbalance,Frishman2020stochasticforce,diterlizzi2024variancesum,battle2016brokendetailedbalance,martinez2019inferring,lucente2022incomplete}. However, many such methods apply to low-dimensional data stemming from biochemical experiments and cannot be scaled to the MVTS emerging from complex networked systems \cite{Frishman2020stochasticforce,martinez2019inferring,battle2016brokendetailedbalance}. Three approaches that can be applied to MVTS are machine-learning \cite{Seif2021machinelearning}, coarse-graining \cite{lynn2021detailedbalance} and the fitting of linear systems \cite{gilson2023OU,benozzo2023linearstatespace}. The machine-learning approach requires much fine-tuning, lots of data and is very specific to each system. Moreover, a coarse-graining approach leads to a poor lower-bound on the true EPR of a system \cite{Esposito2012coarsegraining}. As a consequence, it yields an inaccurate measure of the relative EPR (as shown for the Ising model in Appendix \ref{sec: app: coarsegraining}). Concurrently, a range of methods for inferring directed network structure from data have been developed \cite{timme2014inference,cliff2023pairwise,novelli2019multivariatetransferentropy,zou2009grangercausality,Gilson2016EC,Gilson2020EC,villaverde2014mider}. Some methods require the ability to perturb the system and observe a response \cite{timme2014inference}. Many others employ a directed correlation measure, such as transfer entropy or Granger causality, to infer a \textit{directed functional network} \cite{cliff2023pairwise,novelli2019multivariatetransferentropy,zou2009grangercausality,villaverde2014mider}. Borrowing terminology from network neuroscience, so-called functional networks, obtained from correlation measuring applied to an MVTS, differ conceptually from the structural interactions considered in this paper \cite{friston2011FCEC}. In this section, we will present a simple, intuitive method for both estimating the EPR and inferring a directed structural network from a MVTS using a linear model fit with autoregression \cite{Shumway2017timeseries}}. We fit a first-order multivariate autoregressive model and then associate this model to a corresponding OU process where the EPR can be calculated explicitly \cite{Godreche2018OU}. Furthermore, this fitting will infer a network of interactions between the variables, under the assumed model, whose directedness can be measured. In addition to calculating the overall directedness, we can extract the trophic levels as described in Section \ref{sec: measures} in order to unravel the hierarchical organisation of the system. We will apply this approach to MVTS from human neuroimaging data from the Human Connectome Project (HCP) and stock-prices from the New York Stock Exchange (NYSE) to investigate the relationship between broken detailed balance and structural directedness in dynamic data.\\
\\
\Rev{The key limitation of such an approach is the assumption of the linear model for potentially non-linear data. However, previous studies have found that highly non-linear systems, such as large-scale brain dynamics \cite{nozari2023linearbrain}, or even chaotic systems \cite{brunton2017chaos}, are well approximated by linear models}. Furthermore, autoregressive models have previously found success in a spectrum of areas including finance and economics \cite{Shumway2017timeseries}, neuroscience \cite{harrison2003autoregressionfmri} and beyond. Finally, we note that three recent studies in neuroimaging have considered related approaches, fitting linear models to neural recordings in order to quantify the EPR \cite{gilson2023OU} or the asymmetry of interactions \cite{benozzo2023linearstatespace,deco2023violations} \Rev{but that inferring time-reversibility from incomplete linear data may yield inaccuracies \cite{lucente2022incomplete}}.
\subsection{Linear auto-regression of multivariate time-series}
\noindent Consider an $N-$dimensional MVTS of signals of the form $\bm{X}(t_i)= \{X_1(t_i),...,X_N(t_i)\}$ recorded at equispaced time-points $t_i\in \{t_0,t_1,...,t_T\}$. We assume such signals are discrete, finite observations of either a generalised or network-based OU process,
\begin{align}
\label{eq: linearmodel}
    \frac{d\bm{x}}{dt}&=-\bm{B}\bm{x}(t) + \bm{\xi}(t),\\
    \frac{d\bm{x}}{dt}&= (\Corr{\bm{W}^{\top}}-\bm{I})\bm{x}(t) + \bm{\nu}(t),
\end{align}
with additive noise satisfying,
\begin{align}
\langle \bm{\xi}(t)\bm{\xi}^{\top}(t')\rangle &= 2\bm{D} \delta(t-t'),\\
\langle \bm{\nu}(t)\bm{\nu}^{\top}(t')\rangle &= 2\sigma \bm{I} \delta(t-t'),
\end{align}
respectively, as previously defined in Section \ref{sec: dynamics}.\\\\
In order to find the parameters, $\bm{B}, \bm{D}$ or $\bm{W}, \sigma$, that best explain our observed data, we fit a first-order linear multivariate auto-regressive (MAR) model of the form,
\begin{align}
    \bm{X}(t_{i+1}) = \bm{A}\bm{X}(t_i) + \bm{\chi}(t_i),
\end{align}
where $\bm{A}$ is calculated using least-squares auto-regression and $\bm{\chi}$ is a MVTS of \textit{residuals}. In order to associate the MAR to an OU process, we discretise the continuous-time process with a one-step scheme. Whilst any such discretisation can be applied, we proceed with a Euler-Maruyama discretisation \cite{kloeden1992sdes}, with time-step $\Delta t$ and  obtain,
\begin{align}
\bm{x}(t_{i+1}) &= [\bm{I}-\Delta t\bm{B}]\bm{x}(t_i) + \bm{\Lambda}\bm{\eta}_i,\\
    \bm{x}(t_{i+1}) &= [\bm{I}+\Delta t(\bm{\Corr{W^{\top}}-I})]\bm{x}(t_i) + \bm{\Gamma}\bm{\zeta}_i,
\end{align}
where $\bm{\Lambda}\bm{\Lambda}^{\top} =2\bm{D}$, $\bm{\Gamma}\bm{\Gamma}^{\top} = 2\sigma \bm{I}$ and $\bm{\eta}_i, \bm{\zeta}_i$ are $N-$dimensional independent, identically distributed Gaussian vectors with independent components, each with mean $0$ and variance $\Delta t$. Thus we can associate the discretised OU process to the MAR model using the following relations,
\begin{align}
\bm{A} &= \bm{I}-\Delta t(\bm{B}),\\
    \bm{A} &= \bm{I}+\Delta t(\bm{\Corr{W^{\top}}-I}).
\end{align}
Furthermore, in the case of the network-restricted model, we modify the auto-regressive algorithm to use non-negative least-squares in order to guarantee that $\bm{W}$ is restricted to non-negative entries (see Appendix \ref{sec: app: autoreg}) \cite{Chen2009nnls}. In order to estimate $\bm{D}$ or $\sigma$, we take the covariance of the residual time-series $\bm{\chi}(t)$ and note that,
\begin{align}
    \Cov[\bm{\chi}] & = 2 \Delta t \bm{D},\\
    \Cov[\bm{\chi}] & = 2 \Delta t \sigma \bm{I},
\end{align}
depending on the assumed model. We can estimate $\sigma \approx \langle \frac{1}{2\Delta t} \text{diag}\Cov[\bm{\chi}]\rangle $, where $\langle \cdot \rangle $ is the mean. We note that $\Delta t$ represents the time-scale of the process and is not discernible directly from the time-series data but that $\Delta t <<1$ is an assumption of the discretisation. In the following, we take $\Delta t =0.1$. Once $\bm{B}, \bm{D}$ or $\bm{W}, \sigma$ have been obtained, if $-\bm{B}$ or $\Corr{\bm{W}^{\top}-\bm{I}}$ are stable matrices, one can use the analysis presented in Section \ref{sec: dynamics} to calculate the EPR of the unique steady-state and quantify the degree to which we have broken detailed balance in the time-series. Furthermore, the directedness of the effective network, $\bm{W}$, can be analysed using the measures presented in Section \ref{sec: measures}.
\subsection{Applications to real-world multivariate time-series}
\subsubsection{Human neuroimaging at rest and during task}
\noindent We first apply this approach to neuroimaging data from the HCP \cite{vanessen2013hcp}. We consider BOLD fMRI in the Desikan-Killany (DK80) parcellation \cite{desikan2006dk80} with 62 cortical regions and 18 sub-cortical regions taken from the same 100 (unrelated) participants at rest and during a social and motor task. For further details on the experimental paradigms see Ref. \cite{vanessen2013hcp}. The data was pre-processed following standard HCP protocols and is further described in Ref. \cite{deco2021workspace} and Appendix \ref{sec: app: time-series}. For each participant in each condition, we apply the method described in the previous section to fit both a general linear model and a network-restricted model using auto-regression. We then measure the EPR from the general model and the directedness from the effective network extracted in the restricted model. In addition, we partition the nodes of the network into the 7 canonical Yeo functional sub-networks, each of which is associated with a specific aspect of brain function \cite{yeo2011restingnetworks} (see Appendix \ref{sec: app: time-series}).\\\\ Figure \ref{fig: brainnets} shows the results of the analysis applied to the neuroimaging data. Panel $a)$ shows that the EPR is elevated in task conditions when compared to rest, which coincides with results from previous studies \cite{lynn2021detailedbalance,deco2022insideout,deco2023tenet,deco2023violations}. However, we can also see that the interaction network becomes significantly more directed in the task states which \Rev{is associated with} the increase in the EPR. Panel $b)$ shows that the EPR is \Rev{significantly} positively correlated with the directedness using all measures. Going beyond aggregate quantities, panel $c)$ shows the trophic organisation of a participant-averaged effective network coloured according to the Yeo partition. This representation allows us to see where each sub-network sits in the overall hierarchy and how the network is reorganised by the task stimulus. Similarly, we can plot the mean trophic level of each sub-network in each state, as shown in panel $d)$. We can see that during tasks some sub-networks re-position themselves higher in the hierarchy, such as the dorsal attention network, whilst others are re-positioned lower, such as the sub-cortical regions.
\begin{figure*}
    \centering
    \includegraphics[width=\linewidth]{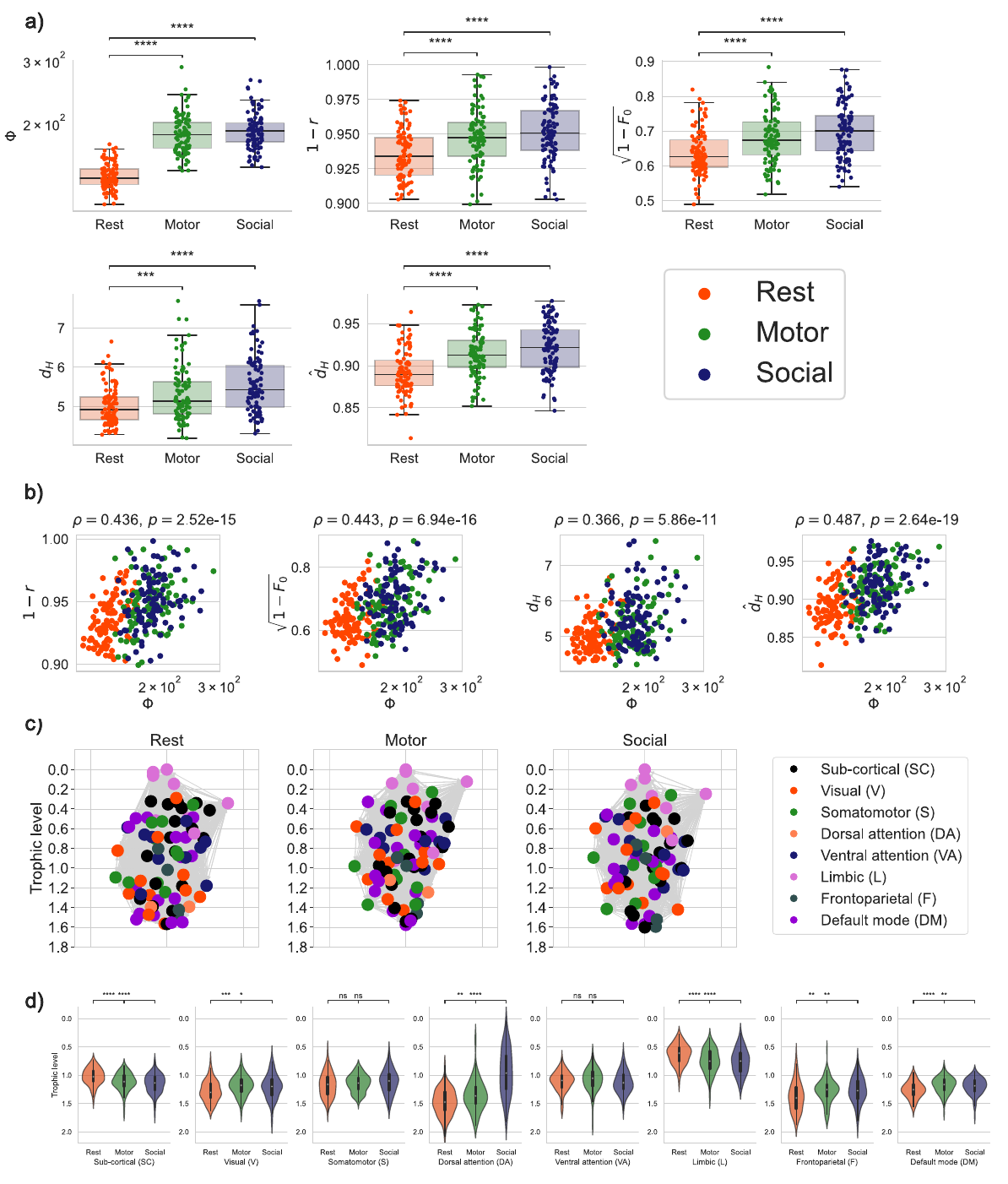}
    \caption{\textbf{Entropy production rate and directedness of effective brain networks:} $a)$ The EPR of the linear model and the directedness of the effective network are significantly elevated during task when compared to rest. $b)$ EPR is positively correlated with each of the directedness measures. $c)$ This panel shows the trophic levels of a participant-averaged effective network in each condition. The hierarchy of brain regions reorganises during tasks. Different functional networks occupy different positions within the processing hierarchy and this position changes depending on the task. $d)$ The distribution among participants of trophic levels for each functional network in each condition. The significant changes in the distribution suggests the reorganisation of the processing hierarchy during task. (ns) = $p>0.05$; (*) = $p>0.01$; (**) = $p>0.001$; (***) = $p>0.0001$; (****) = $p\leq 0.0001$.}
    \label{fig: brainnets}
\end{figure*}
The link between the hierarchical organisation and the overall directedness of the underlying network and the EPR of the dynamics is supported by this empirical analysis.
\begin{figure*}
    \centering
    \includegraphics[width=\linewidth]{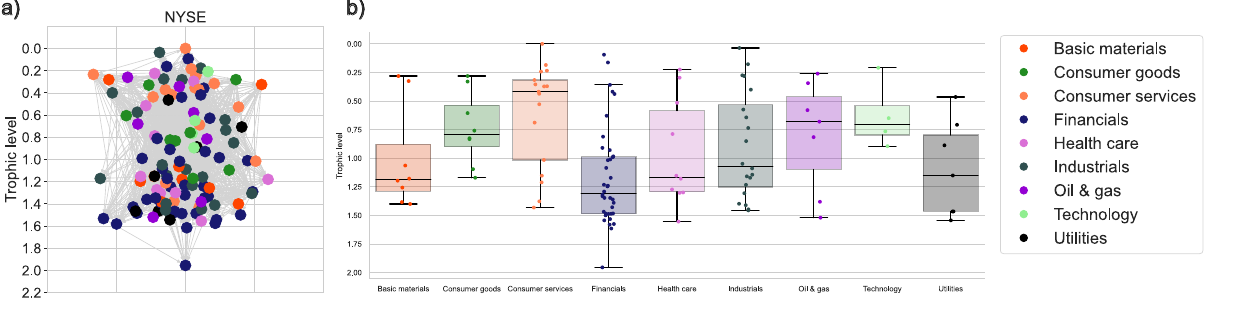}
    \caption{\textbf{Trophic hierarchy in the New York Stock Exchange:} $a)$ The effective network of stock interactions inferred from prices on the NYSE and coloured by industry. The network is hierarchically organised with certain industries like `Consumer services' sitting atop the hierarchy and `Financials' sitting at the bottom. $b)$ Distributions of trophic level by industry. Each industry occupies a different position in the hierarchy with `Consumer goods/services' at the top and `Financials' and `Utilities' at the bottom.}
    \label{fig: NYSE}
\end{figure*}
Furthermore, our results support previous resulting regarding  the hierarchical reorganisation of the brain during tasks \cite{kringelbach2023movie, deco2021workspace,deco2022insideout,deco2023violations,dehane1998globalworkspace} \Rev{but, crucially, link these results to the observed increase in EPR}.
\subsubsection{Stock-prices from the New York Stock Exchange}
\noindent We consider the daily stock-prices of 119 U.S. companies listed on the NYSE over the period 2000-2021, as previously analysed and published in Ref. \cite{santoro2023higherorder}. Each company represents a node whose signal is the fluctuating stock price. Therefore we have a single MVTS in this case-study. Again we fit the general linear model and calculate the EPR and the network-restricted model to obtain an effective network of interactions between the prices. Notably, under this model we restrict to positive connections between nodes, a common assumption for whole-brain modelling, but less common in the modelling of financial time-series where variables may have strong negative correlations \cite{Shumway2017timeseries}. However, our notions of directedness and hierarchy assume positive weights, and so this is an important modelling assumption.\\
\\
Firstly, we note that the financial time-series is out of equilibrium and has EPR $\Phi = 26.2$ (3 s.f.). Furthermore, the effective network of stock interactions is trophically organised with trophic directedness, $\sqrt{1-F_0}=0.405$ (3 s.f.), strongly directed with irreciprocity, $1-r = 0.975$ (3 s.f.) and non-normal ($d_H = 1.53$, $\hat{d}_H= 0.160$ (3 s.f.)). In Figure \ref{fig: NYSE}, panel $a)$ shows the trophic decomposition of the effective network of interactions between stocks, coloured by industry, and panel $b)$ shows the distribution of trophic levels for each industry. We can see that consumer services and goods sit atop the hierarchy, feeding into the dynamics of other stocks, whilst market-sensitive indicators, such as financials and utilities, sit at the bottom, following the trends. The conclusion that the NYSE is operating out of equilibrium can be interpreted both through the lens of thermoeconomics \cite{Sieniutycz1999thermoeconomics,Pokrovskii2020thermodynamics} which argues that the law of statistical mechanics can describe economics systems, or more simply through the lens of economic forces creating a hierarchical interaction structure between stocks and industries \cite{chen1986economicsforces}, that results in non-equilibrium dynamics.
\section{Discussion and conclusions}
\label{sec: dicussion}
\noindent Bridging the gap between the structure and dynamics of complex networks is a fundamental challenge in the modelling of real-world systems \cite{boccaletti2006complexdynamis}. Here, we presented novel results linking the hierarchical, directed structure of a network with the emergence of non-equilibrium dynamics and broken detailed balance. \Rev{We began by introducing a range of directedness measures and prototypical dynamical processes for networked systems. Next, we used analytical results and numerical simulations to show that the EPRs of the OU and Ising models are a function of directedness in a solvable 2-node network.} For a range of dynamical systems and directedness measures, we then showed that the EPR increased with directedness in synthetic hierarchical networks. We then decoupled the notions of directedness using a null ER model and we were able to show how the nature of the dynamics dictated which directedness measure would predict its EPR, highlighting the difference between locally- and globally-evolving processes. Next we considered a vast range of real-world network topologies and showed that the link between the EPR and directedness held in actual network data, \Rev{but noted the influence of overall network size, substructure and weight distribution on the EPR, which resulted in the unnormalised Henrici index being the most strong predictor of the EPR}. Finally, we applied our theory to multivariate time-series using a simple auto-regressive model to measure the EPR directly from dynamic node time-series, but also to unravel the hierarchical structure of the interactions between  variables. Applying this method to human neuroimaging at task and at rest, we found that the brain operates further from equilibrium in task compared to rest indicating that the EPR is a key indicator of cognitive exertion and complexity of neural dynamics which confirms previous results  \cite{lynn2021detailedbalance,deco2022insideout,deco2023tenet,deco2023violations,kringelbach2023movie}. Additionally, our approach extracts a network representation of regional interactions that demonstrates the hierarchical re-organisation of the brain during tasks \cite{kringelbach2023movie,dehane1998globalworkspace}. \Rev{Our analysis links the two previous results, indicating that a hierarchical reorganisation of brain regions during task could explain the increase in the EPR. Thus we identify the key feature of brain network structure associated with non-equilibrium neural dynamics, a hitherto outstanding challenge \cite{papo2024braincomplexnetwork}}. In addition, we  analysed stock-prices from the NYSE to identify the directed influence structure between stocks created by the economic forces at play during speculation \cite{chen1986economicsforces}.\\\\
Our work represents \Rev{an} attempt to link the structure of a network to its non-equilibrium \Rev{dynamics. In particular, as many complex systems can be represented by strongly directed networks, our results indicate that non-equilibrium dynamics could be linked to patterns of hierarchical asymmetries in interaction strengths. As a result,} these results place newfound importance on both the structural \cite{johnson2020digraphs,johnson2014trophiccoherence,johnson2017loops,rodgers2023strongconnectivity, kale2018directed} and the dynamical \cite{asllani2018nonnormal,asllani2014patterns,krakauer2023brokensymmetry,fruchart2021nonreciprocal} phenomena that are unique to complex systems with directed connections. In particular, in neuroimaging, where broken detailed balance emerges consistently \cite{lynn2021detailedbalance}, the traditional assumption of undirected structural connections limits the accuracy of whole-brain models which cannot explain the non-equilibrium nature of the empirical data. Our simple linear modeling approach provides a first-step to simultaneously understanding the hierarchical structure and the non-equilibrium dynamics of multivariate data. These results provide a general framework to study a range of  applications and extensions. For example, hierarchically \textit{modular} network structure has been shown to significantly impact critical dynamics \cite{Moretti2013griffiths}, yet its influence on \Rev{non-equilibrium dynamics} remains to be elucidated. Further, computational techniques for both measuring broken detailed balance and inferring directions interactions directly from data, could allow these results to be extended beyond the limitations of the analytically tractable processes or linear models considered here, to a range of non-linear dynamics and real-world time-series. Finally, our approach provides a \Rev{possible} interpretation for empirically observed violations of detailed balance, as found in the brain \cite{lynn2021detailedbalance}.\\\\
To conclude, we have shown that directedness and non-equilibrium dynamics are intimately linked. The degree to which a system is hierarchically directed in its interactions \Rev{appears closely correlated with the size of its divergence} from thermodynamic equilibrium. As a result, taking into consideration the non-reciprocity of interactions becomes fundamental to understanding the dynamic trajectories of both models and real-world systems.
\subsection*{Code availability}
\noindent The R and Matlab code used in this project will be made available on publication at \url{https://github.com/rnartallo/brokendetailedbalance}.
\subsection*{Data availability}
\noindent The network data used in this project is collated from multiple freely available locations and references are given in Appendix \ref{sec: app: networkdata}. The human neuroimaging data used in this project is freely available from the HCP website \cite{vanessen2013hcp}. The financial time-series used in this project is freely available from the Python package `yfinance' and at Ref. \cite{santoro2023higherorderdata}.
\begin{acknowledgments}
\noindent The authors wish to acknowledge M. Aguilera for his assistance understanding the Ising mean-field approximation \Rev{and R. MacKay for his comments on the manuscript}.\\\\
R.N.K was supported by an EPSRC doctoral scholarship from grants EP/T517811/1 and EP/R513295/1. G.D. was supported by
the AGAUR research support grant (ref. 2021 SGR 00917) funded
by the Department of Research and Universities of the Generalitat
of Catalunya, by the project NEurological MEchanismS of Injury,
and the project Sleep-like cellular dynamics (NEMESIS) (ref.
101071900) funded by the EU ERC Synergy Horizon Europe and
by the project PID2022-136216NB-I00 financed by the MCIN /AEI
/10.13039/501100011033 / FEDER, UE., the Ministry of Science
and Innovation, the State Research Agency and the European
Regional Development Fund. M.L.K was supported by the Centre for Eudaimonia and Human Flourishing funded by the Pettit and Carlsberg Foundations and by the Center
for Music in the Brain (MIB), funded by the Danish National
Research Foundation (project number DNRF117). R.L. was supported by the EPSRC grants EP/V013068/1 and EP/V03474X/1.
\end{acknowledgments}

\appendix

\section{Relationship between trophic directedness and non-normality}
\label{sec: app: trophicdirectednessnn}
\begin{figure}
    \centering
    \includegraphics[width=\linewidth]{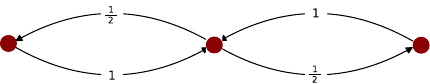}
    \caption{\RevTwo{\textbf{A non-normal random walk obeying detailed balance:} Non-normality is not a sufficient condition for violating detailed balance as shown by this example of a non-normal network where RW dynamics obey detailed balance.}}
    \label{fig: 3node}
\end{figure}

\begin{figure*}
    \centering
\includegraphics[width=\linewidth]{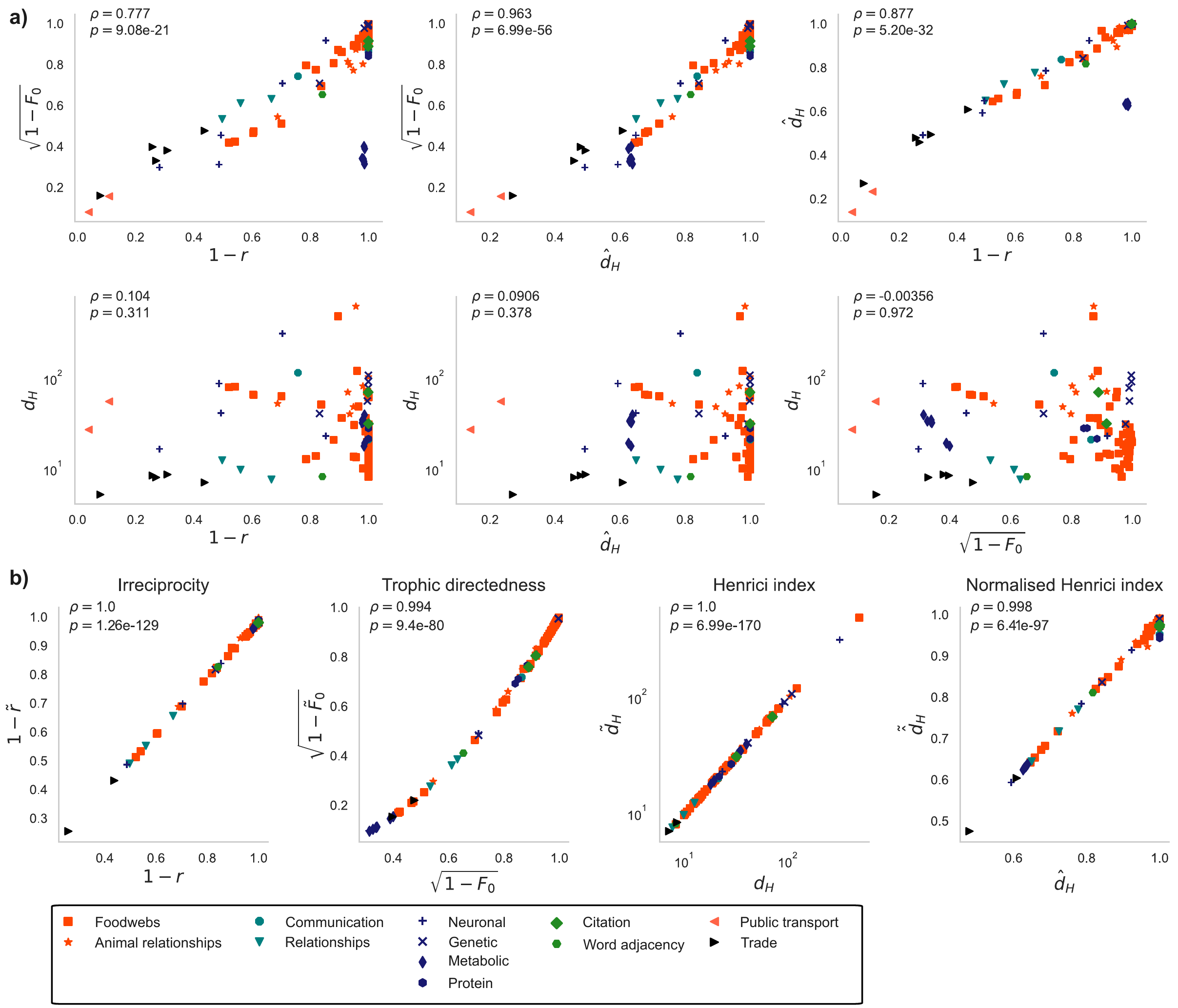}
    \caption{\textbf{Directedness of real-world networks}:
$a)$ A comparison between the directedness measures on real-world networks. Each pane shows a pairwise comparison between two of the four directedness measures on 97 real world networks from different domains. We can see strong correlations between the three normalised measures irreciprocity, trophic directedness and normalised Henrici index with a much weaker correlation to the unbounded Henrici index. We can also identify which fields the strongly, or weakly, directed networks belong to. Ecological networks are in orange, social in teal, biological in blue, language in green, transport in red and economic in black. \Rev{$b)$ The effect of the perturbation on the directedness of the weakly-connected networks is minimal.}}
\label{fig: RW directedness}
\end{figure*}

\noindent In previous studies, the relationship between trophic coherence and non-normality has been discussed \cite{mackay2020directed,johnson2020digraphs}. In this appendix, we build on this by illustrating the situations where they are (in)equivalent, which lends insight into what they are measuring.
\begin{figure}[h]
\centering
% the number in [] of wrapfigure is optional and gives the number of text lines that should be wrapped around the text. Adjust according to your figures height
\includegraphics[width=\linewidth]{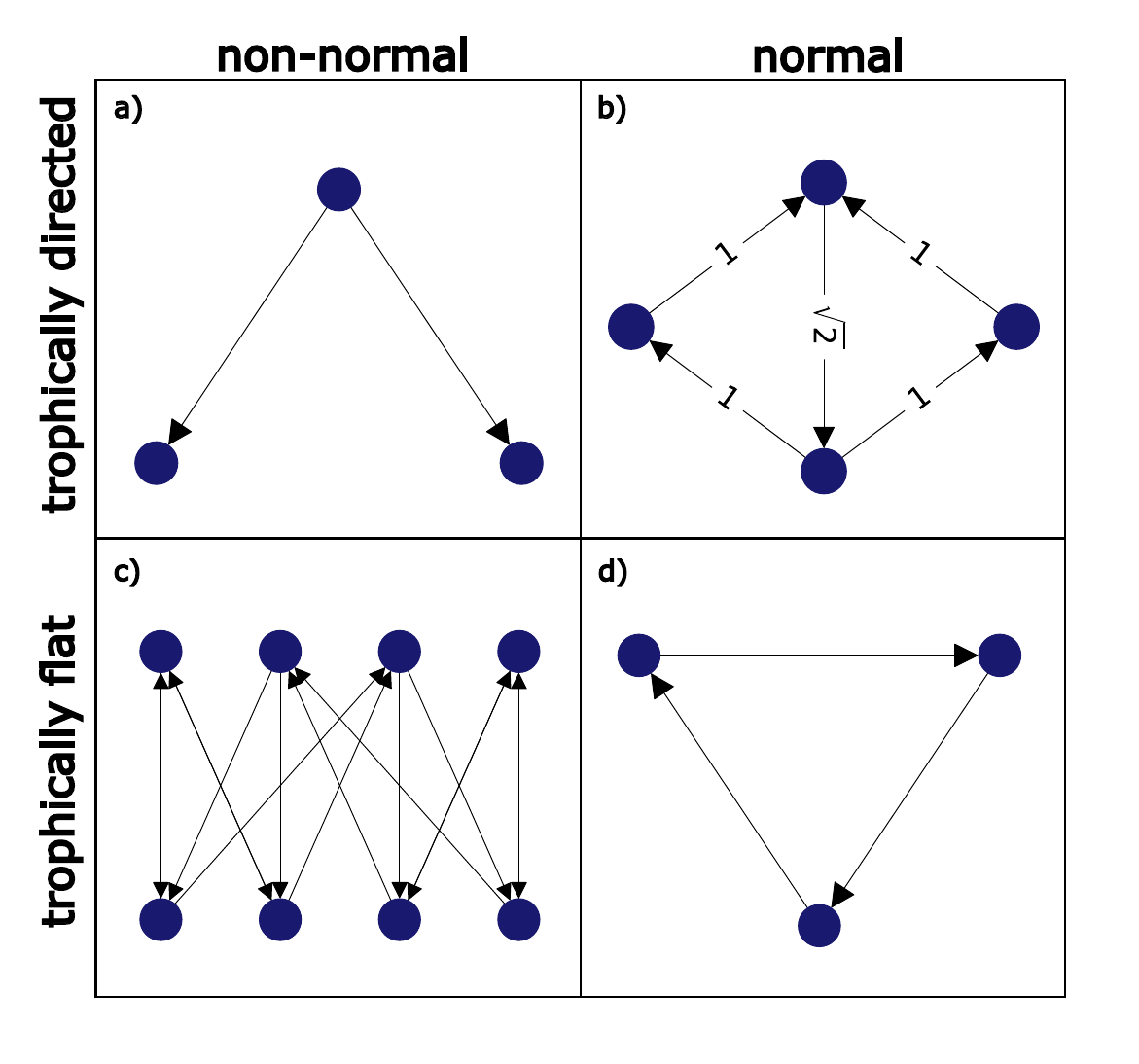}
\caption{\textbf{Non-normality and trophic directedness:} % note that \ref{fig1} refers to the corresponding wrapfigure
Non-normality and trophic directedness are in-equivalent. $a)$ A hierarchical motif which is both non-normal and trophically directed. Most real-world directed networks are in this category. $b)$ A normal but trophically directed network. There are no such unweighted networks. $c)$ A trophically flat but non-normal network. Intuitively, this network has no clear hierarchy but is still non-normal. $d)$ A directed cycle which is both normal and trophically flat. This network has no hierarchy.} 
\label{fig: nonnormaltrophiccoherence} 
\end{figure} % avoid blank space here
\noindent First, we define \textit{perfect trophic directedness} to be $F_0=0$ ($\sqrt{1-F_0}=1$) i.e. maximally trophically directed. This occurs iff the level differences $z_{ji} = h_i -h_j$ are 1 for each pair of levels. Oppositely, we define a network to be \textbf{trophically flat} if $F_0=1$ ($\sqrt{1-F_0}=0$) i.e. maximally trophically undirected. This occurs iff the level differences $z_{ji} = h_i -h_j$ are 0 for each pair of levels. Similarly, a network is defined to be \textit{perfectly (maximally) non-normal} if $d_H(\bm{W}) = ||\bm{W}||_{F}$ ($\hat{d}_H(\bm{W}) = 1$). This means that $\sum_i|\lambda_i|^2 =0$ where $\{\lambda_i\}$ is the eigenspectrum. Figure \ref{fig: nonnormaltrophiccoherence} shows examples of (un)weighted directed networks in all four categories of trophically flat/directed and non-normal/normal.\newline\\
\noindent MacKay et al showed that a normal unweighted network must be trophically flat \cite{mackay2020directed}. In the unweighted case, they begin by noting that $(\bm{W}^{\top}\bm{W})_{ij}$ is the number of common sources between nodes $i$ and $j$, whilst $(\bm{W}\bm{W}^{\top})_{ij}$ is the number of common targets. In particular, $(\bm{W}^{\top}\bm{W})_{ii}=w_{i}^{\text{in}}$ and $(\bm{W}\bm{W}^{\top})_{ii}=w_{i}^{\text{out}}$. If $\bm{W}$ is normal i.e. $\bm{W}^{\top}\bm{W}=\bm{W}\bm{W}^{\top}$, then $w_{i}^{\text{in}}=w_{i}^{\text{out}}$ and $v_i=0$ for all $i$ i.e. the network is trophically flat.\newline\\
\noindent For a weighted network, the same result does not hold. By following the proof of the unweighted case, we can construct a counter example. In the case of a network with non-negative weights, we now have,
\begin{align}
    (\bm{W}\bm{W}^{\top})_{ii} &= \sum_k W_{ik}^2,\\
    (\bm{W}^{\top}\bm{W})_{ii} &= \sum_k W_{ki}^2,
\end{align}
which are no longer simply $w_{i}^{\text{in}}, w_{i}^{\text{out}}$. If $\bm{W}$ is normal we therefore have, $\sum_k W_{ik}^2=\sum_k W_{ki}^2$ for each $i$. On the other hand, a trophically flat network i.e. $v_i=0$ implies $\sum_kW_{ik} = \sum_k W_{ki}$ for each $i$. These two conditions are not equivalent so we can construct a network that is both normal and trophically directed. One such network, shown in panel $b)$ of Fig. \ref{fig: nonnormaltrophiccoherence}, is given by,
\begin{align}
    \bm{W} = \begin{pmatrix}
0 & 1 & 1 &0 \\ 
0 & 0 & 0 & 1\\ 
 0& 0 & 0 & 1\\ 
 \sqrt{2}& 0 & 0 & 0
\end{pmatrix},
\end{align}
 which satisfies that $\sum_k W_{ik}^2=\sum_k W_{ki}^2$ for each $i$ whilst $\sum_kW_{ik} \neq \sum_k W_{ki}$. Therefore, it is normal yet trophically directed ($F_0 = 0.9737$).\newline\\
\noindent On the other hand, trophic flatness does not imply normality, even in the unweighted case. One can consider the network,
\begin{align}
    \bm{W} = \begin{pmatrix}
 0& 0 & 0 & 0 & 1 & 0 & 1 & 0\\ 
  0& 0 & 0 & 0 & 1 & 0 & 1 & 0 \\ 
  0& 0& 0 & 0 & 0 & 1 & 0 & 1 \\ 
 0& 0& 0 & 0 & 0 & 1 & 0 & 1 \\ 
 1 & 1 & 0 & 0 & 0 & 0 & 0 &0 \\ 
 1 & 1 & 0 & 0 & 0 & 0 & 0 &0 \\ 
 0 & 0 & 1 & 1 & 0 & 0 & 0 &0 \\ 
 0 & 0 & 1 & 1 & 0 & 0 & 0 &0
\end{pmatrix},
\end{align}
shown in panel $c)$ of Fig. \ref{fig: nonnormaltrophiccoherence}, which is non-normal $(d_H(\bm{W})=2.8284)$ yet trophically flat. MacKay et al present a simpler example of a trophically flat, non-normal network, ($d_H(\bm{W})=0.6982$), but including a self loop,
\begin{align}
    \bm{W} = \begin{pmatrix}
1 & 1 & 0\\ 
0 & 0 & 1\\ 
1 & 0 & 0
\end{pmatrix}.
\end{align}
MacKay et al further showed that perfect trophic directedness implies implies all eigenvalues are 0 and, consequently, perfect non-normality \cite{mackay2020directed}. On the other hand, perfect non-normality does not imply perfect trophic directedness, as evidenced by the so-called `feed-forward' motif,
\begin{align}
        \bm{W} = \begin{pmatrix}
0 & 1 & 1\\ 
0 & 0 & 1\\ 
0 & 0 & 0
\end{pmatrix},
\end{align}
which has $\hat{d}_H(\bm{W}) = 1$  ($d_H(\bm{W}) = 1.7321$) but $F_0 = 0.1111$.\newline\\
\noindent In real-world directed networks, strong correlations have been found between trophic directedness and non-normality, but the relationship is non-linear \cite{mackay2020directed,johnson2020digraphs}. This relationship can be partially explained by introducing the `loop exponent' of a network which bridges the spectral properties of a network and the trophic decomposition \cite{johnson2017loops,johnson2020digraphs}. Panel $a)$ of Figure \ref{fig: RW directedness} shows the correlations between the four different measures in the 97 real-world networks considered in Section \ref{sec: realworld}. \Rev{In addition, in Section \ref{sec: realworld}, we considered a perturbation to the weakly connected networks in order to render them strongly connected and guarantee the existence of the stationary distribution of the RW. In panel $b)$ of Figure \ref{fig: RW directedness}, we show that the effect of this perturbation on the directedness of the network is minimal. In this case, $1-\Tilde{r}, \sqrt{1-\Tilde{F}_0}, \Tilde{d}_H$ and $\Tilde{\hat{d}}_H$ represent the directedness of the perturbed network in each case.}\\\\
\RevTwo{In the case of the OU and Ising models considered in this study, directedness is a sufficient condition to guarantee that the system breaks detailed balance. In Sec. \ref{sec: dynamics}, using the example of the 2-node network, we showed that this is not true for the RW, where the dynamics can obey detailed balance despite directedness. However, the 2-node network, like all circulant networks, is normal. Nevertheless, non-normality is not a sufficient condition to guarantee that the RW breaks detailed balance either. To illustrate this, we consider the example in Fig. \ref{fig: 3node}, given by the weight matrix,
\begin{align}
        \bm{W} = \begin{pmatrix}
0 & 1 & 0\\ 
1/2 & 0 & 1/2\\ 
0 & 1 & 0
\end{pmatrix}.
\end{align}
The network is non-normal $(d_H=0.7071)$ and yet converges to the stationary distribution $\bm{\pi} = [1/4,1/2,1/4]$ which satisfies $P_{i2}=P_{2i}=1/4$ for $i=1,3$ and $P_{13}=P_{31}=0$, thus the RW obeys detailed balance.\\\\
Whilst this study explores the important correlation between non-normality and EPR. We note that normality is not sufficient for equilibrium, as evidenced by the circulant networks discussed in Appendix \ref{sec: app: circulant}, nor is non-normality sufficient for non-equilibrium, as evidenced by the example in Fig. \ref{fig: 3node}.
}

\section{A preferential attachment scheme for generating hierarchical networks}
\label{sec: app: PA}
\noindent The seminal contributions of de Solla Price \cite{desollaprice1965networksofscience,desollaprice1976bibliometric} and the subsequent development of the preferential attachment (PA) algorithm for generating random networks \cite{barabasi1999pa}, has long been studied for its power-law degree distribution. Modifications to the PA algorithm can cause it to generate non-normal, irreciprocal, trophically directed networks with a strong hierarchy \cite{asllani2018nonnormal}. For the networks considered in Sections \ref{sec: measures} and \ref{sec: dynamics}, we used the following generative algorithm.\\\\
We consider a growing network where, at each step, a new node is added to the network. The new node $j$ connects to an exisiting node $i$ with a probability proportional to $d_i$, the degree of node $i$. We randomly sample the weight of this connection $W_{ji}$ from a uniform distirbution $U(0,1)$, and then introduce a weak reciprocal link $W_{ij}=\frac{W_{ji}}{\gamma}$ for $\gamma \gg 1$ such that $0\leq W_{ij}\ll W_{ji}$.\\\\
As the network begins with no edges and a single node, we must initialise the network in some way. When a new node $j$ is added to the network, if an exisiting node $i$ has $d_i=0$, they are connected with probability $0< p_0\ll 1$. If the exisiting node $i$ has $d_i>0$, then they are connected with probability $\min(1,\frac{d_i}{d_{\text{tot}}} + \mu)$ where $d_{\text{tot}}$ is the total degree of the network. This approach produces a strongly connected graph where each edge is reciprocated, which may not be true for all real-world networks.
\section{Linear interpolation parameterises trophic directedness and irreciprocity}
\label{sec: app: exactinterpolation}
\noindent In Section \ref{sec: measures}, we varied the directedness of networks by first generating a hierarchically directed network using the PA algorithm, as described above, and then linearly interpolating between this network and its Hermitian,
\begin{align}
\label{eq: interpolation2}
    \bm{\hat{W}}(\epsilon) & = (1-\epsilon)\bm{\Tilde{W}} + \epsilon\bm{W},
\end{align}
for $\epsilon\in [0,1]$ where the Hermitian network is given by $\bm{\Tilde{W}} = \frac{1}{2}\left( \bm{W} + \bm{W}^{\top}\right).$ Figure \ref{fig: interpolation} showed that each of the four measures increases almost linearly with $\epsilon$. In this section, we show that the irreciprocity and trophic directedness increase exactly linearly i.e.
\begin{align}
    1-r(\bm{\hat{W}}(\epsilon)) &= \epsilon (1-r(\bm{W})),\\
    \sqrt{1-F_0(\bm{\hat{W}}(\epsilon)} &= \epsilon \sqrt{1-F_0(\bm{W})}.
\end{align}
Unfortunately, non-normality is a spectral measure and the eigenvalues of the matrix $\bm{\hat{W}}(\epsilon)$ are not calculable from $\bm{\Tilde{W}}$ and $\bm{W}$.\newline\\
\noindent We begin with the irreciprocity. We denote the reciprocity of $\bm{W}$ by $r\;(=r(1))$. The irreciprocity of the interpolated network is given by,
\begin{align}
    1-r(\epsilon)&=1- \frac{\sum_i\sum_{j\neq i}\min(\hat{W}_{ij}(\epsilon),\hat{W}_{ji}(\epsilon))}{\sum_i\sum_{j\neq i}\hat{W}_{ij}(\epsilon)},\\
    &=\frac{\sum_i\sum_{j\neq i} \hat{W}_{ij}(\epsilon) - \min(\hat{W}_{ij}(\epsilon),\hat{W}_{ji}(\epsilon))}{\sum_i\sum_{j\neq i}\hat{W}_{ij}(\epsilon)}.
\end{align}
Firstly, we note that the interpolation preserves the row sums (even excluding the diagonal), so the denominator, which we denote $k$, is the same as in the irreciprocity of the original network $\bm{W}$.
\begin{align}
    1-r(\epsilon)&=\frac{\sum_i\sum_{j\neq i} \hat{W}_{ij}(\epsilon) - \min(\hat{W}_{ij}(\epsilon),\hat{W}_{ji}(\epsilon))}{k},\\
    k(1-r(\epsilon))& = \sum_i\sum_{j\neq i} \hat{W}_{ij}(\epsilon) - \min(\hat{W}_{ij}(\epsilon),\hat{W}_{ji}(\epsilon)).
\end{align}
Then, using the defintion of $\hat{W}_{ij}(\epsilon)$ and cancelling the Hermitian terms, we obtain,
\begin{align}
    k(1-r(\epsilon))& = \sum_i\sum_{j\neq i} \epsilon {W}_{ij} - \epsilon \min(W_{ij},W_{ji}).
\end{align}
Finally, we get,
\begin{align}
    1-r(\epsilon)&=\frac{\epsilon \sum_i\sum_{j\neq i} W_{ij} - \min({W}_{ij},W_{ji})}{k},\\
    1-r(\epsilon)& = \epsilon(1-r). \\&\square \notag
\end{align}
Next, we consider the trophic directedness. As defined in Section \ref{sec: measures}, for the original network $\bm{W}$, we have quantities, $w_{i}^{\text{in}}, w_{i}^{\text{out}}$. For the interpolated network these become,
\begin{align}
    w_{i}^{\text{in}}(\epsilon) &= \epsilon w_{i}^{\text{in}} + (1-\epsilon)\frac{1}{2}(w_{i}^{\text{in}} +w_{i}^{\text{out}}),\\
    w_{i}^{\text{out}}(\epsilon) & = \epsilon w_{i}^{\text{out}} + (1-\epsilon)\frac{1}{2}(w_{i}^{\text{in}} +w_{i}^{\text{out}}).
\end{align}
From this, we obtain,
\begin{align}
    u_i(\epsilon) &= w_{i}^{\text{in}}(\epsilon) + w_{i}^{\text{out}}(\epsilon)\\
    &=w_{i}^{\text{in}} + w_{i}^{\text{out}}\\
    &=u_i,
\end{align}
i.e. $u_i$ is preserved under interpolation. On the other hand,
\begin{align}
    v_i(\epsilon) & = w_{i}^{\text{in}}(\epsilon) - w_{i}^{\text{out}}(\epsilon)\\
    &= \epsilon(w_{i}^{\text{in}}-w_{i}^{\text{out}})\\
    &=\epsilon v_i.
\end{align}
Furthermore, the symmetric weighted graph Laplacian of the original network, $\bm{\Lambda}$, is also preserved under the interpolation,
\begin{align}
    \bm{\Lambda} (\epsilon) & = \text{diag}(\bm{u}) - \hat{\bm{W}}(\epsilon) -\hat{\bm{W}}^{\top}(\epsilon)\\
    & = \text{diag}(\bm{u}) - 2\Tilde{\bm{W}}\\
    &=\bm{\Lambda}.
\end{align}
For the original network, $\bm{h}$ is the vector of trophic levels which solves,
\begin{align}
    \bm{\Lambda} \bm{h} = \bm{v}.
\end{align}
The trophic levels of the interpolated network are given by the solution to,
\begin{align}
    \bm{\Lambda} \bm{h}(\epsilon) = \epsilon\bm{v},
\end{align}
i.e. $\bm{h}(\epsilon)=\epsilon \bm{h}$. Finally, we use an alternative, but equivalent, formulation for $1-F_0$, shown by Mackay et al \cite{mackay2020directed}, given by,
\begin{align}
    1-F_0 = \frac{\sum_{i,j}W_{ij}(h_j-h_i)}{\sum_{i,j}W_{ij}}.
\end{align}
Therefore, for the interpolated network, we have,
\begin{align}
    1-F_0(\epsilon)=\frac{\sum_{i,j}\hat{W}_{ij}(\epsilon)(h_j-h_i)\epsilon}{\sum_{i,j}W_{ij}},
\end{align}
as the denominator, which we again denote $k$, is preserved under the interpolation. Expanding the interpolation, we obtain,
\begin{align}
    k(1-F_0(\epsilon)) &= \epsilon(1-\epsilon)\frac{1}{2}\sum_{i,j}(W_{ij}+W_{ji})(h_j-h_i)\\
    &+ \epsilon^2\sum_{i,j}W_{ij}(h_j-h_i).\notag
\end{align}
Notice that the sum $\sum_{i,j}(W_{ij}+W_{ji})(h_j-h_i)$ vanishes as the term for $(i,j)$ cancels with the term for $(j,i)$. Therefore, we get
\begin{align}
    1-F_0(\epsilon)&=\epsilon^2(1-F_0),
\end{align}
or for the trophic directedness defined in Section \ref{sec: measures},
\begin{align}
    \sqrt{1-F_0(\epsilon)}& = \epsilon\sqrt{1-F_0}.\\
    &\square \notag
\end{align}
\section{Deriving the entropy production rate in the Ornstein-Uhlenbeck process}
\label{sec: app: EPROU}
\noindent We recall that the multivariate OU process is given by the Langevin system,
\begin{align}
    \frac{d\bm{x}}{dt}&=-\bm{B}\bm{x}(t) + \bm{\xi}(t),
\end{align}
where $\bm{x}(t)\in \mathbb{R}^N$ is the time-dependent state vector, $\bm{B} \in \mathbb{R}^{N \times N}$ is the \textit{friction matrix}, and $\bm{\xi}(t)\in \mathbb{R}^N$ is additive white noise with covariance given by,
\begin{align}
\langle \bm{\xi}(t)\bm{\xi}^{\top}(t')\rangle = 2\bm{D} \delta(t-t'),
\end{align}
where $\bm{D} \in \mathbb{R}^{N \times N}$ is the \textit{noise covariance matrix} which is symmetric. We follow Godrèche and Luck \cite{Godreche2018OU} to derive the EPR rate of the OU in a steady state. Assuming each eigenvalue of the friction matrix, $\bm{B}$, has positive real part, then the system relaxes exponentially fast to a steady state with Gaussian fluctuations given by,
\begin{align}
\label{eq: sol}
\bm{x}(t) &= e^{\bm{-B}t}\bm{x}(0) + \int_{0}^{t}e^{\bm{-B}(t-s)}\bm{\xi}(s)\;ds,
\end{align}
and covariance,
\begin{align}
    \bm{S} &= \lim_{t\rightarrow \infty}\bm{S}(t) ,\\
&=  \lim_{t\rightarrow \infty}\langle \bm{x}(t)\bm{x}^{\top}(t)\rangle.
\end{align}
The covariance can be written as,
\begin{align}
\bm{S} &= \lim_{t\rightarrow \infty}[ e^{\bm{-B}t}\bm{S}(0)e^{\bm{-B}^{\top}t}, \notag\\ 
&+ 2\int_0^{t} e^{\bm{-B}(t-s)}\bm{D}e^{\bm{-B}^{\top}(t-s)}\;ds],\\
&= 2\int_0^{\infty} e^{\bm{-B}t}\bm{D}e^{\bm{-B}^{\top}t}\;dt.
\end{align}
It can also be shown that $\bm{S}$ satisfies the following Sylvester equation \cite{zabczyk2020mathematicalcontrol},
\begin{align}
\label{eq: sylvester}
\bm{B}\bm{S} + \bm{S}\bm{B}^{\top}&=2\bm{D}.
\end{align}
Next, we define the Onsager matrix, $\bm{L}$, of kinetic coefficients,
\begin{align}
\bm{L}&=\bm{BS}=\bm{D}+\bm{Q},\\
\bm{L}^{\top}&=\bm{SB}^{\top}=\bm{D}-\bm{Q},
\end{align}
parameterising the asymmetries through the matrix $\bm{Q}$, which provides an intuitive measure of the degree of non-equilibrium.\newline\\
\noindent The EPR, $\Phi$, can then be written in the form,
\begin{align}
\Phi &= \langle \bm{x}^{\top}( \bm{D}^{-1} \bm{B} -  \bm{S}^{-1})^{\top} \bm{D}( \bm{D}^{-1} \bm{B}- \bm{S}^{-1}) \bm{x}\rangle,\\
&=-\langle \bm{x}^{\top}\bm{S}^{-1}\bm{Q}\bm{D}^{-1}\bm{Q}\bm{S}^{-1}\bm{x}\rangle,
\end{align}
with the second equation following from the relations $\bm{D}^{-1}\bm{B}-\bm{S}^{-1} = \bm{D}^{-1}\bm{Q}\bm{S}^{-1}$ and $\bm{Q}=-\bm{Q}^{\top}$. Using that the steady state is Gaussian we have that $\langle \bm{x}^{\top}\bm{A}\bm{x}\rangle = \Tr(\bm{SA})$ for a general matrix $\bm{A}$ and thus we have that,
\begin{align}
\Phi &=-\Tr (\bm{Q}\bm{D}^{-1}\bm{Q}\bm{S}^{-1}).
\end{align}
This can be rewritten in the form,
\begin{align}
\Phi &=-\Tr (\bm{D}^{-1}\bm{B}\bm{Q}).
\end{align}
For further details see Ref. \cite{Godreche2018OU}.
\section{2-node networks}
\label{sec: app: 2node}
\noindent In Section \ref{sec: entropydirected}, we considered directed 2-node networks and showed that the EPR of the OU increases with the asymmetry between the two connections. Furthermore, for larger networks, we consider the OU and calculated the EPR as a function of the interpolation parameter $\epsilon$ using the numerical solution of the Sylvester equation (\ref{eq: sylvester}). Here we consider the linear interpolation applied to the case of 2 nodes, where the EPR as a function of $\epsilon$ can be calculated explicitly \cite{Godreche2018OU}. We show, analytically, that the Henrici index increases linearly and the EPR of the OU increases quadratically in $\epsilon$, consistent with the conjectured relationship obtained numerically for hierarchical networks. This varies from the analysis in Section \ref{sec: entropydirected}, where we varied the weights $(W_{12},W_{21})$ as we are now fixing these weights and performing the interpolation between the network and its Hermitian. The 2-node directed network is defined by the weight matrix,
\begin{align}
    \bm{W} = \begin{bmatrix}
0 & W_{12}\\ 
W_{21} & 0
\end{bmatrix}. 
\end{align}
In the previous section, we showed that the irreciprocity and trophic directedness are linearly interpolated by $\epsilon$. Recalling from Section \ref{sec: entropydirected}, that the irreciprocity and trophic directedness of the 2-node network coincides and is given by,
\begin{align}
    1-r= \sqrt{1-F_0}=\frac{|W_{12}-W_{21}|}{W_{12}+W_{21}},
\end{align}
then the irreciprocity and trophic directedness of the interpolated network is given by,
\begin{align}
    1-r(\epsilon)= \sqrt{1-F_0(\epsilon)}=\epsilon\frac{|W_{12}-W_{21}|}{W_{12}+W_{21}}.
\end{align}
On the other hand, we were not able to show that, for general networks, the Henrici indices scaled linearly with $\epsilon$, but we can do so in the case of the 2-node network. Recall that the Henrici indices of the 2-node network were given by,
\begin{align}
       d_H & = |W_{12}-W_{21}|,\\
    \hat{d}_H&=\frac{|W_{12}-W_{21}|}{\sqrt{W_{12}^2+W_{21}^2}}. 
\end{align}
First we define,
\begin{align}
    \Tilde{W}_{12}(\epsilon) &= \epsilon W_{12} + \frac{1}{2}(1-\epsilon)(W_{12}+W_{21})\\
    \Tilde{W}_{21}(\epsilon) &=\epsilon W_{21} + \frac{1}{2}(1-\epsilon)(W_{12}+W_{21})
\end{align}
The eigenvalues of the interpolated network are given by,
\begin{align}
    \lambda_{\pm}(\epsilon) = \pm \sqrt{\Tilde{W}_{12}(\epsilon)\Tilde{W}_{21}(\epsilon)},
\end{align}
thus the Henrici index factorises,
\begin{align}
    d_H^2(\epsilon) & = [\Tilde{W}_{12}(\epsilon)]^2 + [\Tilde{W}_{21}(\epsilon)]^2-2\Tilde{W}_{12}(\epsilon)\Tilde{W}_{21}(\epsilon)\\
    &=\epsilon^2(W_{12}-W_{21})^2\\
    d_H(\epsilon)&= \epsilon|W_{12}-W_{21}|\\
    &= \epsilon d_H.
\end{align}   
Next we consider the OU evolving on an interpolated 2-node network. We recall that for an OU defined by matrices,
\begin{align}
    \bm{B}&=\begin{pmatrix}
a &b \\ 
c &d 
\end{pmatrix}, && \bm{D}=\begin{pmatrix}
u & w \\ 
w & v 
\end{pmatrix},
\end{align}
the EPR is given by,
\begin{align}
    \Phi &= \frac{(cu-bv+(d-a)w)^2}{(a+d)(uv-w^2)}.
\end{align}
For the 2-node network, this expression becomes,
\begin{align}
    \Phi &=\frac{\Theta \gamma^2}{2}(W_{12}-W_{21})^2,
\end{align}
We now fix $W_{12}\neq W_{21}$ which has an associated EPR $\Phi$. Under the interpolation, the network becomes,
\begin{align}
    \bm{\hat{W}}(\epsilon) &= \begin{bmatrix}
0 & \Tilde{W}_{12}(\epsilon)\\ 
\Tilde{W}_{21}(\epsilon) & 0
\end{bmatrix}
\end{align}  
Thus, we have,
\begin{align}
    \bm{B}&=\begin{pmatrix}
\Theta  &\Theta(1-\gamma\Corr{\Tilde{W}_{21}}(\epsilon)) \\ 
\Theta(1-\gamma\Corr{\Tilde{W}_{12}}(\epsilon)) & \Theta  
\end{pmatrix}, \\ \bm{D}&=\begin{pmatrix}
2\sigma & 0 \\ 
0 & 2\sigma 
\end{pmatrix},
\end{align}   
which, after simplification, yields an expression for the EPR of the interpolated system,
\begin{align}
    \Phi(\epsilon) &= \frac{\Theta \gamma^2 \epsilon^2 (W_{12}-W_{21})^2}{2}\\
    &=\epsilon^2 \Phi.\\
    &\square \notag
\end{align}
Linking this to previous result, the EPR in the OU on the interpolated 2-node network is also an exact quadratic function of the unnormalised Henrici index of the corresponding network.
\section{Circulant networks}
\label{sec: app: circulant}
\noindent Next we consider the case of networks with circulant weight matrices. These correspond to $k$-regular directed cyclic networks.
\begin{figure*}
\centering
% the number in [] of wrapfigure is optional and gives the number of text lines that should be wrapped around the text. Adjust according to your figures height
\includegraphics[width=\textwidth]{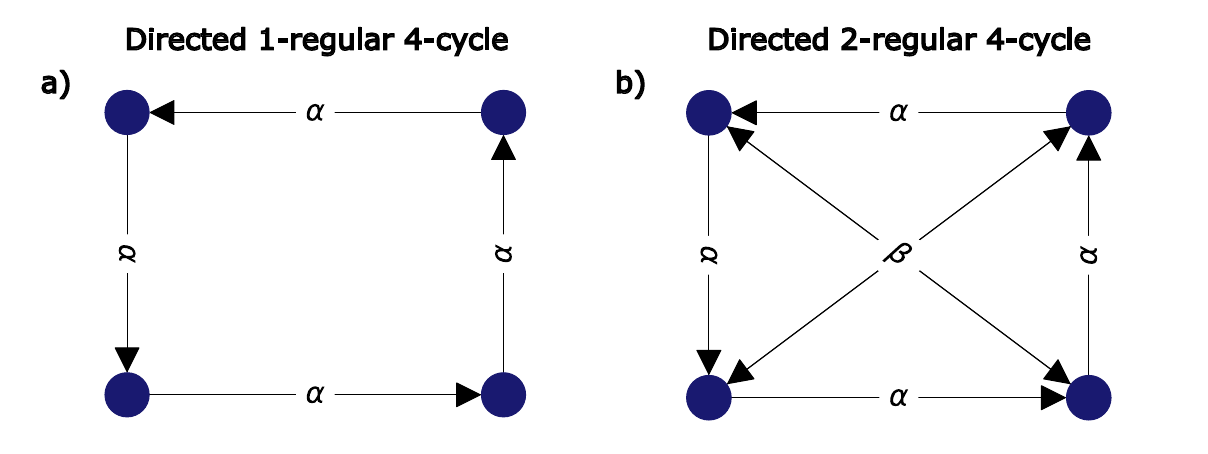}
\caption{\textbf{Circulant matrices and $k-$regular directed cycles:} % note that \ref{fig1} refers to the corresponding wrapfigure
$a)$ A directed 1-regular 4 cycle defined by the vector $\bm{w} = (0,\alpha,0,0)$. $b)$ A directed 2-regular 4 cycle defined by the vector $\bm{w} = (0,\alpha,\beta,0)$.} % add dummy caption - otherwise \label won't work and figure numbering will not count up
\label{fig: circulant} % use \ref{fig1} to reference to this figure
\end{figure*} % avoid blank space here
 \noindent For example, Figure \ref{fig: circulant} shows two $k$-regular 4-cycles ($k=1,2)$. All $k$-regular $N$-cycles are special cases of the $(N-1)-$regular $N$-cycle which can be written as,
\begin{align}
    \bm{W} & = \begin{pmatrix}
0 &w_1  & w_2 & \hdots &w_{N-1} \\ 
 w_{N-1}& 0 & w_1 & \hdots & w_{N-2}\\ 
 \vdots& \vdots & \vdots & \ddots & \vdots\\ 
w_1 & w_2 & w_3 & \hdots & 0
\end{pmatrix}.
\end{align}
This is uniquely defined by the top row,
\begin{align}
    \bm{w}= (0,w_1,...,w_{N-1}).
\end{align}
It is important to note that $\bm{W}$ is circulant and therefore normal. However, as it is directed, it breaks detailed balance for all the systems considered in this paper (with the exception of the RW for $N=2$). More importantly, as we will show, applying the interpolation to the cyclic networks, $\epsilon$ increases the EPR of the corresponding OU. We found in Sections \ref{sec: er} and \ref{sec: realworld} that the non-normality of the underlying network was a extremely strong indicator of the EPR of the OU, yet in this case, the non-normality is 0 and the EPR still varies with $\epsilon$. This represents an important special case where hierarchical asymmetry or a global direction are not necessary for the OU to progressively break detailed balance. The lack of these cyclic structures in real-world networks \cite{johnson2017loops} explains why, despite this special case, the non-normality correlates so closely with the EPR for the OU on real-world networks, as shown in Section \ref{sec: realworld}.\\\\
Firstly, as mentioned, the Henrici indices are 0 for these networks as their weight matrices are normal. Furthermore, for all $i$
\begin{align}
    w^{\text{in}}_i = w^{\text{out}}_i = \sum_j w_j,
\end{align}
meaning $\bm{v} = 0$ and therefore $F_0 = 1$ ($\sqrt{1-F_0}=0$) so the network is trophically flat. This implies that the network has no global direction or hierarchy as both global measures vanish. On the other hand, it is locally asymmetric and so has non-zero irreciprocity, 
\begin{align}
    1-r& = 1 - \frac{\sum_j \min(w_j,w_{N-j})}{\sum_j w_j}\\
    &=\frac{\sum_j |w_j - w_{N-j}|}{2\sum_j w_j}.
\end{align}
Next, we consider the EPR of the OU on circulant networks. The interpolated network is also circulant and so we can write it in terms of the top row. First, we define,
\begin{align}
    \Tilde{w}_i(\epsilon) = \epsilon w_i + \frac{1}{2}(1-\epsilon)(w_i + w_{N-i}),
\end{align}
then the top row of the interpolated weight matrix is given by
\begin{align}
    \bm{\hat{w}}(\epsilon) = (0,\Tilde{w}_1(\epsilon),.\hdots ,\Tilde{w}_{N-1}(\epsilon)).
\end{align}  
Therefore, the friction matrix $\bm{B}= \Theta (\bm{I} - \gamma\bm{\Corr{\hat{W}^{\top}}}(\epsilon))$ is also circulant and defined by the vector,
\begin{align}
    \bm{b}(\epsilon) = (\Theta, -\Theta\gamma\Corr{\Tilde{w}_{N-1}}(\epsilon),\hdots, -\Theta\gamma\Corr{\Tilde{w}_{1}}(\epsilon)).
\end{align} 
The matrix $\bm{D}$ is diagonal for the OU and therefore circulant. Godrèche and Luck \cite{Godreche2018OU} showed that the EPR of the OU with circulant $\bm{B,D}$ is given by,
\begin{align}
    \Phi & = \sum_{k=0}^{N-1}\frac{(\Im(\Tilde{b}_k))^2}{\Re(\Tilde{b}_k)},
\end{align}
where $\bm{\Tilde{b}}=(\Tilde{b}_0,...,\Tilde{b}_{N-1})$ is the discrete Fourier transform of the vector $\bm{b}$, $\Im(\cdot)$ is the imaginary part and $\Re(\cdot)$ is the real part. Furthermore, $\bm{\Tilde{b}}$ is also the vector of eigenvalues of the circulant matrix $\bm{B}$.\newline\\
\noindent The discrete Fourier transform gives us,
\begin{align}
    \Tilde{b}_k&=\Theta -\Theta \gamma\sum_{j=1}^{N-1}\Corr{\Tilde{w}_{N-j}}(\epsilon)\exp{\bm{i}\frac{2\pi j k}{N}}.
\end{align}  
Using the periodicity and oddness/evenness of sine/cosine we have that,
\begin{align}
    \sin(\frac{2\pi j k}{N}) + \sin(\frac{2\pi (N-j) k}{N}) &= 0,\\
    \cos(\frac{2\pi j k}{N}) + \cos(\frac{2\pi (N-j) k}{N}) &= 2,
\end{align}
which allows us to simplify our expression to,
\begin{widetext}
   \begin{align}
    \Tilde{b}_k&=\Theta -\Theta \gamma\sum_{j=1}^{N-1}(\epsilon \Corr{w_{N-j}})(\bm{i}\sin(\frac{2\pi j k}{N}) + \cos(\frac{2\pi j k}{N})) -\Theta \gamma\sum_{j=1}^{\lfloor\frac{N}{2}\rfloor}(1-\epsilon)(w_j + w_{N-j})\cos(\frac{2\pi j k}{N}),
\end{align} 
\end{widetext}
which we can then substitute into the formula for $\Phi(\epsilon)$. Doing so, we get,
\begin{widetext}
\begin{align}
    \Phi(\epsilon)&= \sum_{k=0}^{N-1}\frac{\Theta^2\gamma^2\left(\sum_{j=1}^{N-1}\epsilon \Corr{w_{N-j}} \sin(\frac{2\pi j k}{N})\right)^2}{\Theta - \Theta\gamma\sum_{j=1}^{\lfloor\frac{N}{2}\rfloor}(1-\epsilon)(w_j + w_{N-j})\cos(\frac{2\pi j k}{N}) - \Theta \gamma \sum_{j=1}^{N-1}\epsilon \Corr{w_{N-j}}\cos(\frac{2\pi j k}{N})},\\
    &=\epsilon^2 \sum_{k=0}^{N-1} \frac{\Theta^2\gamma^2\left(\sum_{j=1}^{N-1} \Corr{w_{N-j}} \sin(\frac{2\pi j k}{N})\right)^2}{\Theta - \Theta\gamma\sum_{j=1}^{\lfloor\frac{N}{2}\rfloor}(w_j + w_{N-j})\cos(\frac{2\pi j k}{N})},\\
    &= \epsilon^2\Phi,
    \end{align}    
\end{widetext}
where $\Phi$ is the EPR of the fully directed network ($\Phi(1)$). Again, we can see that the EPR of the OU scales quadratically with $\epsilon$, but in this case all the interpolated networks are normal.
\section{Coarse-graining fails to capture entropy production in small Ising models}
\label{sec: app: coarsegraining}
\noindent In an Ising model with $N$ variables, there are $2^N$ possible configurations, meaning the state space expands exponentially with the system size. For small Ising networks, using the Glauber dynamics \cite{Newman1999MonteCarlo}, we were able to sample trajectories from the Ising model and estimate the steady state probabilities. Then, with the conditional transition probability defined by the model, we were able to estimate the joint transition probability and the EPR. However, for $N>10$, the state space is so large that estimating the steady state probabilities required too many samples and sorting samples into distinct states became computational infeasible. An alternative approach that has been applied for the Ising model and in empirical time-series \cite{lynn2021detailedbalance}, is to measure the EPR in a coarse-grained state-space \cite{Esposito2012coarsegraining,roldan2010dissipation}.
\begin{figure}[h]
\centering
% the number in [] of wrapfigure is optional and gives the number of text lines that should be wrapped around the text. Adjust according to your figures height
\includegraphics[width=\linewidth]{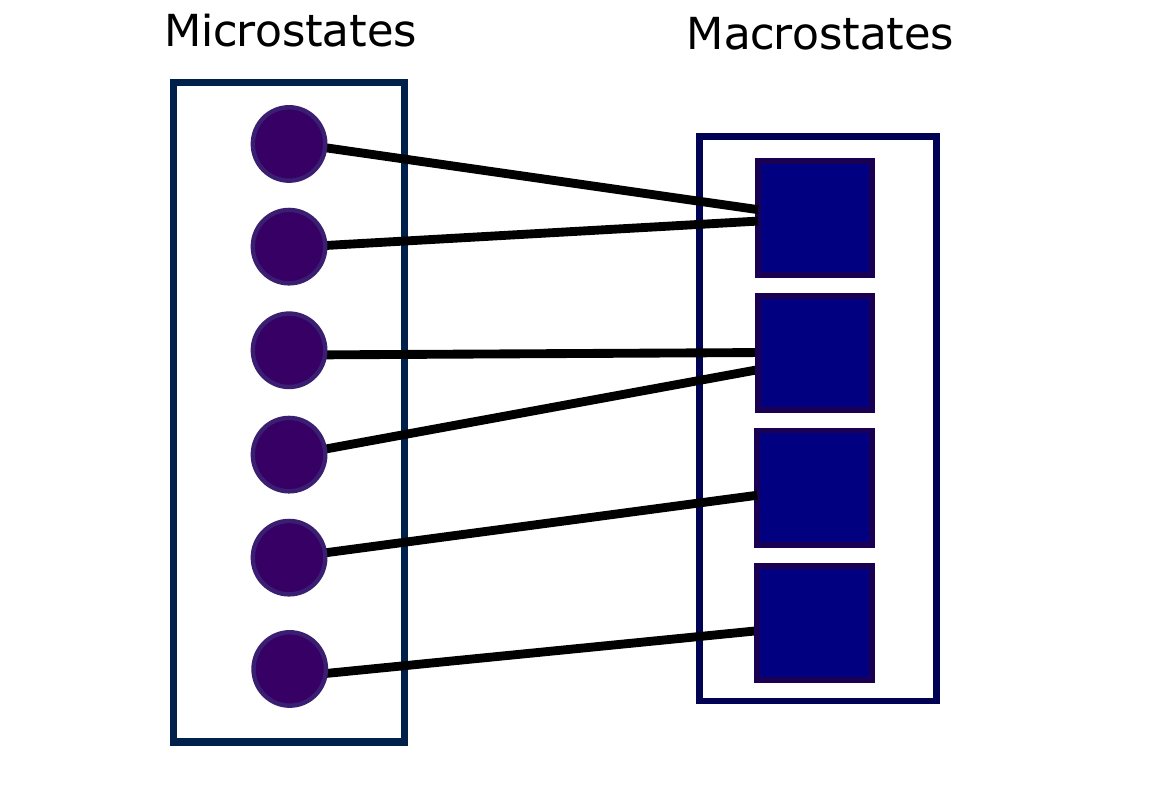}
\caption{\textbf{Coarse-graining of state space:} % note that \ref{fig1} refers to the corresponding wrapfigure
High dimensional state space can be coarse-grained into a lower one by mapping multiple microstates into a lower number of macrostates. This by done via a clustering approach. The EPR in the coarse-grained state space is a lower bound on the entropy production in the original state-space.} % add dummy caption - otherwise \label won't work and figure numbering will not count up
\label{fig: coarsegrainschematic} % use \ref{fig1} to reference to this figure
\end{figure} % avoid blank space here
\noindent Coarse-graining the state-space is achieved by mapping a number of \textit{microstates}, in the original high-dimensional state-space (micro-space), to a lower number of \textit{macrostates} in a lower dimensional space (macro-space). This can be achieved by clustering, in particular hierarchical clustering, which maps `similar' microstates into a single macrostate, as shown by the schematic in Figure \ref{fig: coarsegrainschematic}. Then, the EPR can be estimated by measuring the divergence between forward and backward joint transition probabilities in the coarse-grained space \cite{roldan2010dissipation}. This measurement is a lower-bound on the `true' entropy production in micro-space meaning that broken detailed balance at the coarse-grained level implies non-equilibrium dynamics at the micro-level \cite{lynn2021detailedbalance}. However, as we show here, this lower bound may be inaccurate as a relative estimate of the EPR.\newline\\
\noindent In order to illustrate this, we consider an Ising model with 10 spins. This system has a state space of size $2^{10}=1024$ which is close to the size-limit where we can estimate the steady-state probabilities, yet large enough to cluster states sensibly, meaning we can compare the results in micro- and macro-space. We generate a hierarchical 10-node network and apply the interpolation described previously to vary the directedness of the network. At each value of $\epsilon$, we sample from the Ising model using Glauber dynamics. We then estimate the steady-state probabilities in micro-space and use the conditional transition probability that defines the dynamics to calculate the joint transition probabilities and the EPR, as described in Section \ref{sec: dynamics}. Concurrently, at each $\epsilon$, we apply bisecting, hierarchical $k-$means clustering \cite{lamrous2006hierarchicalkmeans} as applied by Lynn et al \cite{lynn2021detailedbalance} to the samples to coarsen the state-space into $k=10$ macrostates. We note that, while we present the results for $k=10$, this result remained consistent over a reasonable number of macro-states. In the macro-space, we no longer have the conditional transition probability and so we directly estimate the joint transition probability by counting the occurrences of each transition, following Lynn et al \cite{lynn2021detailedbalance}, which can then be used to estimate the EPR.
\begin{figure}[h]
\centering
% the number in [] of wrapfigure is optional and gives the number of text lines that should be wrapped around the text. Adjust according to your figures height
\includegraphics[width=\linewidth]{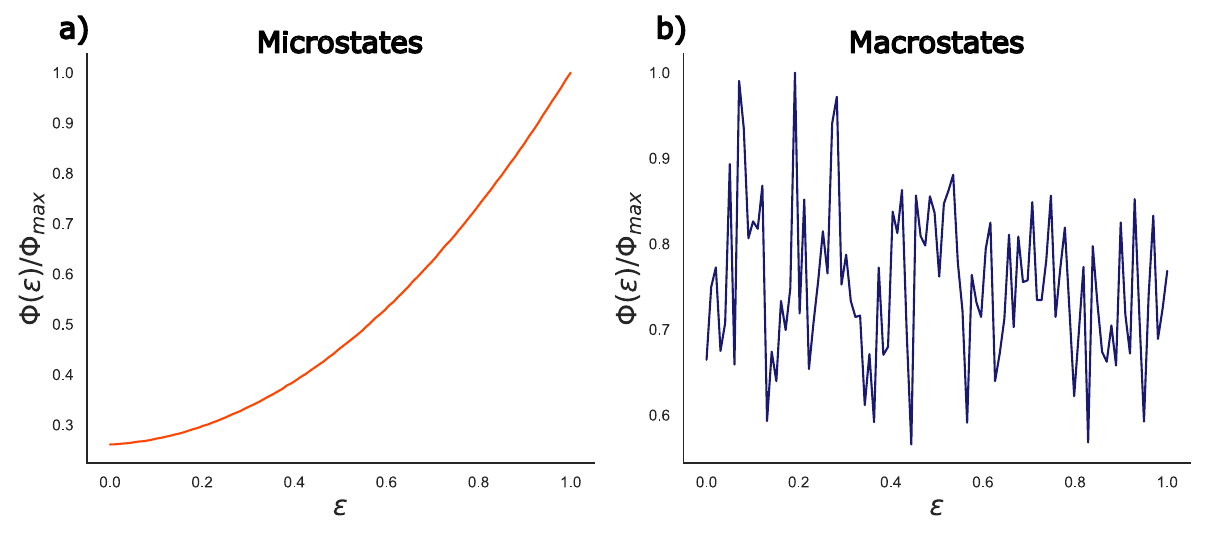}
\caption{\textbf{Entropy production in micro- and macro-space:} % note that \ref{fig1} refers to the corresponding wrapfigure
Normalised EPR in the asymmetric Ising model with 10 nodes, calculated from samples, as a function of the interpolation parameter $\epsilon$. $a)$ The normalised EPR calculated in micro-space which follows the expected behaviour found in Sections \ref{sec: dynamics}, \ref{sec: er} and \ref{sec: realworld}. $b)$ The normalised EPR calculated in a coarse-grained macro-space. There is no correlation between the directedness of the network, $\epsilon$, and the EPR $\frac{\Phi(\epsilon)}{\Phi_{\text{max}}}$.} % add dummy caption - otherwise \label won't work and figure numbering will not count up
\label{fig: coarsegrained} % use \ref{fig1} to reference to this figure
\end{figure} % avoid blank space here
\noindent Figure \ref{fig: coarsegrained} shows the normalised EPR in both state-spaces as a function of $\epsilon$. Panel $a)$ shows that the EPR in micro-space follows the expected behaviour, increasing non-linearly with $\epsilon$, as was found in Sections \ref{sec: entropydirected}, \ref{sec: er} and \ref{sec: realworld}. On the other hand, Panel $b)$ shows that the coarse-graining procedure causes an inaccurate relative measurement of the EPR and there is no correlation between the directedness of the network and the EPR.
\section{Derivation of the naive mean field for the asymmetric Ising model}
\label{sec: app: nMF}
\noindent The naive mean-field (NMF) and the Thouless-Anderson-Palmer (TAP) mean-field are standard approaches to solving the so-called `inverse Ising problem' in equilibrium Ising models \cite{aguilera2021meanfield}. For non-equilibrium (asymmetric) Ising systems, an information-geometric approach can be used to approximate the mean-field solution \cite{kappen2000meanfield}. Here, we will derive the NMF for the asymmetric Ising model following a recent framework that unifies a number of mean-field approaches to the Ising model \cite{aguilera2021meanfield}.\newline\\
\noindent We recall that the Ising model is defined by a discrete time Markov chain where
the spins at time $t+1$ are updated according to,
\begin{align}
\label{eq: conditionalising}
    P(\bm{x}(t+1)|\bm{x}(t)) &= \prod_i \frac{e^{x_i(t+1)h_i(t+1)}}{2\cosh h_{i}(t+1)},\\
    h_i(t+1)&= \frac{1}{T} (H_i +\sum_j\Corr{W_{ji}}x_j(t)),
\end{align}
where $T$ is the thermodynamic temperature, $H_i$ are external fields and $\bm{W} = (W_{ij})$ are the pairwise coupling strengths defined by a weighted network.\newline\\
\noindent Furthermore, we recall that the sufficient thermodynamic quantities we aim to approximate are,
\begin{widetext}
\begin{align}
    m_{i}(t)&=\sum_{\bm{x}(t)}x_{i}(t)P(\bm{x}(t)),\\
    D_{il}(t)&=\sum_{\bm{x}(t),\bm{x}(t-1)}x_{i}(t)x_{l}(t-1)P(\bm{x}(t),\bm{x}(t-1))-m_{i}(t)m_{l}(t-1).
\end{align}
\end{widetext}
where $\bm{m}(t)=(m_1(t),...,m_N(t))$ is the average activation rate of the system
and $\bm{D}(t)$ is the delayed correlation matrix.\newline\\
\noindent Using the language of information geometry \cite{tanaka2001InformationGeometry,amari2001alphaprojection}, we define $\mathcal{P}(t)$ to be the manifold of $P(\bm{x}(t))$ where each point on the manifold corresponds to a set of parameter values. Within $\mathcal{P}(t)$, there are sub-manifolds $\mathcal{Q}(t)$ that are `analytically tractable'. The simplest such manifold $\mathcal{Q}(t)$ is the manifold of models where each spin is independent. Each point on the sub-manifold is defined by a vector of parameters $\bm{\Theta}(t)=\{\Theta_i(t)\}$, and the distribution on this sub-manifold is given by,
\begin{align}
    P(\bm{x}(t)|\bm{\Theta}(t)) = \prod_i\frac{e^{x_i(t)\Theta_i(t)}}{2\cosh \Theta_i(t)}.
\end{align}
The average activation rate of the spins is therefore given by,
\begin{align}
    m_i(t)=\tanh \Theta_i(t).
\end{align}
Given local fields $\bm{H}$ and a network $\bm{W}$, we aim to approximate the thermodynamic quantities of the intractable target distribution $P(\bm{x}(t)|\bm{H},\bm{W}) \in \mathcal{P}(t)$ with a tractable distribution from $\mathcal{Q}(t)$. To do this, we aim to find a distribution $Q(\bm{x}(t))\in \mathcal{Q}(t)$ that minimises the KL-divergence to $P(\bm{x}(t)|\bm{H},\bm{W})$. An important result is that the independent model that minimises the KL divergence, which we denote $Q(\bm{x}(t)|\bm{\Theta}^*(t))$, has identical activation rates to the target \cite{kappen2000meanfield}.\newline\\
\noindent Next, we perform the so-called $\alpha-$projection and the Plefka expansion \cite{amari2001alphaprojection}. We parameterise a curve between a tractable distribution $Q(\bm{x}(t)|\bm{\Theta}(t))$ and the target distribution $P(\bm{x}(t)|\bm{H},\bm{W})$ with a parameter $\alpha\in [0,1]$ such that we have a family of distributions,
\begin{align}
      P_{\alpha}(x_i(t+1)&|\bm{x}(t)) = \prod_i \frac{e^{x_i(t+1)h^{\alpha}_i(t+1)}}{2\cosh h^{\alpha}_{i}(t+1)},\\
    h^{\alpha}_i(t+1)= (1-&\alpha)\Theta_i(t) + \alpha \left( \frac{1}{T} (H_i +\sum_j\Corr{W_{ji}}x_j(t))\right).
\end{align}
Therefore, at $\alpha=0$, we have $P_{0}(\bm{x}(t+1)|\bm{x}(t))=Q(\bm{x}(t+1)|\bm{\Theta}(t+1))$ and at $\alpha =1$, we have $P_{1}(\bm{x}(t+1)|\bm{x}(t))=P(\bm{x}(t+1)|\bm{x}(t))$. We can write the thermodynamic quantities of the distribution at each value of $\alpha$ as $\bm{m}^{\alpha}(t), \bm{D}^{\alpha}(t)$, functions of $\alpha$.\newline\\
\noindent The Plefka expansion is a Taylor expansion around $\alpha=0$,
\begin{align}
    \bm{m}^{\alpha}(t) = \bm{m}^{0}(t) + \sum_{k=1}^{n}\frac{\alpha^k}{k!}\frac{\partial^k\bm{m}^{0}(t)}{\partial \alpha^k}+ \mathcal{O}(\alpha^{(n+1)}),
\end{align}
but we note that $ \bm{m}^{1}(t)=\bm{m}^{0}(t)$ as mentioned earlier \cite{kappen2000meanfield}. Thus, the optimal tractable distribution, $Q(\bm{x}(t)|\bm{\Theta}^*(t))$ satisfies,
\begin{align}
    \sum_{k=1}^{n}\frac{\alpha^k}{k!}\frac{\partial^k\bm{m}^{0}(t)}{\partial \alpha^k}=0,
\end{align}
which should be solved with respect to $\bm{\Theta}(t)$. The approximation is defined first by the number of terms in this sum that we set equal to 0 and then solve for, but also by the time-points at which the model is assumed to have independent units \cite{aguilera2021meanfield}. The NMF is obtained by setting only the first derivative to 0 and by assuming that the model at both $t$ and $t-1$ have independent spins. The first derivative at 0 is given by,
\begin{widetext}
    \begin{align}
    \frac{\partial m_i^{\alpha=0}(t)}{\partial \alpha} =(1-m_i(t)^2)\left(-\Theta_i(t)+H_i + \sum_j \Corr{W_{ji}}m_j(t-1) \right),
\end{align}
\end{widetext}
where we direct the reader to the appendices of \cite{aguilera2021meanfield,kappen2000meanfield} for further details. This gives an approximation of the optimal parameter setting,
\begin{align}
    \bm{\Theta}^*(t) &\approx H_i + \sum_j \Corr{W_{ji}}m_j(t-1),
\end{align}
and the NMF,
\begin{align}
    m_i(t)\approx \tanh \left(H_i + \sum_j \Corr{W_{ji}}m_j(t-1)\right).
\end{align}
Similarly, one can expand $D_{il}^{\alpha}(t)$ around $\alpha=0$ and set $\Theta_i(t)=\Theta^*_i(t)$ to obtain the approximation,
\begin{align}
    D_{il}(t)\approx \Corr{W_{li}}(1-m^2_{i}(t))((1-m^2_{l}(t-1)),
\end{align}
where we direct again to the appendices of \cite{aguilera2021meanfield,kappen2000meanfield} for further details.
\section{Auto-regression with (un)constrained least-squares}
\label{sec: app: autoreg}
\noindent In Section \ref{sec: timeseries}, we defined a method to quantify the EPR and infer an interaction network directly from a MVTS using a linear model. This model is either an unconstrained
\begin{align}
\label{eq: linearmodel2}
\frac{d\bm{x}}{dt}&= -\bm{B}\bm{x}(t) + \bm{\nu}(t),
\end{align}
multivariate OU process, or one that was constrained to the edges of a network,
\begin{align}
\label{eq: networklinearmodel2}
\frac{d\bm{x}}{dt}&= (\bm{\Corr{W^{\top}}-I})\bm{x}(t) + \bm{\nu}(t).
\end{align}
Using the time-discretisation presented in Section \ref{sec: timeseries}, we associate either model with an auto-regressive process of the form \cite{Shumway2017timeseries},
\begin{align}
    \bm{X}(t_{i+1}) = \bm{A}\bm{X}(t_i) + \bm{\chi}(t_i),
\end{align}
where we find the coefficient matrix $\bm{A}$ using least-square regression. In the case of the unconstrained model, we simply solve the convex optimisation problem,
\begin{align}
    \min_{\bm{A}\in \mathbb{R}^N\times \mathbb{R}^N}||\bm{X}_{1:T}-\bm{AX}_{0:T-1}||^2,
\end{align}
where $\bm{X}_{i_1:i_T}$ is an $N \times T$ matrix of data-points where each column is the multivariate observation taken from a time-point $t=i_1,...,i_T$. Such a problem can be solved using any standard convex optimisation solver but is also available as a stand alone function in most scientific programming languages.\\\\
From $\bm{A}$, we can calculate the residuals at each time-step,
\begin{align}
    \bm{\chi}(t_{i-1}) = \bm{X}(t_i)-\bm{AX}(t_{i-1}).
\end{align}
We then estimate $\bm{B}$ and the noise covariance $\bm{D}$ by performing,
\begin{align}
    \bm{B}&= \frac{1}{\Delta t}(\bm{I}-\bm{A}),\;\;
    \bm{D}= \frac{1}{2\Delta t}\Cov[\bm{\chi}]
\end{align}
for some choice of $\Delta t <<1$. Clearly, $\Delta t$ simply scales the process and as such can be chosen arbitrarily to be 0.1.\\\\
In the case of the network-constrained model, we assume $\bm{W}$ has non-negative entries which means we must restrict our solution space. Using the relation,
\begin{align}
    \bm{W} = \frac{1}{\Delta t}(\Corr{\bm{A}^{\top}}-\bm{I})+\bm{I},
\end{align}
we solve the following optimisation problem,
\begin{align}
    \min_{\bm{W}: W_{ij}\geq 0}||\bm{X}_{1:T}-((1-\Delta t)\bm{I}+\Delta t\Corr{\bm{W}^{\top}})\bm{X}_{0:T-1}||^2,
\end{align}
which remains convex. Additional constraints such as no self-loops or restriction to the existence of particular edges can be added without breaking the convexity of the problem. In addition, this particular problem can also be solved using built in non-negative least-squares algorithms in most programming languages subject to appropriate modification. We estimate the noise intensity using,
\begin{align}
    \sigma = \langle \frac{1}{2\Delta t} \text{diag}\Cov[\bm{\chi}]\rangle, 
\end{align}
where $\langle \cdot \rangle $ is the mean. Again, the $\Delta t$ simply scales the process and only affects the diagonal entries (self-loop) of the network, but does not affect the asymmetries.
\\\\
Finally, we note that in order to define the EPR, we require that the process converges to a steady-state. This requires that $\bm{-B}$ or $\bm{\Corr{W^{\top}}-I}$ is a stable matrix i.e. all eigenvalues have negative real part. This constraint is non-convex and so cannot be enforced as part of the algorithm without resorting to more heuristic and complex optimisation methods. In this study, we found that all time-series considered, were best fit with an unconstrained model that converged to a stationary state, which did not always hold for the constrained model. As a result the EPR estimates are obtained with the unconstrained model, whilst the network inference is done with the constrained model.
\section{Empirical network data}
\label{sec: app: networkdata}
\noindent In this study we consider 97 real-world directed networks from a range of different fields including ecology, sociology, biology, language, transport and economics. These network were compiled from a range of different sources and are reported in Table \ref{tab: networkdata}. \Rev{Entries denoted with $\textsuperscript{\textdagger}$ are strongly connected whilst those denoted with $^*$ are not weakly-connected (disconnected).}
\begin{longtable*}[c]{ l  l  l  l}

 \caption{Real-world directed network data from various sources.\label{tab: networkdata}}\\

 \hline
 \hline
 Name & Ref. & Nodes & Edges\\
 \hline
 \cr
 \textbf{Ecological} \cr 
 Marine Foodweb in Bahia Falsa, Mexico & \cite{dunne2013fooweb} & 166 & 9576\\
 Marine Foodweb in Estero de Punta Banda, Mexico & \cite{dunne2013fooweb} & 143 & 3696\\
 Marine Foodweb in Flensburg Fjord, Germany/Denmark & \cite{dunne2013fooweb} & 77 & 576\\
 Marine Foodweb 1 in Ythan Estuary, Scotland & \cite{dunne2013fooweb} & 166 & 9029\\
 Marine Foodweb in Carpinteria Salt Marsh Reserve, USA & \cite{dunne2013fooweb} & 166 & 7682\\
 Marine Foodweb in Sylt Tidal Basin, Germany & \cite{dunne2013fooweb} & 215 & 14963\\
 Marine Foodweb in Otago Harbour, New Zealand & \cite{dunne2013fooweb} & 215 & 15266\\
 River Foodweb in Berwick Stream, New Zealand & \cite{thompson2003foodweb} & 77	& 240\\ 
 River Foodweb 1 in Coweeta, USA & \cite{thompson2003foodweb} & 58	& 126\\ 
 River Foodweb 2 in Coweeta, USA & \cite{thompson2003foodweb} & 71	& 148\\ 
 River Foodweb in Martins Stream, USA & \cite{thompson2003foodweb} & 105 & 343\\
 River Foodweb in Powder Stream, New Zealand & \cite{thompson2003foodweb} & 78	& 268\\
 River Foodweb in Troy Stream, USA & \cite{thompson2003foodweb} & 77	& 181\\
 River Foodweb in Venlaw Stream, New Zealand & \cite{thompson2003foodweb} & 66	& 187\\
 River Foodweb in Black Rock Stream, New Zealand & \cite{thompson2002foodweb} & 86	& 375\\
 River Foodweb in Broad Stream, New Zealand & \cite{thompson2002foodweb} & 94	& 564\\
River Foodweb in Dempsters Stream during summer, New Zealand & \cite{thompson2002foodweb} & 107 &	965\\
River Foodweb in German Creek, New Zealand & \cite{thompson2002foodweb} & 84	& 352\\
River Foodweb in Healy Creek, New Zealand & \cite{thompson2002foodweb} & 96	& 634\\
River Foodweb in Kye Burn, New Zealand& \cite{thompson2002foodweb} & 98	& 629\\
River Foodweb in Little Kye Burn, New Zealand& \cite{thompson2002foodweb} & 78	& 375\\
River Foodweb in Stony Stream, New Zealand& \cite{thompson2002foodweb} & 109 & 827\\
River Foodweb in Sutton Stream during summer, New Zealand & \cite{thompson2002foodweb} & 87 &	424\\
River Foodweb in Canton Creek, New Zealand& \cite{thompson2002foodweb} & 102 & 696\\
River Foodweb in Catlins Stream, New Zealand & \cite{klaise2017foodwebs} & 48 & 110\\
River Foodweb in Dempsters Stream during autumn, New Zealand & \cite{klaise2017foodwebs} & 83 & 414\\
River Foodweb in Dempsters Stream during spring, New Zealand& \cite{klaise2017foodwebs} & 93	& 538\\
River Foodweb in Sutton Stream during autumn, New Zealand& \cite{klaise2017foodwebs} & 80	& 335\\
River Foodweb in Sutton Stream during spring, New Zealand& \cite{klaise2017foodwebs} & 74	& 391\\
River Foodweb in Narrowdale Stream, New Zealand& \cite{thompson2004energy} & 71 & 154\\
River Foodweb in North Col Stream, New Zealand & \cite{thompson2004energy} & 78 & 241\\
$^*$Terrestrial Foodweb in Scotch Broom, England & \cite{memmott2001predators} & 86	& 219\\
Marine Foodweb in Cayman Islands & \cite{bascompte2005marinefoodweb} & 242	& 3764\\
Marine Foodweb in Chesapeake Bay, USA & \cite{ulanowicz1999nutrients} & 31	&67\\
Dominance amongst ants & \cite{cole1981ants} & 16	& 36\\
Dominance amongst kangaroos & \cite{grant1973kangaroos} & 17 & 91\\
Marine Foodweb in St. Marks Estuary, US & \cite{christian1999foodweb} & 48 & 218\\
Terrestrial Foodweb in Saint-Martin Island, Lesser Antilles & \cite{goldwasser1993foodweb} & 42 & 205\\
Marine Foodweb 2 in Ythan Estuary, Scotland & \cite{huxham1996parasites} & 82 & 391\\ 
Lake Foodweb in Lough Hyne, Ireland & \cite{eklof2013ecology} & 349 & 5102\\ 
Marine Foodweb in Weddel Sea, Antarctica & \cite{eklof2013ecology} & 483 & 15317\\ 
Fossil Assemblage Foodweb from Chengjiang Shale, China & \cite{dunne2008cambrianfood} & 33 & 90\\ 
Fossil Assemblage Foodweb from  Burgess Shale, Canada & \cite{dunne2008cambrianfood} & 48 & 243\\ 
Lake Foodweb in Bridge Broom Lake & \cite{havens1992foodwebs} & 25	& 104\\
Dominance amongst wolves & \cite{vanhooff1987wolf} & 16	& 148\\
Lake Foodweb in Little Rock Lake, USA 1 & \cite{martinez1991littlerock} & 183	& 2476\\
Lake Foodweb in Little Rock Lake, USA 2 & \cite{johnsondata} & 92	& 997\\
Marine Foodweb in Northeast United States Shelf & \cite{link2002marine} & 79 & 1378\\
Lake Foodweb in Skipwith Common, England & \cite{warren1989foodweb} & 25	& 189\\
Marine Foodweb in Benguela Current, South Africa & \cite{yodzis1998benguela} & 29	& 196\\
Marine Foodweb in Florida Bay during dry season & \cite{ulanowicz1999southflorida} & 128	& 2137\\
Dominance among ponies & \cite{cluttonbrock1976ponies} & 17&146\\
Dominance among cattle & \cite{schein1955cattle} & 28 & 217\\
Dominance among sheep & \cite{hass1991sheep} & 28 & 250\\
\textsuperscript{\textdagger}Dominance among bison & \cite{lott1979bison} & 26	& 314\\
Dominance among macaques & \cite{strayer1980monkeys} & 62 & 1187\\
Terrestrial Foodweb in grasslands of the United Kingdom & \cite{johnsondata} & 61 & 97\\
Terrestrial Foodweb in El Verde Field Station, Puerto Rico & \cite{johnsondata} & 155 & 1507\\
Terrestrial Foodweb in Coachella Valley, USA & \cite{polis1991coachella} & 29 & 262\\
Marine Foodweb in the Caribbean & \cite{opitz1996caribbean} & 155 & 1507\\
\cr
\textbf{Sociological} \cr
$^*$Political Blogs Network & \cite{adamic2005blogs} & 1224 & 18957\\
Friendship among college students in a course about leadership & \cite{milo2004networks} & 32 & 96\\
Friendship among high-school students & \cite{johnsondata} & 70 & 366\\
Co-purchased political books on Amazon & \cite{clauset2016icon} & 105 & 441\\
Social interactions between inmates in prison & \cite{macrae1960prison} & 67 & 182\\

\cr
\textbf{Biological}\cr
Protein network for 1A4J & \cite{milo2004networks} & 95 & 404\\
Protein network for 1AOR & \cite{milo2004networks} & 96 & 406\\
Protein network for 1EAW & \cite{milo2004networks} & 53 & 236\\
Gene regulatory network for \textit{Saccharomyces
cerevisiae} & \cite{harbison2004genome} & 2933	& 6152\\
Human gene regulatory network for a healthy person & \cite{gerstein2012encode} & 4071 &	8466\\
Human gene regulatory network for a person with cancer & \cite{gerstein2012encode} & 4049 & 11707\\
Gene regulatory network for \textit{Pseudomonas aeruginosa} & \cite{galan2011regnetwork}  & 691 & 991\\
Gene regulatory network for Mycobacterium tuberculosis & \cite{sans2011regnetwork} & 1624 & 3169\\
\textsuperscript{\textdagger}Neuronal network for a mouse brain & \cite{johnsondata} & 213 & 21654\\
\textsuperscript{\textdagger}Connectome of the Rhesus brain, extracted from tract tracing & \cite{harriger2012rhesus} & 242	& 4090\\
Connectome of the Rhesus brain via retrograde tracer & \cite{markov2013macaque} & 91 & 628\\
\textsuperscript{\textdagger}Neuronal network for \textit{Caenorhabditis elegans} & \cite{watts1998smallworld} & 297 & 2345\\
Connectome of the cat brain & \cite{dereus2013cat} & 65 & 1139\\
Connectome of the rat brain & \cite{bota2006neuralnetwork} & 503 & 47329\\
Metabolic network of \textit{Archaeoglobus fulgidus} & \cite{jeong2000metabolic} & 1267 & 3011\\
Metabolic network of \textit{Caenorhabditis elegans} & \cite{jeong2000metabolic} & 1172 & 2864\\
Metabolic network of \textit{Chlamydia pneumoniae} & \cite{jeong2000metabolic} & 386 & 792\\
Metabolic network of \textit{Chlamydia trachomatis} & \cite{jeong2000metabolic} & 446 & 941\\
Metabolic network of \textit{Methanococcus jannaschii} & \cite{jeong2000metabolic} & 1081 & 2589\\
Metabolic network of \textit{Saccharomyces cerevisiae} & \cite{jeong2000metabolic} & 1510 & 3833\\
Metabolic network of \textit{Methanobacterium thermoautotrophicum} & \cite{jeong2000metabolic} & 1111 & 2705\\
\cr
\textbf{Language}\cr
Citations from papers that cite ``Small World Problem” & \cite{garfieldhistcite} & 233 & 994\\
Citations to Small, Griffith and descendants & \cite{garfieldhistcite} & 1024 & 4918\\
Word adjacency network for Dr. Seuss’s Green Eggs and Ham book & \cite{johnsondata} & 50 & 101\\
\cr
\textbf{Trade}\cr
International trade network of minerals & \cite{denooy2018socialnetworks} & 24 & 135\\
\textsuperscript{\textdagger}International trade network of manufactured food products & \cite{denooy2018socialnetworks} & 24 & 307\\
International trade network of manufactured goods & \cite{denooy2018socialnetworks} & 24 & 310\\
\textsuperscript{\textdagger}International trade network of crude animal and vegetable material & \cite{denooy2018socialnetworks} & 24 & 307\\
\textsuperscript{\textdagger}International trade network of diplomatic exchanges & \cite{denooy2018socialnetworks} & 24 & 369\\
\cr
\textbf{Transport}\cr
\textsuperscript{\textdagger}London tube network & \cite{william2016spatiotempnetworks} & 270 & 628\\
\textsuperscript{\textdagger}Paris metropolitan train grid & \cite{william2016spatiotempnetworks} & 302 & 705
\end{longtable*}

\section{Null-modelling of network data with the configuration model}
\label{sec: app: configuration}
\begin{figure*}
    \centering
    \includegraphics[width=\linewidth]{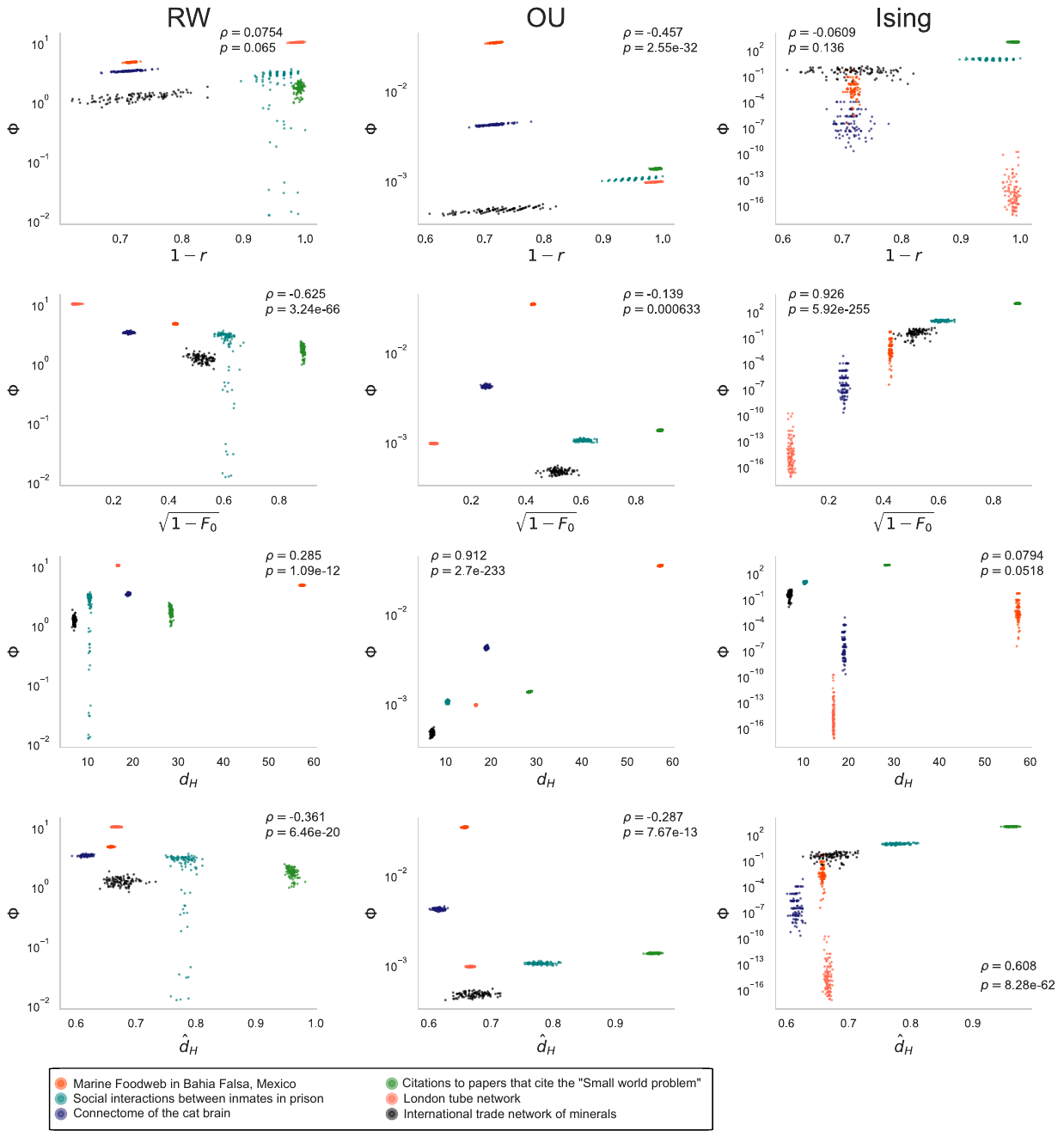}
    \caption{\RevTwo{\textbf{\Corr{Null-modelling of network data with the configuration model:}} We plot the EPR of the three dynamical systems against four directedness measures on 100 reconfigurations of 6 real world networks from each of 6 fields. The first column shows the EPR of the
RW. We again use a perturbation to ensure the reconfigurations are strongly connected and guarantee a stationary distribution. The second column shows the EPR of the OU and the third the EPR of the Ising.}}
    \label{fig: config}
\end{figure*}
\RevTwo{We study the correlation between directedness and EPR in a range of models. First we consider a hierarchical network generated with PA followed by the ER model, real world networks and, finally, networks inferred from MVTS. In this appendix we expand on this further by considering a null model of our empirical network data in order to investigate the relationship between directedness and EPR when other topological features are broken. We employ the directed configuration model \cite{newman2001randomgraphs}, a common null model that preserves exact in- and out- degree sequences whilst randomly rewiring the network. More explicitly, the directed configuration algorithm creates stubs, half-edges, for every edge that exists in the original network. It then randomly connects in- and out-stubs by sampling from a uniform distribution. In doing so, the reconfigured networks preserve the in- and out-degree sequences whilst, on average, destroying other topological features. We consider a single empirical network from each of the six categories detailed in Appendix \ref{sec: app: networkdata} and generate 100 reconfigurations of each. Next, we calculate the EPR of the dynamics on each of these reconfigurations and plot the correlations in Fig. \ref{fig: config} mirroring Fig. \ref{fig: RW entropy}. Firstly, we note that the effect of the configuration model on network directedness is unclear and its action appears to vary each of the directedness measures in a nuanced fashion. Firstly, the global measures of $\sqrt{1-F_0}, d_H, \hat{d}_H$ appear to be \textit{relatively} unaffected by the reconfiguration as evidenced by the dense clustering of reconfigurations from the same original network. As a result, we see a preservation of the main correlations observed in Fig. \ref{fig: RW entropy} such as between the OU process and $d_H$ which confirms that it is network directedness, not other topological features that drive the EPR. On the other hand, it is clear that irreciprocity is more variable under reconfiguration. This leads to each of the clusters being spread out over the $x-$axis. This results in a very weak overall correlation between irreciprocity and EPR for the RW. However, within clusters, it appears that there is a very strong correlation between irreciprocity and EPR for both the RW and the OU process. This indicates that for `comparable networks', of similar size or degree distribution, the irreciprocity is predictive of the EPR but not when comparing networks with a wide range of different sizes and structures.\\\\
Nevertheless, the relationship between global directedness, EPR and other topological features such as modularity, small-worldness etc. is an interesting one that remains to be explored. Whilst directedness is intrinsically linked to other topological features, thus cannot be perfectly decoupled, the observation that non-normality and trophic directedness remain relatively constant under the configuration model opens up avenues to investigate this relationship further.}
\section{Empirical time-series data}
\label{sec: app: time-series}
\subsection{Human neuroimaging}
\begin{figure}[t!]
    \centering
    \includegraphics[width=0.5\textwidth]{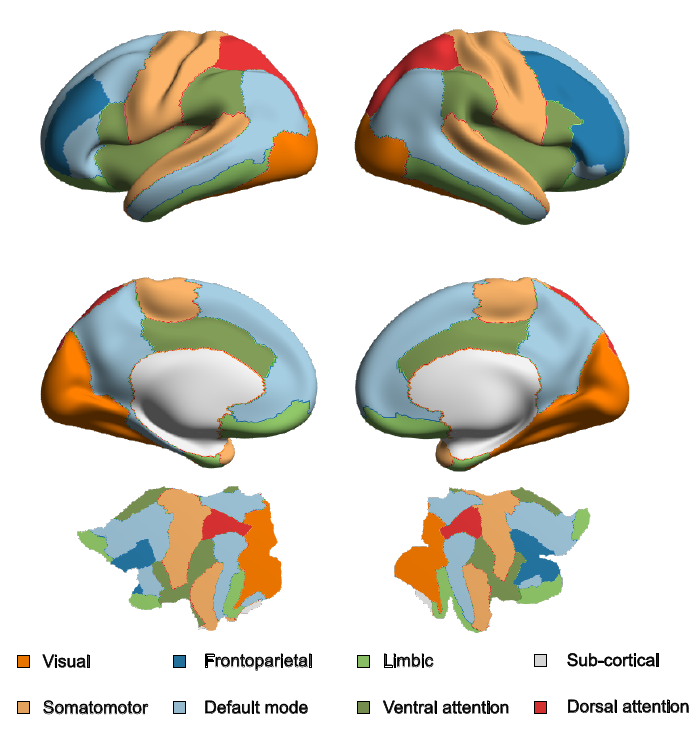}
    \caption{\textbf{Yeo parcellation projected onto the connectome:} The 7 canonical resting state networks are show projected onto the DK80 parcellation with the sub-cortical regions also labelled.}
    \label{fig: dk80yeo}
\end{figure}
\noindent In Section \ref{sec: timeseries} we analysed functional magnetic resonance imaging (fMRI) from 100 unrelated participants at rest and during task. This data is freely available as part of the Human Connectome Project HCP1003 release. The data used here is the same as that analysed previously in Ref. \cite{deco2021workspace}. In brief, we used the Desikan-Killany parcellation \cite{desikan2006dk80} made up of 62 cortical regions and 18 sub-cortical regions for a total of 80 regions of interest (DK80). The data was pre-processed using the HCP pipeline using standard software packages from the FMRIB Software Library, FreeSurfer and the Connectome Workbench \cite{glasser2013HCP,smith2013restinghcp}. This included correcting for spatial and gradient distortions, head motions and further included bias-field removal, intensity normalisation, registration to a T1-structural image, transformation to 2mm MNI (Montreal Neurological Institute) space and application of the FIX artefact removal procedure \cite{smith2013restinghcp,schroder2015topography}. Head motion was regressed out and, using independent component analysis, artefacts were removed using ICA+FIX processing \cite{salimi2014autodenoise,griffanti2014ica}. Using the Fieldtrip toolbox \cite{oostenveld2011FT}, the average time-series of the grayordinates in each region of the DK80 parcellation was extracted.\\\\
We further parcellate the DK80 into 8 sub-networks, the 7 Yeo resting-state networks \cite{yeo2011restingnetworks}, and the sub-cortical regions. The projection of the 8 sub-networks onto the DK80 parcellation is shown in Figure \ref{fig: dk80yeo}.
\subsection{Stock-prices from the New York Stock Exchange}
\noindent In Section \ref{sec: timeseries}, we analysed stock prices from 119 U.S. companies from the New York Stock Exchange (NYSE) in the period 1 January 2000 to 17 June 2021. This is the same financial dataset previously studied in Ref. \cite{santoro2023higherorder}. It was obtained from the Yahoo! finance historical data application programming interface (`yfinance' Python library) but is freely deposited online in Ref. \cite{santoro2023higherorderdata}.  
\bibliography{Report}% Produces the bibliography via BibTeX.

\providecommand{\noopsort}[1]{}\providecommand{\singleletter}[1]{#1}%
\begin{thebibliography}{186}%
\makeatletter
\providecommand \@ifxundefined [1]{%
 \@ifx{#1\undefined}
}%
\providecommand \@ifnum [1]{%
 \ifnum #1\expandafter \@firstoftwo
 \else \expandafter \@secondoftwo
 \fi
}%
\providecommand \@ifx [1]{%
 \ifx #1\expandafter \@firstoftwo
 \else \expandafter \@secondoftwo
 \fi
}%
\providecommand \natexlab [1]{#1}%
\providecommand \enquote  [1]{``#1''}%
\providecommand \bibnamefont  [1]{#1}%
\providecommand \bibfnamefont [1]{#1}%
\providecommand \citenamefont [1]{#1}%
\providecommand \href@noop [0]{\@secondoftwo}%
\providecommand \href [0]{\begingroup \@sanitize@url \@href}%
\providecommand \@href[1]{\@@startlink{#1}\@@href}%
\providecommand \@@href[1]{\endgroup#1\@@endlink}%
\providecommand \@sanitize@url [0]{\catcode `\\12\catcode `\$12\catcode `\&12\catcode `\#12\catcode `\^12\catcode `\_12\catcode `\%12\relax}%
\providecommand \@@startlink[1]{}%
\providecommand \@@endlink[0]{}%
\providecommand \url  [0]{\begingroup\@sanitize@url \@url }%
\providecommand \@url [1]{\endgroup\@href {#1}{\urlprefix }}%
\providecommand \urlprefix  [0]{URL }%
\providecommand \Eprint [0]{\href }%
\providecommand \doibase [0]{https://doi.org/}%
\providecommand \selectlanguage [0]{\@gobble}%
\providecommand \bibinfo  [0]{\@secondoftwo}%
\providecommand \bibfield  [0]{\@secondoftwo}%
\providecommand \translation [1]{[#1]}%
\providecommand \BibitemOpen [0]{}%
\providecommand \bibitemStop [0]{}%
\providecommand \bibitemNoStop [0]{.\EOS\space}%
\providecommand \EOS [0]{\spacefactor3000\relax}%
\providecommand \BibitemShut  [1]{\csname bibitem#1\endcsname}%
\let\auto@bib@innerbib\@empty
%</preamble>
\bibitem [{\citenamefont {Dunne}\ \emph {et~al.}(2002)\citenamefont {Dunne}, \citenamefont {Williams},\ and\ \citenamefont {Martinez}}]{dunne2002ecology}%
  \BibitemOpen
  \bibfield  {author} {\bibinfo {author} {\bibfnamefont {J.~A.}\ \bibnamefont {Dunne}}, \bibinfo {author} {\bibfnamefont {R.~J.}\ \bibnamefont {Williams}},\ and\ \bibinfo {author} {\bibfnamefont {N.~D.}\ \bibnamefont {Martinez}},\ }\bibfield  {title} {\bibinfo {title} {Food-web structure and network theory: The role of connectance and size},\ }\href@noop {} {\bibfield  {journal} {\bibinfo  {journal} {Proceedings of the National Academy of Sciences of the United States of America}\ }\textbf {\bibinfo {volume} {99}},\ \bibinfo {pages} {12917} (\bibinfo {year} {2002})}\BibitemShut {NoStop}%
\bibitem [{\citenamefont {Jackson}(2010)}]{jackson2010socialeconomic}%
  \BibitemOpen
  \bibfield  {author} {\bibinfo {author} {\bibfnamefont {M.~O.}\ \bibnamefont {Jackson}},\ }\href@noop {} {\emph {\bibinfo {title} {Social and Economic Networks}}}\ (\bibinfo  {publisher} {Princeton University Press},\ \bibinfo {year} {2010})\BibitemShut {NoStop}%
\bibitem [{\citenamefont {Wasserman}(1994)}]{wasserman2012socialnetwork}%
  \BibitemOpen
  \bibfield  {author} {\bibinfo {author} {\bibfnamefont {S.}~\bibnamefont {Wasserman}},\ }\href@noop {} {\emph {\bibinfo {title} {Social Network Analysis}}}\ (\bibinfo  {publisher} {Cambridge University Press},\ \bibinfo {year} {1994})\BibitemShut {NoStop}%
\bibitem [{\citenamefont {Barabási}\ \emph {et~al.}(2011)\citenamefont {Barabási}, \citenamefont {Gulbahce},\ and\ \citenamefont {Loscalzo}}]{barabasi2011networkmedicine}%
  \BibitemOpen
  \bibfield  {author} {\bibinfo {author} {\bibfnamefont {A.-L.}\ \bibnamefont {Barabási}}, \bibinfo {author} {\bibfnamefont {N.}~\bibnamefont {Gulbahce}},\ and\ \bibinfo {author} {\bibfnamefont {J.}~\bibnamefont {Loscalzo}},\ }\bibfield  {title} {\bibinfo {title} {Network medicine: A network-based approach to human disease},\ }\href@noop {} {\bibfield  {journal} {\bibinfo  {journal} {Nature Reviews Genetics}\ }\textbf {\bibinfo {volume} {12}},\ \bibinfo {pages} {56} (\bibinfo {year} {2011})}\BibitemShut {NoStop}%
\bibitem [{\citenamefont {Bullmore}\ and\ \citenamefont {Sporns}(2009)}]{bullmore2009brainnetwork}%
  \BibitemOpen
  \bibfield  {author} {\bibinfo {author} {\bibfnamefont {E.}~\bibnamefont {Bullmore}}\ and\ \bibinfo {author} {\bibfnamefont {O.}~\bibnamefont {Sporns}},\ }\bibfield  {title} {\bibinfo {title} {Complex brain networks: graph theoretical analysis of structural and functional systems},\ }\href@noop {} {\bibfield  {journal} {\bibinfo  {journal} {Nature Reviews Neuroscience}\ }\textbf {\bibinfo {volume} {10}},\ \bibinfo {pages} {186} (\bibinfo {year} {2009})}\BibitemShut {NoStop}%
\bibitem [{\citenamefont {Bassett}\ and\ \citenamefont {Sporns}(2017)}]{basset2017networkneuro}%
  \BibitemOpen
  \bibfield  {author} {\bibinfo {author} {\bibfnamefont {D.~S.}\ \bibnamefont {Bassett}}\ and\ \bibinfo {author} {\bibfnamefont {O.}~\bibnamefont {Sporns}},\ }\bibfield  {title} {\bibinfo {title} {Network neuroscience},\ }\href@noop {} {\bibfield  {journal} {\bibinfo  {journal} {Nature Neuroscience}\ }\textbf {\bibinfo {volume} {20}},\ \bibinfo {pages} {353–364} (\bibinfo {year} {2017})}\BibitemShut {NoStop}%
\bibitem [{\citenamefont {Newman}(2018{\natexlab{a}})}]{newman2018networks}%
  \BibitemOpen
  \bibfield  {author} {\bibinfo {author} {\bibfnamefont {M.}~\bibnamefont {Newman}},\ }\href@noop {} {\emph {\bibinfo {title} {Networks}}}\ (\bibinfo  {publisher} {Oxford University Press},\ \bibinfo {year} {2018})\BibitemShut {NoStop}%
\bibitem [{\citenamefont {Newman}(2003)}]{newman2003structurefunction}%
  \BibitemOpen
  \bibfield  {author} {\bibinfo {author} {\bibfnamefont {M.~E.~J.}\ \bibnamefont {Newman}},\ }\bibfield  {title} {\bibinfo {title} {The structure and function of complex networks},\ }\href@noop {} {\bibfield  {journal} {\bibinfo  {journal} {SIAM Review}\ }\textbf {\bibinfo {volume} {45}},\ \bibinfo {pages} {167} (\bibinfo {year} {2003})}\BibitemShut {NoStop}%
\bibitem [{\citenamefont {Boccaletti}\ \emph {et~al.}(2006)\citenamefont {Boccaletti}, \citenamefont {Latora}, \citenamefont {Moreno}, \citenamefont {Chavez},\ and\ \citenamefont {Hwang}}]{boccaletti2006complexdynamis}%
  \BibitemOpen
  \bibfield  {author} {\bibinfo {author} {\bibfnamefont {S.}~\bibnamefont {Boccaletti}}, \bibinfo {author} {\bibfnamefont {V.}~\bibnamefont {Latora}}, \bibinfo {author} {\bibfnamefont {Y.}~\bibnamefont {Moreno}}, \bibinfo {author} {\bibfnamefont {M.}~\bibnamefont {Chavez}},\ and\ \bibinfo {author} {\bibfnamefont {D.-U.}\ \bibnamefont {Hwang}},\ }\bibfield  {title} {\bibinfo {title} {Complex networks: Structure and dynamics},\ }\href@noop {} {\bibfield  {journal} {\bibinfo  {journal} {Physics Reports}\ }\textbf {\bibinfo {volume} {424}},\ \bibinfo {pages} {175} (\bibinfo {year} {2006})}\BibitemShut {NoStop}%
\bibitem [{\citenamefont {Barrat}\ \emph {et~al.}(2008)\citenamefont {Barrat}, \citenamefont {Barthélemy},\ and\ \citenamefont {Vespignani}}]{barrat2008dynamical}%
  \BibitemOpen
  \bibfield  {author} {\bibinfo {author} {\bibfnamefont {A.}~\bibnamefont {Barrat}}, \bibinfo {author} {\bibfnamefont {M.}~\bibnamefont {Barthélemy}},\ and\ \bibinfo {author} {\bibfnamefont {A.}~\bibnamefont {Vespignani}},\ }\href@noop {} {\emph {\bibinfo {title} {Dynamical Processes on Complex Networks}}}\ (\bibinfo  {publisher} {Cambridge University Press},\ \bibinfo {year} {2008})\BibitemShut {NoStop}%
\bibitem [{\citenamefont {Prigogine}(1968)}]{Prigogine1968Thermodynamics}%
  \BibitemOpen
  \bibfield  {author} {\bibinfo {author} {\bibfnamefont {I.}~\bibnamefont {Prigogine}},\ }\href@noop {} {\emph {\bibinfo {title} {Introduction to Thermodynamics of Irreversible Processes}}}\ (\bibinfo  {publisher} {John Wiley \& Sons},\ \bibinfo {address} {California, United States},\ \bibinfo {year} {1968})\BibitemShut {NoStop}%
\bibitem [{\citenamefont {Schrödinger}(1944)}]{Schrodinger1944whatislife}%
  \BibitemOpen
  \bibfield  {author} {\bibinfo {author} {\bibfnamefont {E.}~\bibnamefont {Schrödinger}},\ }\href@noop {} {\emph {\bibinfo {title} {What is Life? The Physical Aspect of the Living Cell and Mind}}}\ (\bibinfo  {publisher} {Cambridge University Press},\ \bibinfo {address} {Cambridge, United Kingdom},\ \bibinfo {year} {1944})\BibitemShut {NoStop}%
\bibitem [{\citenamefont {Seifert}(2012)}]{seifert2012thermodynamics}%
  \BibitemOpen
  \bibfield  {author} {\bibinfo {author} {\bibfnamefont {U.}~\bibnamefont {Seifert}},\ }\bibfield  {title} {\bibinfo {title} {Stochastic thermodynamics, fluctuation theorems and molecular machines},\ }\href@noop {} {\bibfield  {journal} {\bibinfo  {journal} {Reports of Progress in Physics}\ }\textbf {\bibinfo {volume} {475}} (\bibinfo {year} {2012})}\BibitemShut {NoStop}%
\bibitem [{\citenamefont {Roldán}\ and\ \citenamefont {Parrondo}(2010)}]{roldan2010dissipation}%
  \BibitemOpen
  \bibfield  {author} {\bibinfo {author} {\bibfnamefont {E.}~\bibnamefont {Roldán}}\ and\ \bibinfo {author} {\bibfnamefont {J.~M.~R.}\ \bibnamefont {Parrondo}},\ }\bibfield  {title} {\bibinfo {title} {Estimating dissipation from single stationary trajectories},\ }\href@noop {} {\bibfield  {journal} {\bibinfo  {journal} {Physical Review Letters}\ }\textbf {\bibinfo {volume} {105}} (\bibinfo {year} {2010})}\BibitemShut {NoStop}%
\bibitem [{\citenamefont {Gnesotto}\ \emph {et~al.}(2018)\citenamefont {Gnesotto}, \citenamefont {Mura}, \citenamefont {Gladrow},\ and\ \citenamefont {Broedersz}}]{Gnesotto2018brokendetailedbalance}%
  \BibitemOpen
  \bibfield  {author} {\bibinfo {author} {\bibfnamefont {F.~S.}\ \bibnamefont {Gnesotto}}, \bibinfo {author} {\bibfnamefont {F.}~\bibnamefont {Mura}}, \bibinfo {author} {\bibfnamefont {J.}~\bibnamefont {Gladrow}},\ and\ \bibinfo {author} {\bibfnamefont {C.~P.}\ \bibnamefont {Broedersz}},\ }\bibfield  {title} {\bibinfo {title} {Broken detailed balance and non-equilibrium dynamics in living systems: a review},\ }\href@noop {} {\bibfield  {journal} {\bibinfo  {journal} {Reports on Progress in Physics}\ }\textbf {\bibinfo {volume} {81}} (\bibinfo {year} {2018})}\BibitemShut {NoStop}%
\bibitem [{\citenamefont {Brangwynne}\ \emph {et~al.}(2008)\citenamefont {Brangwynne}, \citenamefont {Koenderink}, \citenamefont {MacKintosh},\ and\ \citenamefont {Weitz}}]{brangwynne2008cytoplasmicdiffusion}%
  \BibitemOpen
  \bibfield  {author} {\bibinfo {author} {\bibfnamefont {C.~P.}\ \bibnamefont {Brangwynne}}, \bibinfo {author} {\bibfnamefont {G.~H.}\ \bibnamefont {Koenderink}}, \bibinfo {author} {\bibfnamefont {F.~C.}\ \bibnamefont {MacKintosh}},\ and\ \bibinfo {author} {\bibfnamefont {D.~A.}\ \bibnamefont {Weitz}},\ }\bibfield  {title} {\bibinfo {title} {Cytoplasmic diffusion: molecular motors mix it up},\ }\href@noop {} {\bibfield  {journal} {\bibinfo  {journal} {Journal of Cell Biology}\ }\textbf {\bibinfo {volume} {183}},\ \bibinfo {pages} {583} (\bibinfo {year} {2008})}\BibitemShut {NoStop}%
\bibitem [{\citenamefont {Yin}\ \emph {et~al.}(1999)\citenamefont {Yin}, \citenamefont {Artsimovitch}, \citenamefont {Landick},\ and\ \citenamefont {Gelles}}]{yin1999nonequilibriumRNA}%
  \BibitemOpen
  \bibfield  {author} {\bibinfo {author} {\bibfnamefont {H.}~\bibnamefont {Yin}}, \bibinfo {author} {\bibfnamefont {I.}~\bibnamefont {Artsimovitch}}, \bibinfo {author} {\bibfnamefont {R.}~\bibnamefont {Landick}},\ and\ \bibinfo {author} {\bibfnamefont {J.}~\bibnamefont {Gelles}},\ }\bibfield  {title} {\bibinfo {title} {Nonequilibrium mechanism of transcription termination from observations of single {RNA} polymerase molecules},\ }\href@noop {} {\bibfield  {journal} {\bibinfo  {journal} {Proceedings of the National Academy of Sciences of the United States of America}\ }\textbf {\bibinfo {volume} {96}},\ \bibinfo {pages} {13124} (\bibinfo {year} {1999})}\BibitemShut {NoStop}%
\bibitem [{\citenamefont {Huang}\ \emph {et~al.}(2003)\citenamefont {Huang}, \citenamefont {Meir},\ and\ \citenamefont {Wingreen}}]{Huang2003ecoli}%
  \BibitemOpen
  \bibfield  {author} {\bibinfo {author} {\bibfnamefont {K.~C.}\ \bibnamefont {Huang}}, \bibinfo {author} {\bibfnamefont {Y.}~\bibnamefont {Meir}},\ and\ \bibinfo {author} {\bibfnamefont {N.~S.}\ \bibnamefont {Wingreen}},\ }\bibfield  {title} {\bibinfo {title} {Dynamic structures in escherichia coli: Spontaneous formation of {MinE} rings and {MinD} polar zones},\ }\href@noop {} {\bibfield  {journal} {\bibinfo  {journal} {Proceedings of the National Academy of Sciences of the United States of America}\ }\textbf {\bibinfo {volume} {100}},\ \bibinfo {pages} {12724} (\bibinfo {year} {2003})}\BibitemShut {NoStop}%
\bibitem [{\citenamefont {Mehta}\ and\ \citenamefont {Schwab}(2012)}]{mehta2012cellularcomputation}%
  \BibitemOpen
  \bibfield  {author} {\bibinfo {author} {\bibfnamefont {P.}~\bibnamefont {Mehta}}\ and\ \bibinfo {author} {\bibfnamefont {D.~J.}\ \bibnamefont {Schwab}},\ }\bibfield  {title} {\bibinfo {title} {Energetic costs of cellular computation},\ }\href@noop {} {\bibfield  {journal} {\bibinfo  {journal} {Proceedings of the National Academy of Sciences of the United States of America}\ }\textbf {\bibinfo {volume} {109}},\ \bibinfo {pages} {17978} (\bibinfo {year} {2012})}\BibitemShut {NoStop}%
\bibitem [{\citenamefont {Stuhrmann}\ \emph {et~al.}(2012)\citenamefont {Stuhrmann}, \citenamefont {Silva}, \citenamefont {Depken}, \citenamefont {Mackintosh},\ and\ \citenamefont {Koenderink}}]{stuhrman2012cytoskeleton}%
  \BibitemOpen
  \bibfield  {author} {\bibinfo {author} {\bibfnamefont {B.}~\bibnamefont {Stuhrmann}}, \bibinfo {author} {\bibfnamefont {M.~S.~E.}\ \bibnamefont {Silva}}, \bibinfo {author} {\bibfnamefont {M.}~\bibnamefont {Depken}}, \bibinfo {author} {\bibfnamefont {F.~C.}\ \bibnamefont {Mackintosh}},\ and\ \bibinfo {author} {\bibfnamefont {G.~H.}\ \bibnamefont {Koenderink}},\ }\bibfield  {title} {\bibinfo {title} {Nonequilibrium fluctuations of a remodeling in vitro cytoskeleton},\ }\href@noop {} {\bibfield  {journal} {\bibinfo  {journal} {Physical Review E}\ }\textbf {\bibinfo {volume} {86}} (\bibinfo {year} {2012})}\BibitemShut {NoStop}%
\bibitem [{\citenamefont {Battle}\ \emph {et~al.}(2016)\citenamefont {Battle}, \citenamefont {Broedersz}, \citenamefont {Fakhri}, \citenamefont {Geyer}, \citenamefont {Howard}, \citenamefont {Schmidt},\ and\ \citenamefont {Mackintosh}}]{battle2016brokendetailedbalance}%
  \BibitemOpen
  \bibfield  {author} {\bibinfo {author} {\bibfnamefont {C.}~\bibnamefont {Battle}}, \bibinfo {author} {\bibfnamefont {C.~P.}\ \bibnamefont {Broedersz}}, \bibinfo {author} {\bibfnamefont {N.}~\bibnamefont {Fakhri}}, \bibinfo {author} {\bibfnamefont {V.~F.}\ \bibnamefont {Geyer}}, \bibinfo {author} {\bibfnamefont {J.}~\bibnamefont {Howard}}, \bibinfo {author} {\bibfnamefont {C.~F.}\ \bibnamefont {Schmidt}},\ and\ \bibinfo {author} {\bibfnamefont {F.~C.}\ \bibnamefont {Mackintosh}},\ }\bibfield  {title} {\bibinfo {title} {Broken detailed balance at mesoscopic scales in active biological systems},\ }\href@noop {} {\bibfield  {journal} {\bibinfo  {journal} {Science}\ }\textbf {\bibinfo {volume} {352}},\ \bibinfo {pages} {604} (\bibinfo {year} {2016})}\BibitemShut {NoStop}%
\bibitem [{\citenamefont {England}(2013)}]{England2013selfreplication}%
  \BibitemOpen
  \bibfield  {author} {\bibinfo {author} {\bibfnamefont {J.~L.}\ \bibnamefont {England}},\ }\bibfield  {title} {\bibinfo {title} {Statistical physics of self-replication},\ }\href@noop {} {\bibfield  {journal} {\bibinfo  {journal} {Journal of Chemical Physics}\ }\textbf {\bibinfo {volume} {139}} (\bibinfo {year} {2013})}\BibitemShut {NoStop}%
\bibitem [{\citenamefont {Lynn}\ \emph {et~al.}(2021)\citenamefont {Lynn}, \citenamefont {Cornblath}, \citenamefont {Papadopoulos},\ and\ \citenamefont {Bassett}}]{lynn2021detailedbalance}%
  \BibitemOpen
  \bibfield  {author} {\bibinfo {author} {\bibfnamefont {C.~W.}\ \bibnamefont {Lynn}}, \bibinfo {author} {\bibfnamefont {E.~J.}\ \bibnamefont {Cornblath}}, \bibinfo {author} {\bibfnamefont {L.}~\bibnamefont {Papadopoulos}},\ and\ \bibinfo {author} {\bibfnamefont {D.~S.}\ \bibnamefont {Bassett}},\ }\bibfield  {title} {\bibinfo {title} {Broken detailed balance and entropy production in the human brain},\ }\href@noop {} {\bibfield  {journal} {\bibinfo  {journal} {Proceedings of the National Academy of Sciences of the United States of America}\ }\textbf {\bibinfo {volume} {118}} (\bibinfo {year} {2021})}\BibitemShut {NoStop}%
\bibitem [{\citenamefont {Deco}\ \emph {et~al.}(2023{\natexlab{a}})\citenamefont {Deco}, \citenamefont {Sanz-Perl}, \citenamefont {de~la Fuente}, \citenamefont {Sitt}, \citenamefont {Yeo}, \citenamefont {Tagliazucchi},\ and\ \citenamefont {Kringelbach}}]{deco2023tenet}%
  \BibitemOpen
  \bibfield  {author} {\bibinfo {author} {\bibfnamefont {G.}~\bibnamefont {Deco}}, \bibinfo {author} {\bibfnamefont {Y.}~\bibnamefont {Sanz-Perl}}, \bibinfo {author} {\bibfnamefont {L.}~\bibnamefont {de~la Fuente}}, \bibinfo {author} {\bibfnamefont {J.~D.}\ \bibnamefont {Sitt}}, \bibinfo {author} {\bibfnamefont {B.~T.~T.}\ \bibnamefont {Yeo}}, \bibinfo {author} {\bibfnamefont {E.}~\bibnamefont {Tagliazucchi}},\ and\ \bibinfo {author} {\bibfnamefont {M.~L.}\ \bibnamefont {Kringelbach}},\ }\bibfield  {title} {\bibinfo {title} {The arrow of time of brain signals in cognition: Potential intriguing role of parts of the default mode network},\ }\href@noop {} {\bibfield  {journal} {\bibinfo  {journal} {Network Neuroscience}\ }\textbf {\bibinfo {volume} {7}},\ \bibinfo {pages} {966–998} (\bibinfo {year} {2023}{\natexlab{a}})}\BibitemShut {NoStop}%
\bibitem [{\citenamefont {Bolton}\ \emph {et~al.}(2023)\citenamefont {Bolton}, \citenamefont {Van De~Ville}, \citenamefont {Amico}, \citenamefont {Preti},\ and\ \citenamefont {Liégeois}}]{bolton2023AoT}%
  \BibitemOpen
  \bibfield  {author} {\bibinfo {author} {\bibfnamefont {T.}~\bibnamefont {Bolton}}, \bibinfo {author} {\bibfnamefont {D.}~\bibnamefont {Van De~Ville}}, \bibinfo {author} {\bibfnamefont {E.}~\bibnamefont {Amico}}, \bibinfo {author} {\bibfnamefont {M.}~\bibnamefont {Preti}},\ and\ \bibinfo {author} {\bibfnamefont {R.}~\bibnamefont {Liégeois}},\ }\bibfield  {title} {\bibinfo {title} {The arrow-of-time in neuroimaging time series identifies causal triggers of brain function},\ }\href@noop {} {\bibfield  {journal} {\bibinfo  {journal} {Human Brain Mapping}\ }\textbf {\bibinfo {volume} {44}},\ \bibinfo {pages} {4077} (\bibinfo {year} {2023})}\BibitemShut {NoStop}%
\bibitem [{\citenamefont {Aguilera}\ \emph {et~al.}(2023)\citenamefont {Aguilera}, \citenamefont {Igarashi},\ and\ \citenamefont {Shimazaki}}]{aguilera2023sherrington}%
  \BibitemOpen
  \bibfield  {author} {\bibinfo {author} {\bibfnamefont {M.}~\bibnamefont {Aguilera}}, \bibinfo {author} {\bibfnamefont {M.}~\bibnamefont {Igarashi}},\ and\ \bibinfo {author} {\bibfnamefont {H.}~\bibnamefont {Shimazaki}},\ }\bibfield  {title} {\bibinfo {title} {Nonequilibrium thermodynamics of the asymmetric {Sherrington-Kirkpatrick} model},\ }\href@noop {} {\bibfield  {journal} {\bibinfo  {journal} {Nature Communications}\ }\textbf {\bibinfo {volume} {14}} (\bibinfo {year} {2023})}\BibitemShut {NoStop}%
\bibitem [{\citenamefont {Herpich}\ \emph {et~al.}(2018)\citenamefont {Herpich}, \citenamefont {Thingna},\ and\ \citenamefont {Esposito}}]{herpich2018collective}%
  \BibitemOpen
  \bibfield  {author} {\bibinfo {author} {\bibfnamefont {T.}~\bibnamefont {Herpich}}, \bibinfo {author} {\bibfnamefont {J.}~\bibnamefont {Thingna}},\ and\ \bibinfo {author} {\bibfnamefont {M.}~\bibnamefont {Esposito}},\ }\bibfield  {title} {\bibinfo {title} {Collective power: minimal model for thermodynamics of nonequilibrium phase transitions},\ }\href@noop {} {\bibfield  {journal} {\bibinfo  {journal} {Physical Review X}\ }\textbf {\bibinfo {volume} {8}} (\bibinfo {year} {2018})}\BibitemShut {NoStop}%
\bibitem [{\citenamefont {Herpich}\ \emph {et~al.}(2020)\citenamefont {Herpich}, \citenamefont {Cossetto}, \citenamefont {Falasco},\ and\ \citenamefont {Esposito}}]{Herphic2020manybody}%
  \BibitemOpen
  \bibfield  {author} {\bibinfo {author} {\bibfnamefont {T.}~\bibnamefont {Herpich}}, \bibinfo {author} {\bibfnamefont {T.}~\bibnamefont {Cossetto}}, \bibinfo {author} {\bibfnamefont {G.}~\bibnamefont {Falasco}},\ and\ \bibinfo {author} {\bibfnamefont {M.}~\bibnamefont {Esposito}},\ }\bibfield  {title} {\bibinfo {title} {Stochastic thermodynamics of all-to-all interacting many-body systems},\ }\href@noop {} {\bibfield  {journal} {\bibinfo  {journal} {New Journal of Physics}\ }\textbf {\bibinfo {volume} {22}} (\bibinfo {year} {2020})}\BibitemShut {NoStop}%
\bibitem [{\citenamefont {Suñé}\ and\ \citenamefont {Imparato}(2019)}]{sune2019clock}%
  \BibitemOpen
  \bibfield  {author} {\bibinfo {author} {\bibfnamefont {M.}~\bibnamefont {Suñé}}\ and\ \bibinfo {author} {\bibfnamefont {A.}~\bibnamefont {Imparato}},\ }\bibfield  {title} {\bibinfo {title} {Out-of-equilibrium clock model at the verge of criticality},\ }\href@noop {} {\bibfield  {journal} {\bibinfo  {journal} {Physical Review Letters}\ }\textbf {\bibinfo {volume} {123}} (\bibinfo {year} {2019})}\BibitemShut {NoStop}%
\bibitem [{\citenamefont {J.Schnakenberg}(1976)}]{schnakenberg1976networktheory}%
  \BibitemOpen
  \bibfield  {author} {\bibinfo {author} {\bibnamefont {J.Schnakenberg}},\ }\bibfield  {title} {\bibinfo {title} {Network theory of microscopic and macroscopic behavior of master equation systems},\ }\href@noop {} {\bibfield  {journal} {\bibinfo  {journal} {Reviews of Modern Physics}\ }\textbf {\bibinfo {volume} {48}} (\bibinfo {year} {1976})}\BibitemShut {NoStop}%
\bibitem [{\citenamefont {Oster}\ \emph {et~al.}(1971)\citenamefont {Oster}, \citenamefont {Perelson},\ and\ \citenamefont {Katchalsky}}]{oster1971networkthermodynamics}%
  \BibitemOpen
  \bibfield  {author} {\bibinfo {author} {\bibfnamefont {G.}~\bibnamefont {Oster}}, \bibinfo {author} {\bibfnamefont {A.}~\bibnamefont {Perelson}},\ and\ \bibinfo {author} {\bibfnamefont {A.}~\bibnamefont {Katchalsky}},\ }\bibfield  {title} {\bibinfo {title} {Network thermodynamics},\ }\href@noop {} {\bibfield  {journal} {\bibinfo  {journal} {Nature}\ }\textbf {\bibinfo {volume} {234}},\ \bibinfo {pages} {393} (\bibinfo {year} {1971})}\BibitemShut {NoStop}%
\bibitem [{\citenamefont {Rao}\ and\ \citenamefont {Esposito}(2016)}]{rao2016crn}%
  \BibitemOpen
  \bibfield  {author} {\bibinfo {author} {\bibfnamefont {R.}~\bibnamefont {Rao}}\ and\ \bibinfo {author} {\bibfnamefont {M.}~\bibnamefont {Esposito}},\ }\bibfield  {title} {\bibinfo {title} {Nonequilibrium thermodynamics of chemical reaction networks: Wisdom from stochastic thermodynamics},\ }\href@noop {} {\bibfield  {journal} {\bibinfo  {journal} {Physical Review X}\ }\textbf {\bibinfo {volume} {6}} (\bibinfo {year} {2016})}\BibitemShut {NoStop}%
\bibitem [{\citenamefont {Polettini}\ \emph {et~al.}(2015)\citenamefont {Polettini}, \citenamefont {Wachtel},\ and\ \citenamefont {Esposito}}]{polettini2015crn}%
  \BibitemOpen
  \bibfield  {author} {\bibinfo {author} {\bibfnamefont {M.}~\bibnamefont {Polettini}}, \bibinfo {author} {\bibfnamefont {A.}~\bibnamefont {Wachtel}},\ and\ \bibinfo {author} {\bibfnamefont {M.}~\bibnamefont {Esposito}},\ }\bibfield  {title} {\bibinfo {title} {Dissipation in noisy chemical networks: The role of deficiency},\ }\href@noop {} {\bibfield  {journal} {\bibinfo  {journal} {Journal of Chemical Physics}\ }\textbf {\bibinfo {volume} {143}} (\bibinfo {year} {2015})}\BibitemShut {NoStop}%
\bibitem [{\citenamefont {Cengio}\ \emph {et~al.}(2023)\citenamefont {Cengio}, \citenamefont {Lecomte},\ and\ \citenamefont {Polettini}}]{DalCengio2023geometrycrn}%
  \BibitemOpen
  \bibfield  {author} {\bibinfo {author} {\bibfnamefont {S.~D.}\ \bibnamefont {Cengio}}, \bibinfo {author} {\bibfnamefont {V.}~\bibnamefont {Lecomte}},\ and\ \bibinfo {author} {\bibfnamefont {M.}~\bibnamefont {Polettini}},\ }\bibfield  {title} {\bibinfo {title} {Geometry of nonequilibrium reaction networks},\ }\href@noop {} {\bibfield  {journal} {\bibinfo  {journal} {Physical Review X}\ }\textbf {\bibinfo {volume} {13}} (\bibinfo {year} {2023})}\BibitemShut {NoStop}%
\bibitem [{\citenamefont {Vaidya}\ \emph {et~al.}(2021)\citenamefont {Vaidya}, \citenamefont {Chotibut},\ and\ \citenamefont {Piliouras}}]{vaidya2021sociallearning}%
  \BibitemOpen
  \bibfield  {author} {\bibinfo {author} {\bibfnamefont {T.}~\bibnamefont {Vaidya}}, \bibinfo {author} {\bibfnamefont {T.}~\bibnamefont {Chotibut}},\ and\ \bibinfo {author} {\bibfnamefont {G.}~\bibnamefont {Piliouras}},\ }\bibfield  {title} {\bibinfo {title} {Broken detailed balance and non-equilibrium dynamics in noisy social learning models},\ }\href@noop {} {\bibfield  {journal} {\bibinfo  {journal} {Physica A: Statistical Mechanics and its Applications}\ }\textbf {\bibinfo {volume} {570}} (\bibinfo {year} {2021})}\BibitemShut {NoStop}%
\bibitem [{\citenamefont {Lynn}\ \emph {et~al.}(2022)\citenamefont {Lynn}, \citenamefont {Holmes}, \citenamefont {Bialek},\ and\ \citenamefont {Schwab}}]{lynn2023decomposing}%
  \BibitemOpen
  \bibfield  {author} {\bibinfo {author} {\bibfnamefont {C.~W.}\ \bibnamefont {Lynn}}, \bibinfo {author} {\bibfnamefont {C.~M.}\ \bibnamefont {Holmes}}, \bibinfo {author} {\bibfnamefont {W.}~\bibnamefont {Bialek}},\ and\ \bibinfo {author} {\bibfnamefont {D.~J.}\ \bibnamefont {Schwab}},\ }\bibfield  {title} {\bibinfo {title} {Decomposing the local arrow of time in interacting systems},\ }\href@noop {} {\bibfield  {journal} {\bibinfo  {journal} {Physical Review Letters}\ }\textbf {\bibinfo {volume} {129}} (\bibinfo {year} {2022})}\BibitemShut {NoStop}%
\bibitem [{\citenamefont {Papo}\ and\ \citenamefont {Buldú}(2024)}]{papo2024braincomplexnetwork}%
  \BibitemOpen
  \bibfield  {author} {\bibinfo {author} {\bibfnamefont {D.}~\bibnamefont {Papo}}\ and\ \bibinfo {author} {\bibfnamefont {J.}~\bibnamefont {Buldú}},\ }\bibfield  {title} {\bibinfo {title} {Does the brain behave like a (complex) network? {I. D}ynamics},\ }\href@noop {} {\bibfield  {journal} {\bibinfo  {journal} {Physics of Life Reviews}\ }\textbf {\bibinfo {volume} {48}},\ \bibinfo {pages} {47} (\bibinfo {year} {2024})}\BibitemShut {NoStop}%
\bibitem [{\citenamefont {Krakauer}(2023)}]{krakauer2023brokensymmetry}%
  \BibitemOpen
  \bibfield  {author} {\bibinfo {author} {\bibfnamefont {D.~C.}\ \bibnamefont {Krakauer}},\ }\bibfield  {title} {\bibinfo {title} {Symmetry–simplicity, broken symmetry–complexity},\ }\href@noop {} {\bibfield  {journal} {\bibinfo  {journal} {Interface Focus}\ }\textbf {\bibinfo {volume} {13}} (\bibinfo {year} {2023})}\BibitemShut {NoStop}%
\bibitem [{\citenamefont {Johnson}(2020)}]{johnson2020digraphs}%
  \BibitemOpen
  \bibfield  {author} {\bibinfo {author} {\bibfnamefont {S.}~\bibnamefont {Johnson}},\ }\bibfield  {title} {\bibinfo {title} {Digraphs are different: why directionality matters in complex systems},\ }\href@noop {} {\bibfield  {journal} {\bibinfo  {journal} {Journal of Physics: Complexity}\ }\textbf {\bibinfo {volume} {1}} (\bibinfo {year} {2020})}\BibitemShut {NoStop}%
\bibitem [{\citenamefont {Fruchart}\ \emph {et~al.}(2021)\citenamefont {Fruchart}, \citenamefont {Hanai}, \citenamefont {Littlewood},\ and\ \citenamefont {Vitelli}}]{fruchart2021nonreciprocal}%
  \BibitemOpen
  \bibfield  {author} {\bibinfo {author} {\bibfnamefont {M.}~\bibnamefont {Fruchart}}, \bibinfo {author} {\bibfnamefont {R.}~\bibnamefont {Hanai}}, \bibinfo {author} {\bibfnamefont {P.~B.}\ \bibnamefont {Littlewood}},\ and\ \bibinfo {author} {\bibfnamefont {V.}~\bibnamefont {Vitelli}},\ }\bibfield  {title} {\bibinfo {title} {Non-reciprocal phase transitions},\ }\href@noop {} {\bibfield  {journal} {\bibinfo  {journal} {Nature}\ }\textbf {\bibinfo {volume} {592}},\ \bibinfo {pages} {363} (\bibinfo {year} {2021})}\BibitemShut {NoStop}%
\bibitem [{\citenamefont {Muolo}\ \emph {et~al.}(2020)\citenamefont {Muolo}, \citenamefont {Carletti}, \citenamefont {Gleeson},\ and\ \citenamefont {Asllani}}]{muolo2020sychnronisation}%
  \BibitemOpen
  \bibfield  {author} {\bibinfo {author} {\bibfnamefont {R.}~\bibnamefont {Muolo}}, \bibinfo {author} {\bibfnamefont {T.}~\bibnamefont {Carletti}}, \bibinfo {author} {\bibfnamefont {J.~P.}\ \bibnamefont {Gleeson}},\ and\ \bibinfo {author} {\bibfnamefont {M.}~\bibnamefont {Asllani}},\ }\bibfield  {title} {\bibinfo {title} {Synchronization dynamics in non-normal networks: The trade-off for optimality},\ }\href@noop {} {\bibfield  {journal} {\bibinfo  {journal} {Entropy}\ }\textbf {\bibinfo {volume} {23}} (\bibinfo {year} {2020})}\BibitemShut {NoStop}%
\bibitem [{\citenamefont {Muolo}\ \emph {et~al.}(2024)\citenamefont {Muolo}, \citenamefont {O’Brien}, \citenamefont {Carletti},\ and\ \citenamefont {Asllani}}]{muolo_persistence_2024}%
  \BibitemOpen
  \bibfield  {author} {\bibinfo {author} {\bibfnamefont {R.}~\bibnamefont {Muolo}}, \bibinfo {author} {\bibfnamefont {J.~D.}\ \bibnamefont {O’Brien}}, \bibinfo {author} {\bibfnamefont {T.}~\bibnamefont {Carletti}},\ and\ \bibinfo {author} {\bibfnamefont {M.}~\bibnamefont {Asllani}},\ }\bibfield  {title} {\bibinfo {title} {Persistence of chimera states and the challenge for synchronization in real-world networks},\ }\href@noop {} {\bibfield  {journal} {\bibinfo  {journal} {The European Physical Journal B}\ }\textbf {\bibinfo {volume} {97}},\ \bibinfo {pages} {6} (\bibinfo {year} {2024})}\BibitemShut {NoStop}%
\bibitem [{\citenamefont {Asllani}\ and\ \citenamefont {Carletti}(2018)}]{asllani2018topologicalresilience}%
  \BibitemOpen
  \bibfield  {author} {\bibinfo {author} {\bibfnamefont {M.}~\bibnamefont {Asllani}}\ and\ \bibinfo {author} {\bibfnamefont {T.}~\bibnamefont {Carletti}},\ }\bibfield  {title} {\bibinfo {title} {Topological resilience in non-normal networked systems},\ }\href@noop {} {\bibfield  {journal} {\bibinfo  {journal} {Physical Review E}\ }\textbf {\bibinfo {volume} {97}} (\bibinfo {year} {2018})}\BibitemShut {NoStop}%
\bibitem [{\citenamefont {Nicoletti}\ \emph {et~al.}(2019)\citenamefont {Nicoletti}, \citenamefont {Fanelli}, \citenamefont {Zagli}, \citenamefont {Asllani}, \citenamefont {Battistelli}, \citenamefont {Carletti}, \citenamefont {Chisci}, \citenamefont {Innocenti},\ and\ \citenamefont {Livi}}]{nicoletti_resilience_2019}%
  \BibitemOpen
  \bibfield  {author} {\bibinfo {author} {\bibfnamefont {S.}~\bibnamefont {Nicoletti}}, \bibinfo {author} {\bibfnamefont {D.}~\bibnamefont {Fanelli}}, \bibinfo {author} {\bibfnamefont {N.}~\bibnamefont {Zagli}}, \bibinfo {author} {\bibfnamefont {M.}~\bibnamefont {Asllani}}, \bibinfo {author} {\bibfnamefont {G.}~\bibnamefont {Battistelli}}, \bibinfo {author} {\bibfnamefont {T.}~\bibnamefont {Carletti}}, \bibinfo {author} {\bibfnamefont {L.}~\bibnamefont {Chisci}}, \bibinfo {author} {\bibfnamefont {G.}~\bibnamefont {Innocenti}},\ and\ \bibinfo {author} {\bibfnamefont {R.}~\bibnamefont {Livi}},\ }\bibfield  {title} {\bibinfo {title} {Resilience for stochastic systems interacting via a quasi-degenerate network},\ }\href@noop {} {\bibfield  {journal} {\bibinfo  {journal} {Chaos: An Interdisciplinary Journal of Nonlinear Science}\ }\textbf {\bibinfo {volume} {29}},\ \bibinfo {pages} {083123} (\bibinfo {year} {2019})}\BibitemShut {NoStop}%
\bibitem [{\citenamefont {Asllani}\ \emph {et~al.}(2014)\citenamefont {Asllani}, \citenamefont {Challenger}, \citenamefont {Pavone}, \citenamefont {Sacconi},\ and\ \citenamefont {Fanelli}}]{asllani2014patterns}%
  \BibitemOpen
  \bibfield  {author} {\bibinfo {author} {\bibfnamefont {M.}~\bibnamefont {Asllani}}, \bibinfo {author} {\bibfnamefont {J.~D.}\ \bibnamefont {Challenger}}, \bibinfo {author} {\bibfnamefont {F.~S.}\ \bibnamefont {Pavone}}, \bibinfo {author} {\bibfnamefont {L.}~\bibnamefont {Sacconi}},\ and\ \bibinfo {author} {\bibfnamefont {D.}~\bibnamefont {Fanelli}},\ }\bibfield  {title} {\bibinfo {title} {The theory of pattern formation on directed networks},\ }\href@noop {} {\bibfield  {journal} {\bibinfo  {journal} {Nature Communications}\ }\textbf {\bibinfo {volume} {5}} (\bibinfo {year} {2014})}\BibitemShut {NoStop}%
\bibitem [{\citenamefont {Muolo}\ \emph {et~al.}(2019)\citenamefont {Muolo}, \citenamefont {Asllani}, \citenamefont {Fanelli}, \citenamefont {Maini},\ and\ \citenamefont {Carletti}}]{muolo2019patterns}%
  \BibitemOpen
  \bibfield  {author} {\bibinfo {author} {\bibfnamefont {R.}~\bibnamefont {Muolo}}, \bibinfo {author} {\bibfnamefont {M.}~\bibnamefont {Asllani}}, \bibinfo {author} {\bibfnamefont {D.}~\bibnamefont {Fanelli}}, \bibinfo {author} {\bibfnamefont {P.~K.}\ \bibnamefont {Maini}},\ and\ \bibinfo {author} {\bibfnamefont {T.}~\bibnamefont {Carletti}},\ }\bibfield  {title} {\bibinfo {title} {Patterns of non-normality in networked systems},\ }\href@noop {} {\bibfield  {journal} {\bibinfo  {journal} {Journal of Theoretical Biology}\ }\textbf {\bibinfo {volume} {480}},\ \bibinfo {pages} {81} (\bibinfo {year} {2019})}\BibitemShut {NoStop}%
\bibitem [{\citenamefont {Johnson}\ \emph {et~al.}(2014)\citenamefont {Johnson}, \citenamefont {Domínguez-García}, \citenamefont {Donetti},\ and\ \citenamefont {Muñoz}}]{johnson2014trophiccoherence}%
  \BibitemOpen
  \bibfield  {author} {\bibinfo {author} {\bibfnamefont {S.}~\bibnamefont {Johnson}}, \bibinfo {author} {\bibfnamefont {V.}~\bibnamefont {Domínguez-García}}, \bibinfo {author} {\bibfnamefont {L.}~\bibnamefont {Donetti}},\ and\ \bibinfo {author} {\bibfnamefont {M.~A.}\ \bibnamefont {Muñoz}},\ }\bibfield  {title} {\bibinfo {title} {Trophic coherence determines food-web stability},\ }\href@noop {} {\bibfield  {journal} {\bibinfo  {journal} {Proceedings of the National Academy of Sciences}\ }\textbf {\bibinfo {volume} {111}},\ \bibinfo {pages} {17923} (\bibinfo {year} {2014})}\BibitemShut {NoStop}%
\bibitem [{\citenamefont {Asllani}\ \emph {et~al.}(2018)\citenamefont {Asllani}, \citenamefont {Lambiotte},\ and\ \citenamefont {Carletti}}]{asllani2018nonnormal}%
  \BibitemOpen
  \bibfield  {author} {\bibinfo {author} {\bibfnamefont {M.}~\bibnamefont {Asllani}}, \bibinfo {author} {\bibfnamefont {R.}~\bibnamefont {Lambiotte}},\ and\ \bibinfo {author} {\bibfnamefont {T.}~\bibnamefont {Carletti}},\ }\bibfield  {title} {\bibinfo {title} {Structure and dynamical behavior of non-normal networks},\ }\href@noop {} {\bibfield  {journal} {\bibinfo  {journal} {Science Advances}\ }\textbf {\bibinfo {volume} {4}} (\bibinfo {year} {2018})}\BibitemShut {NoStop}%
\bibitem [{\citenamefont {MacKay}\ \emph {et~al.}(2020)\citenamefont {MacKay}, \citenamefont {Johnson},\ and\ \citenamefont {Sansom}}]{mackay2020directed}%
  \BibitemOpen
  \bibfield  {author} {\bibinfo {author} {\bibfnamefont {R.}~\bibnamefont {MacKay}}, \bibinfo {author} {\bibfnamefont {S.}~\bibnamefont {Johnson}},\ and\ \bibinfo {author} {\bibfnamefont {B.}~\bibnamefont {Sansom}},\ }\bibfield  {title} {\bibinfo {title} {How directed is a directed network?},\ }\href@noop {} {\bibfield  {journal} {\bibinfo  {journal} {Royal Society Open Science}\ }\textbf {\bibinfo {volume} {7}} (\bibinfo {year} {2020})}\BibitemShut {NoStop}%
\bibitem [{\citenamefont {O'Brien}\ \emph {et~al.}(2021)\citenamefont {O'Brien}, \citenamefont {Oliveira}, \citenamefont {Gleeson},\ and\ \citenamefont {Asllani}}]{obrien2021hierarchical}%
  \BibitemOpen
  \bibfield  {author} {\bibinfo {author} {\bibfnamefont {J.}~\bibnamefont {O'Brien}}, \bibinfo {author} {\bibfnamefont {K.}~\bibnamefont {Oliveira}}, \bibinfo {author} {\bibfnamefont {J.}~\bibnamefont {Gleeson}},\ and\ \bibinfo {author} {\bibfnamefont {M.}~\bibnamefont {Asllani}},\ }\bibfield  {title} {\bibinfo {title} {Hierarchical route to the emergence of leader nodes in real-world networks},\ }\href@noop {} {\bibfield  {journal} {\bibinfo  {journal} {Physical Review Research}\ }\textbf {\bibinfo {volume} {3}} (\bibinfo {year} {2021})}\BibitemShut {NoStop}%
\bibitem [{\citenamefont {Trefethen}\ and\ \citenamefont {Embree}(2005)}]{trefethen2001nonnormal}%
  \BibitemOpen
  \bibfield  {author} {\bibinfo {author} {\bibfnamefont {L.}~\bibnamefont {Trefethen}}\ and\ \bibinfo {author} {\bibfnamefont {M.}~\bibnamefont {Embree}},\ }\href@noop {} {\emph {\bibinfo {title} {Spectra and Pseudospectra: The Behaviour of Non-Normal Matrices and Operators}}}\ (\bibinfo  {publisher} {Princeton University Press},\ \bibinfo {year} {2005})\BibitemShut {NoStop}%
\bibitem [{\citenamefont {Masuda}\ \emph {et~al.}(2017)\citenamefont {Masuda}, \citenamefont {Porter},\ and\ \citenamefont {Lambiotte}}]{Masuda2017randomwalks}%
  \BibitemOpen
  \bibfield  {author} {\bibinfo {author} {\bibfnamefont {N.}~\bibnamefont {Masuda}}, \bibinfo {author} {\bibfnamefont {M.}~\bibnamefont {Porter}},\ and\ \bibinfo {author} {\bibfnamefont {R.}~\bibnamefont {Lambiotte}},\ }\bibfield  {title} {\bibinfo {title} {Random walks and diffusion on networks},\ }\href@noop {} {\bibfield  {journal} {\bibinfo  {journal} {Physics Reports}\ }\textbf {\bibinfo {volume} {716-717}},\ \bibinfo {pages} {1} (\bibinfo {year} {2017})}\BibitemShut {NoStop}%
\bibitem [{\citenamefont {Lambiotte}(2023)}]{lambiotte2023continuoustimerandomwalk}%
  \BibitemOpen
  \bibfield  {author} {\bibinfo {author} {\bibfnamefont {R.}~\bibnamefont {Lambiotte}},\ }\bibfield  {title} {\bibinfo {title} {Continuous-time random walks and temporal networks},\ }in\ \href@noop {} {\emph {\bibinfo {booktitle} {Temporal Network Theory}}}\ (\bibinfo  {publisher} {Springer},\ \bibinfo {year} {2023})\ pp.\ \bibinfo {pages} {225--239}\BibitemShut {NoStop}%
\bibitem [{\citenamefont {Godrèche}\ and\ \citenamefont {Luck}(2018)}]{Godreche2018OU}%
  \BibitemOpen
  \bibfield  {author} {\bibinfo {author} {\bibfnamefont {C.}~\bibnamefont {Godrèche}}\ and\ \bibinfo {author} {\bibfnamefont {J.}~\bibnamefont {Luck}},\ }\bibfield  {title} {\bibinfo {title} {Characterising the nonequilibrium stationary states of {O}rnstein–{U}hlenbeck processes},\ }\href@noop {} {\bibfield  {journal} {\bibinfo  {journal} {Journal of Physics A: Mathematical and Theoretical}\ }\textbf {\bibinfo {volume} {52}} (\bibinfo {year} {2018})}\BibitemShut {NoStop}%
\bibitem [{\citenamefont {Nishimori}(2001)}]{Nishimori2001statphys}%
  \BibitemOpen
  \bibfield  {author} {\bibinfo {author} {\bibfnamefont {H.}~\bibnamefont {Nishimori}},\ }\href@noop {} {\emph {\bibinfo {title} {Statistical physics of spin glasses and information processing: an introduction}}}\ (\bibinfo  {publisher} {Clarendon Press},\ \bibinfo {year} {2001})\BibitemShut {NoStop}%
\bibitem [{\citenamefont {Erdös}\ and\ \citenamefont {Rényi}(1959)}]{erdos1959random}%
  \BibitemOpen
  \bibfield  {author} {\bibinfo {author} {\bibfnamefont {P.}~\bibnamefont {Erdös}}\ and\ \bibinfo {author} {\bibfnamefont {A.}~\bibnamefont {Rényi}},\ }\bibfield  {title} {\bibinfo {title} {On random graphs {I}},\ }\href@noop {} {\bibfield  {journal} {\bibinfo  {journal} {Publicationes Mathematicae Debrecen}\ }\textbf {\bibinfo {volume} {6}},\ \bibinfo {pages} {290} (\bibinfo {year} {1959})}\BibitemShut {NoStop}%
\bibitem [{\citenamefont {Shumway}\ and\ \citenamefont {Stoffer}(2017)}]{Shumway2017timeseries}%
  \BibitemOpen
  \bibfield  {author} {\bibinfo {author} {\bibfnamefont {R.~H.}\ \bibnamefont {Shumway}}\ and\ \bibinfo {author} {\bibfnamefont {D.~S.}\ \bibnamefont {Stoffer}},\ }\href@noop {} {\emph {\bibinfo {title} {Time Series Analysis and Its Applications}}}\ (\bibinfo  {publisher} {Springer},\ \bibinfo {year} {2017})\BibitemShut {NoStop}%
\bibitem [{\citenamefont {Deco}\ \emph {et~al.}(2022)\citenamefont {Deco}, \citenamefont {Sanz-Perl}, \citenamefont {Bocaccio}, \citenamefont {Tagliazucchi},\ and\ \citenamefont {Kringelbach}}]{deco2022insideout}%
  \BibitemOpen
  \bibfield  {author} {\bibinfo {author} {\bibfnamefont {G.}~\bibnamefont {Deco}}, \bibinfo {author} {\bibfnamefont {Y.}~\bibnamefont {Sanz-Perl}}, \bibinfo {author} {\bibfnamefont {H.}~\bibnamefont {Bocaccio}}, \bibinfo {author} {\bibfnamefont {E.}~\bibnamefont {Tagliazucchi}},\ and\ \bibinfo {author} {\bibfnamefont {M.~L.}\ \bibnamefont {Kringelbach}},\ }\bibfield  {title} {\bibinfo {title} {The {INSIDEOUT} framework provides precise signatures of the balance of intrinsic and extrinsic dynamics in brain states},\ }\href@noop {} {\bibfield  {journal} {\bibinfo  {journal} {Communications Biology}\ }\textbf {\bibinfo {volume} {5}} (\bibinfo {year} {2022})}\BibitemShut {NoStop}%
\bibitem [{\citenamefont {Deco}\ \emph {et~al.}(2023{\natexlab{b}})\citenamefont {Deco}, \citenamefont {Lynn}, \citenamefont {Sanz-Perl},\ and\ \citenamefont {Kringelbach}}]{deco2023violations}%
  \BibitemOpen
  \bibfield  {author} {\bibinfo {author} {\bibfnamefont {G.}~\bibnamefont {Deco}}, \bibinfo {author} {\bibfnamefont {C.}~\bibnamefont {Lynn}}, \bibinfo {author} {\bibfnamefont {Y.}~\bibnamefont {Sanz-Perl}},\ and\ \bibinfo {author} {\bibfnamefont {M.~L.}\ \bibnamefont {Kringelbach}},\ }\bibfield  {title} {\bibinfo {title} {Violations of the fluctuation-dissipation theorem reveal distinct non-equilibrium dynamics of brain states},\ }\href@noop {} {\bibfield  {journal} {\bibinfo  {journal} {Physical Review E}\ }\textbf {\bibinfo {volume} {108}} (\bibinfo {year} {2023}{\natexlab{b}})}\BibitemShut {NoStop}%
\bibitem [{\citenamefont {Kale}\ \emph {et~al.}(2018)\citenamefont {Kale}, \citenamefont {Zalesky},\ and\ \citenamefont {Gollo}}]{kale2018directed}%
  \BibitemOpen
  \bibfield  {author} {\bibinfo {author} {\bibfnamefont {P.}~\bibnamefont {Kale}}, \bibinfo {author} {\bibfnamefont {A.}~\bibnamefont {Zalesky}},\ and\ \bibinfo {author} {\bibfnamefont {L.}~\bibnamefont {Gollo}},\ }\bibfield  {title} {\bibinfo {title} {Estimating the impact of structural directionality: How reliable are undirected connectomes?},\ }\href@noop {} {\bibfield  {journal} {\bibinfo  {journal} {Network Neuroscience}\ }\textbf {\bibinfo {volume} {2}},\ \bibinfo {pages} {259} (\bibinfo {year} {2018})}\BibitemShut {NoStop}%
\bibitem [{\citenamefont {Friston}\ \emph {et~al.}(2021)\citenamefont {Friston}, \citenamefont {Fagerholm}, \citenamefont {Zarghami}, \citenamefont {Parr}, \citenamefont {Hipólito}, \citenamefont {Magrou},\ and\ \citenamefont {Razi}}]{friston2021parcels}%
  \BibitemOpen
  \bibfield  {author} {\bibinfo {author} {\bibfnamefont {K.~J.}\ \bibnamefont {Friston}}, \bibinfo {author} {\bibfnamefont {E.~D.}\ \bibnamefont {Fagerholm}}, \bibinfo {author} {\bibfnamefont {T.~S.}\ \bibnamefont {Zarghami}}, \bibinfo {author} {\bibfnamefont {T.}~\bibnamefont {Parr}}, \bibinfo {author} {\bibfnamefont {I.}~\bibnamefont {Hipólito}}, \bibinfo {author} {\bibfnamefont {L.}~\bibnamefont {Magrou}},\ and\ \bibinfo {author} {\bibfnamefont {A.}~\bibnamefont {Razi}},\ }\bibfield  {title} {\bibinfo {title} {Parcels and particles: {M}arkov blankets in the brain},\ }\href@noop {} {\bibfield  {journal} {\bibinfo  {journal} {Network Neuroscience}\ }\textbf {\bibinfo {volume} {5}},\ \bibinfo {pages} {211} (\bibinfo {year} {2021})}\BibitemShut {NoStop}%
\bibitem [{Note1()}]{Note1}%
  \BibitemOpen
  \bibinfo {note} {\textcolor {black}{The definition of $W_{ij}$ to represent the link from $i$ to $j$ is one of two conventions. We choose this convention to be consistent with the definition of trophic coherence \cite {mackay2020directed}. The opposite, is the more typical convention for network dynamics \cite {newman2018networks}.}}\BibitemShut {Stop}%
\bibitem [{\citenamefont {Corominas-Murtra}\ \emph {et~al.}(2013)\citenamefont {Corominas-Murtra}, \citenamefont {Goñi}, \citenamefont {Solé},\ and\ \citenamefont {Rodríguez-Caso}}]{corominas2013hierarchy}%
  \BibitemOpen
  \bibfield  {author} {\bibinfo {author} {\bibfnamefont {B.}~\bibnamefont {Corominas-Murtra}}, \bibinfo {author} {\bibfnamefont {J.}~\bibnamefont {Goñi}}, \bibinfo {author} {\bibfnamefont {R.~V.}\ \bibnamefont {Solé}},\ and\ \bibinfo {author} {\bibfnamefont {C.}~\bibnamefont {Rodríguez-Caso}},\ }\bibfield  {title} {\bibinfo {title} {On the origins of hierarchy in complex networks},\ }\href@noop {} {\bibfield  {journal} {\bibinfo  {journal} {Proceedings of the National Academy of Sciences of the United States of America}\ }\textbf {\bibinfo {volume} {110}},\ \bibinfo {pages} {13316} (\bibinfo {year} {2013})}\BibitemShut {NoStop}%
\bibitem [{Note2()}]{Note2}%
  \BibitemOpen
  \bibinfo {note} {We consider the (ir)reciprocity of weighted networks as defined in Ref. \cite {squartini2013weightedreciprocity}. Alternative definitions of reciprocity for unweighted graphs are given in Ref. \cite {garlaschelli2004reciprocity,newman2018networks}.}\BibitemShut {Stop}%
\bibitem [{\citenamefont {Garlaschelli}\ and\ \citenamefont {Loffredo}(2004)}]{garlaschelli2004reciprocity}%
  \BibitemOpen
  \bibfield  {author} {\bibinfo {author} {\bibfnamefont {D.}~\bibnamefont {Garlaschelli}}\ and\ \bibinfo {author} {\bibfnamefont {M.~I.}\ \bibnamefont {Loffredo}},\ }\bibfield  {title} {\bibinfo {title} {Patterns of link reciprocity in directed networks},\ }\href@noop {} {\bibfield  {journal} {\bibinfo  {journal} {Physical Review Letters}\ }\textbf {\bibinfo {volume} {93}} (\bibinfo {year} {2004})}\BibitemShut {NoStop}%
\bibitem [{\citenamefont {Squartini}\ \emph {et~al.}(2013)\citenamefont {Squartini}, \citenamefont {Picciolo}, \citenamefont {Ruzzenenti},\ and\ \citenamefont {Garlaschelli}}]{squartini2013weightedreciprocity}%
  \BibitemOpen
  \bibfield  {author} {\bibinfo {author} {\bibfnamefont {T.}~\bibnamefont {Squartini}}, \bibinfo {author} {\bibfnamefont {F.}~\bibnamefont {Picciolo}}, \bibinfo {author} {\bibfnamefont {F.}~\bibnamefont {Ruzzenenti}},\ and\ \bibinfo {author} {\bibfnamefont {D.}~\bibnamefont {Garlaschelli}},\ }\bibfield  {title} {\bibinfo {title} {Reciprocity of weighted networks},\ }\href@noop {} {\bibfield  {journal} {\bibinfo  {journal} {Scientific Reports}\ }\textbf {\bibinfo {volume} {3}} (\bibinfo {year} {2013})}\BibitemShut {NoStop}%
\bibitem [{\citenamefont {Jiang}\ \emph {et~al.}(2011)\citenamefont {Jiang}, \citenamefont {Lim}, \citenamefont {Yao},\ and\ \citenamefont {Ye}}]{Jiang2011Hodge}%
  \BibitemOpen
  \bibfield  {author} {\bibinfo {author} {\bibfnamefont {X.}~\bibnamefont {Jiang}}, \bibinfo {author} {\bibfnamefont {L.-H.}\ \bibnamefont {Lim}}, \bibinfo {author} {\bibfnamefont {Y.}~\bibnamefont {Yao}},\ and\ \bibinfo {author} {\bibfnamefont {Y.}~\bibnamefont {Ye}},\ }\bibfield  {title} {\bibinfo {title} {Statistical ranking and combinatorial hodge theory},\ }\href@noop {} {\bibfield  {journal} {\bibinfo  {journal} {Mathematical Programming}\ }\textbf {\bibinfo {volume} {127}},\ \bibinfo {pages} {203} (\bibinfo {year} {2011})}\BibitemShut {NoStop}%
\bibitem [{Note3()}]{Note3}%
  \BibitemOpen
  \bibinfo {note} {Equivalently, the `SpringRank' formulation considers directed springs between nodes and aims to find a ranking that minimizes the total energy of these springs \cite {debacco2018springrank}}\BibitemShut {NoStop}%
\bibitem [{\citenamefont {Kichikawa}\ \emph {et~al.}(2019)\citenamefont {Kichikawa}, \citenamefont {Iyetomi}, \citenamefont {Iino},\ and\ \citenamefont {Inoue}}]{kichikawa2019community}%
  \BibitemOpen
  \bibfield  {author} {\bibinfo {author} {\bibfnamefont {Y.}~\bibnamefont {Kichikawa}}, \bibinfo {author} {\bibfnamefont {H.}~\bibnamefont {Iyetomi}}, \bibinfo {author} {\bibfnamefont {T.}~\bibnamefont {Iino}},\ and\ \bibinfo {author} {\bibfnamefont {H.}~\bibnamefont {Inoue}},\ }\bibfield  {title} {\bibinfo {title} {Community structure based on circular flow in a large-scale transaction network},\ }\href@noop {} {\bibfield  {journal} {\bibinfo  {journal} {Applied Network Science}\ }\textbf {\bibinfo {volume} {4}} (\bibinfo {year} {2019})}\BibitemShut {NoStop}%
\bibitem [{\citenamefont {Hennequin}\ \emph {et~al.}(2012)\citenamefont {Hennequin}, \citenamefont {Vogels},\ and\ \citenamefont {Gerstner}}]{hennequin2012nonnormal}%
  \BibitemOpen
  \bibfield  {author} {\bibinfo {author} {\bibfnamefont {G.}~\bibnamefont {Hennequin}}, \bibinfo {author} {\bibfnamefont {T.~P.}\ \bibnamefont {Vogels}},\ and\ \bibinfo {author} {\bibfnamefont {W.}~\bibnamefont {Gerstner}},\ }\bibfield  {title} {\bibinfo {title} {Non-normal amplification in random balanced neuronal networks},\ }\href@noop {} {\bibfield  {journal} {\bibinfo  {journal} {Physical Review E}\ }\textbf {\bibinfo {volume} {86}} (\bibinfo {year} {2012})}\BibitemShut {NoStop}%
\bibitem [{\citenamefont {Baggio}\ \emph {et~al.}(2020)\citenamefont {Baggio}, \citenamefont {Rutten}, \citenamefont {Hennequin},\ and\ \citenamefont {Zampieri}}]{baggio2020communication}%
  \BibitemOpen
  \bibfield  {author} {\bibinfo {author} {\bibfnamefont {G.}~\bibnamefont {Baggio}}, \bibinfo {author} {\bibfnamefont {V.}~\bibnamefont {Rutten}}, \bibinfo {author} {\bibfnamefont {G.}~\bibnamefont {Hennequin}},\ and\ \bibinfo {author} {\bibfnamefont {S.}~\bibnamefont {Zampieri}},\ }\bibfield  {title} {\bibinfo {title} {Efficient communication over complex dynamical networks: The role of matrix non-normality},\ }\href@noop {} {\bibfield  {journal} {\bibinfo  {journal} {Science Advances}\ }\textbf {\bibinfo {volume} {6}} (\bibinfo {year} {2020})}\BibitemShut {NoStop}%
\bibitem [{\citenamefont {Lindmark}\ and\ \citenamefont {Altafini}(2021)}]{lindmark2021control}%
  \BibitemOpen
  \bibfield  {author} {\bibinfo {author} {\bibfnamefont {G.}~\bibnamefont {Lindmark}}\ and\ \bibinfo {author} {\bibfnamefont {C.}~\bibnamefont {Altafini}},\ }\bibfield  {title} {\bibinfo {title} {Centrality measures and the role of non-normality for network control energy reduction},\ }\href@noop {} {\bibfield  {journal} {\bibinfo  {journal} {IEEE Control Systems Letters}\ }\textbf {\bibinfo {volume} {5}},\ \bibinfo {pages} {1013} (\bibinfo {year} {2021})}\BibitemShut {NoStop}%
\bibitem [{\citenamefont {Duan}\ \emph {et~al.}(2022)\citenamefont {Duan}, \citenamefont {Nishikawa}, \citenamefont {Eroglu},\ and\ \citenamefont {Motter}}]{duan2022instability}%
  \BibitemOpen
  \bibfield  {author} {\bibinfo {author} {\bibfnamefont {C.}~\bibnamefont {Duan}}, \bibinfo {author} {\bibfnamefont {T.}~\bibnamefont {Nishikawa}}, \bibinfo {author} {\bibfnamefont {D.}~\bibnamefont {Eroglu}},\ and\ \bibinfo {author} {\bibfnamefont {A.~E.}\ \bibnamefont {Motter}},\ }\bibfield  {title} {\bibinfo {title} {Network structural origin of instabilities in large complex systems},\ }\href@noop {} {\bibfield  {journal} {\bibinfo  {journal} {Science Advances}\ }\textbf {\bibinfo {volume} {8}} (\bibinfo {year} {2022})}\BibitemShut {NoStop}%
\bibitem [{\citenamefont {Schwarze}\ and\ \citenamefont {Porter}(2021)}]{scwarze2021motifs}%
  \BibitemOpen
  \bibfield  {author} {\bibinfo {author} {\bibfnamefont {A.}~\bibnamefont {Schwarze}}\ and\ \bibinfo {author} {\bibfnamefont {M.}~\bibnamefont {Porter}},\ }\bibfield  {title} {\bibinfo {title} {Motifs for processes on networks},\ }\href@noop {} {\bibfield  {journal} {\bibinfo  {journal} {SIAM Journal on Applied Dynamical Systems}\ }\textbf {\bibinfo {volume} {20}} (\bibinfo {year} {2021})}\BibitemShut {NoStop}%
\bibitem [{\citenamefont {Häggström}(2010)}]{Haggstrom2020MarkovChain}%
  \BibitemOpen
  \bibfield  {author} {\bibinfo {author} {\bibfnamefont {O.}~\bibnamefont {Häggström}},\ }\href@noop {} {\emph {\bibinfo {title} {Finite {M}arkov Chains and Algorithmic Applications}}}\ (\bibinfo  {publisher} {Cambridge University Press},\ \bibinfo {year} {2010})\BibitemShut {NoStop}%
\bibitem [{\citenamefont {Busiello}\ \emph {et~al.}(2020)\citenamefont {Busiello}, \citenamefont {Gupta},\ and\ \citenamefont {Maritan}}]{busiello2020undirectional}%
  \BibitemOpen
  \bibfield  {author} {\bibinfo {author} {\bibfnamefont {D.~M.}\ \bibnamefont {Busiello}}, \bibinfo {author} {\bibfnamefont {D.}~\bibnamefont {Gupta}},\ and\ \bibinfo {author} {\bibfnamefont {A.}~\bibnamefont {Maritan}},\ }\bibfield  {title} {\bibinfo {title} {Entropy production in systems with unidirectional transitions},\ }\href@noop {} {\bibfield  {journal} {\bibinfo  {journal} {Physical Review Research}\ }\textbf {\bibinfo {volume} {2}} (\bibinfo {year} {2020})}\BibitemShut {NoStop}%
\bibitem [{\citenamefont {Manzano}\ \emph {et~al.}(2024)\citenamefont {Manzano}, \citenamefont {Karde\ifmmode~\mbox{\c{s}}\else \c{s}\fi{}}, \citenamefont {Rold\'an},\ and\ \citenamefont {Wolpert}}]{Mazano2024Thermocomputation}%
  \BibitemOpen
  \bibfield  {author} {\bibinfo {author} {\bibfnamefont {G.}~\bibnamefont {Manzano}}, \bibinfo {author} {\bibfnamefont {G.}~\bibnamefont {Karde\ifmmode~\mbox{\c{s}}\else \c{s}\fi{}}}, \bibinfo {author} {\bibfnamefont {E.}~\bibnamefont {Rold\'an}},\ and\ \bibinfo {author} {\bibfnamefont {D.~H.}\ \bibnamefont {Wolpert}},\ }\bibfield  {title} {\bibinfo {title} {Thermodynamics of computations with absolute irreversibility, unidirectional transitions, and stochastic computation times},\ }\href@noop {} {\bibfield  {journal} {\bibinfo  {journal} {Physical Review X}\ }\textbf {\bibinfo {volume} {14}},\ \bibinfo {pages} {021026} (\bibinfo {year} {2024})}\BibitemShut {NoStop}%
\bibitem [{\citenamefont {Nielsen}(2019)}]{nielsen2019jsd}%
  \BibitemOpen
  \bibfield  {author} {\bibinfo {author} {\bibfnamefont {F.}~\bibnamefont {Nielsen}},\ }\bibfield  {title} {\bibinfo {title} {On the {J}ensen–{S}hannon symmetrization of distances relying on abstract means},\ }\href@noop {} {\bibfield  {journal} {\bibinfo  {journal} {Entropy}\ }\textbf {\bibinfo {volume} {21}} (\bibinfo {year} {2019})}\BibitemShut {NoStop}%
\bibitem [{\citenamefont {Uhlenbeck}\ and\ \citenamefont {Ornstein}(1930)}]{uhlenbeck1930brownian}%
  \BibitemOpen
  \bibfield  {author} {\bibinfo {author} {\bibfnamefont {G.}~\bibnamefont {Uhlenbeck}}\ and\ \bibinfo {author} {\bibfnamefont {L.}~\bibnamefont {Ornstein}},\ }\bibfield  {title} {\bibinfo {title} {On the theory of the {B}rownian motion},\ }\href@noop {} {\bibfield  {journal} {\bibinfo  {journal} {Physical Review}\ }\textbf {\bibinfo {volume} {36}} (\bibinfo {year} {1930})}\BibitemShut {NoStop}%
\bibitem [{\citenamefont {Zabczyk}(2020)}]{zabczyk2020mathematicalcontrol}%
  \BibitemOpen
  \bibfield  {author} {\bibinfo {author} {\bibfnamefont {J.}~\bibnamefont {Zabczyk}},\ }\href@noop {} {\emph {\bibinfo {title} {Mathematical Control Theory: An Introduction}}}\ (\bibinfo  {publisher} {Springer},\ \bibinfo {year} {2020})\BibitemShut {NoStop}%
\bibitem [{\citenamefont {Lax}(1960)}]{lax1960fluctuationsnonequilibrium}%
  \BibitemOpen
  \bibfield  {author} {\bibinfo {author} {\bibfnamefont {M.}~\bibnamefont {Lax}},\ }\bibfield  {title} {\bibinfo {title} {Fluctuations from the nonequilibrium steady state},\ }\href@noop {} {\bibfield  {journal} {\bibinfo  {journal} {Reviews of Modern Physics}\ }\textbf {\bibinfo {volume} {32}} (\bibinfo {year} {1960})}\BibitemShut {NoStop}%
\bibitem [{\citenamefont {Simoncini}(2016)}]{simoncini2016matrix}%
  \BibitemOpen
  \bibfield  {author} {\bibinfo {author} {\bibfnamefont {V.}~\bibnamefont {Simoncini}},\ }\bibfield  {title} {\bibinfo {title} {Computational methods for linear matrix equations},\ }\href@noop {} {\bibfield  {journal} {\bibinfo  {journal} {SIAM Review}\ }\textbf {\bibinfo {volume} {58}} (\bibinfo {year} {2016})}\BibitemShut {NoStop}%
\bibitem [{\citenamefont {Huang}\ and\ \citenamefont {Kabashima}(2014)}]{huang2014asymmetric}%
  \BibitemOpen
  \bibfield  {author} {\bibinfo {author} {\bibfnamefont {H.}~\bibnamefont {Huang}}\ and\ \bibinfo {author} {\bibfnamefont {Y.}~\bibnamefont {Kabashima}},\ }\bibfield  {title} {\bibinfo {title} {Dynamics of asymmetric kinetic {I}sing systems revisited},\ }\href@noop {} {\bibfield  {journal} {\bibinfo  {journal} {Journal of Statistical Mechanics: Theory and Experiment}\ } (\bibinfo {year} {2014})}\BibitemShut {NoStop}%
\bibitem [{\citenamefont {Aguilera}\ \emph {et~al.}(2021)\citenamefont {Aguilera}, \citenamefont {Moosavi},\ and\ \citenamefont {Shimazaki}}]{aguilera2021meanfield}%
  \BibitemOpen
  \bibfield  {author} {\bibinfo {author} {\bibfnamefont {M.}~\bibnamefont {Aguilera}}, \bibinfo {author} {\bibfnamefont {S.~A.}\ \bibnamefont {Moosavi}},\ and\ \bibinfo {author} {\bibfnamefont {H.}~\bibnamefont {Shimazaki}},\ }\bibfield  {title} {\bibinfo {title} {A unifying framework for mean-field theories of asymmetric kinetic {Ising} systems},\ }\href@noop {} {\bibfield  {journal} {\bibinfo  {journal} {Nature Communications}\ }\textbf {\bibinfo {volume} {58}} (\bibinfo {year} {2021})}\BibitemShut {NoStop}%
\bibitem [{\citenamefont {Newman}\ and\ \citenamefont {Barkema}(1999)}]{Newman1999MonteCarlo}%
  \BibitemOpen
  \bibfield  {author} {\bibinfo {author} {\bibfnamefont {M.}~\bibnamefont {Newman}}\ and\ \bibinfo {author} {\bibfnamefont {G.~T.}\ \bibnamefont {Barkema}},\ }\href@noop {} {\emph {\bibinfo {title} {Monte Carlo Methods in Statistical Physics}}}\ (\bibinfo  {publisher} {Oxford University Press},\ \bibinfo {year} {1999})\BibitemShut {NoStop}%
\bibitem [{\citenamefont {Esposito}(2012)}]{Esposito2012coarsegraining}%
  \BibitemOpen
  \bibfield  {author} {\bibinfo {author} {\bibfnamefont {M.}~\bibnamefont {Esposito}},\ }\bibfield  {title} {\bibinfo {title} {Stochastic thermodynamics under coarse graining},\ }\href@noop {} {\bibfield  {journal} {\bibinfo  {journal} {Physical Review E}\ }\textbf {\bibinfo {volume} {85}} (\bibinfo {year} {2012})}\BibitemShut {NoStop}%
\bibitem [{\citenamefont {Martynec}\ \emph {et~al.}(2020)\citenamefont {Martynec}, \citenamefont {Klapp1},\ and\ \citenamefont {Loos}}]{martynec2020entropycriticality}%
  \BibitemOpen
  \bibfield  {author} {\bibinfo {author} {\bibfnamefont {T.}~\bibnamefont {Martynec}}, \bibinfo {author} {\bibfnamefont {S.~H.~L.}\ \bibnamefont {Klapp1}},\ and\ \bibinfo {author} {\bibfnamefont {S.~A.~M.}\ \bibnamefont {Loos}},\ }\bibfield  {title} {\bibinfo {title} {Entropy production at criticality in a nonequilibrium {P}otts model},\ }\href@noop {} {\bibfield  {journal} {\bibinfo  {journal} {New Journal of Physics}\ }\textbf {\bibinfo {volume} {22}} (\bibinfo {year} {2020})}\BibitemShut {NoStop}%
\bibitem [{\citenamefont {Bianconi}\ \emph {et~al.}(2008)\citenamefont {Bianconi}, \citenamefont {Gulbahce},\ and\ \citenamefont {Motter}}]{bianconi2008localstructuredirected}%
  \BibitemOpen
  \bibfield  {author} {\bibinfo {author} {\bibfnamefont {G.}~\bibnamefont {Bianconi}}, \bibinfo {author} {\bibfnamefont {N.}~\bibnamefont {Gulbahce}},\ and\ \bibinfo {author} {\bibfnamefont {A.~E.}\ \bibnamefont {Motter}},\ }\bibfield  {title} {\bibinfo {title} {Local structure of directed networks},\ }\href@noop {} {\bibfield  {journal} {\bibinfo  {journal} {Physical Review Letters}\ }\textbf {\bibinfo {volume} {100}} (\bibinfo {year} {2008})}\BibitemShut {NoStop}%
\bibitem [{\citenamefont {Johnson}\ and\ \citenamefont {Jones}(2017)}]{johnson2017loops}%
  \BibitemOpen
  \bibfield  {author} {\bibinfo {author} {\bibfnamefont {S.}~\bibnamefont {Johnson}}\ and\ \bibinfo {author} {\bibfnamefont {N.~S.}\ \bibnamefont {Jones}},\ }\bibfield  {title} {\bibinfo {title} {Looplessness in networks is linked to trophic coherence},\ }\href@noop {} {\bibfield  {journal} {\bibinfo  {journal} {Proceedings of the National Academy of Sciences of the United States of America}\ }\textbf {\bibinfo {volume} {114}},\ \bibinfo {pages} {5618} (\bibinfo {year} {2017})}\BibitemShut {NoStop}%
\bibitem [{\citenamefont {Newman}(2018{\natexlab{b}})}]{Newman2018noisy}%
  \BibitemOpen
  \bibfield  {author} {\bibinfo {author} {\bibfnamefont {M.}~\bibnamefont {Newman}},\ }\bibfield  {title} {\bibinfo {title} {Network structure from rich but noisy data},\ }\href@noop {} {\bibfield  {journal} {\bibinfo  {journal} {Nature Physics}\ }\textbf {\bibinfo {volume} {14}},\ \bibinfo {pages} {542} (\bibinfo {year} {2018}{\natexlab{b}})}\BibitemShut {NoStop}%
\bibitem [{\citenamefont {Newman}\ \emph {et~al.}(2001)\citenamefont {Newman}, \citenamefont {Strogatz},\ and\ \citenamefont {Watts}}]{newman2001randomgraphs}%
  \BibitemOpen
  \bibfield  {author} {\bibinfo {author} {\bibfnamefont {M.~E.~J.}\ \bibnamefont {Newman}}, \bibinfo {author} {\bibfnamefont {S.~H.}\ \bibnamefont {Strogatz}},\ and\ \bibinfo {author} {\bibfnamefont {D.~J.}\ \bibnamefont {Watts}},\ }\bibfield  {title} {\bibinfo {title} {Random graphs with arbitrary degree distributions and their applications},\ }\href@noop {} {\bibfield  {journal} {\bibinfo  {journal} {Physical Review E}\ }\textbf {\bibinfo {volume} {64}} (\bibinfo {year} {2001})}\BibitemShut {NoStop}%
\bibitem [{\citenamefont {Seifert}(2019)}]{seifert2019inference}%
  \BibitemOpen
  \bibfield  {author} {\bibinfo {author} {\bibfnamefont {U.}~\bibnamefont {Seifert}},\ }\bibfield  {title} {\bibinfo {title} {From stochastic thermodynamics to thermodynamic inference},\ }\href@noop {} {\bibfield  {journal} {\bibinfo  {journal} {Annual Review of Condensed Matter Physics}\ }\textbf {\bibinfo {volume} {10}},\ \bibinfo {pages} {171} (\bibinfo {year} {2019})}\BibitemShut {NoStop}%
\bibitem [{\citenamefont {Seif}\ \emph {et~al.}(2021)\citenamefont {Seif}, \citenamefont {Hafezi},\ and\ \citenamefont {Jarzynski}}]{Seif2021machinelearning}%
  \BibitemOpen
  \bibfield  {author} {\bibinfo {author} {\bibfnamefont {A.}~\bibnamefont {Seif}}, \bibinfo {author} {\bibfnamefont {M.}~\bibnamefont {Hafezi}},\ and\ \bibinfo {author} {\bibfnamefont {C.}~\bibnamefont {Jarzynski}},\ }\bibfield  {title} {\bibinfo {title} {Machine learning the thermodynamic arrow of time},\ }\href@noop {} {\bibfield  {journal} {\bibinfo  {journal} {Nature Physics}\ }\textbf {\bibinfo {volume} {17}},\ \bibinfo {pages} {105} (\bibinfo {year} {2021})}\BibitemShut {NoStop}%
\bibitem [{\citenamefont {Frishman}\ and\ \citenamefont {Ronceray}(2020)}]{Frishman2020stochasticforce}%
  \BibitemOpen
  \bibfield  {author} {\bibinfo {author} {\bibfnamefont {A.}~\bibnamefont {Frishman}}\ and\ \bibinfo {author} {\bibfnamefont {P.}~\bibnamefont {Ronceray}},\ }\bibfield  {title} {\bibinfo {title} {Learning force fields from stochastic trajectories},\ }\href@noop {} {\bibfield  {journal} {\bibinfo  {journal} {Physical Review X}\ }\textbf {\bibinfo {volume} {10}} (\bibinfo {year} {2020})}\BibitemShut {NoStop}%
\bibitem [{\citenamefont {Terlizzi}\ \emph {et~al.}(2024)\citenamefont {Terlizzi}, \citenamefont {Gironella}, \citenamefont {Herraez-Aguilar}, \citenamefont {Betz}, \citenamefont {Monroy}, \citenamefont {Baiesi},\ and\ \citenamefont {Ritort}}]{diterlizzi2024variancesum}%
  \BibitemOpen
  \bibfield  {author} {\bibinfo {author} {\bibfnamefont {I.~D.}\ \bibnamefont {Terlizzi}}, \bibinfo {author} {\bibfnamefont {M.}~\bibnamefont {Gironella}}, \bibinfo {author} {\bibfnamefont {D.}~\bibnamefont {Herraez-Aguilar}}, \bibinfo {author} {\bibfnamefont {T.}~\bibnamefont {Betz}}, \bibinfo {author} {\bibfnamefont {F.}~\bibnamefont {Monroy}}, \bibinfo {author} {\bibfnamefont {M.}~\bibnamefont {Baiesi}},\ and\ \bibinfo {author} {\bibfnamefont {F.}~\bibnamefont {Ritort}},\ }\bibfield  {title} {\bibinfo {title} {Variance sum rule for entropy production},\ }\href@noop {} {\bibfield  {journal} {\bibinfo  {journal} {Science}\ }\textbf {\bibinfo {volume} {383}},\ \bibinfo {pages} {971} (\bibinfo {year} {2024})}\BibitemShut {NoStop}%
\bibitem [{\citenamefont {Martínez}\ \emph {et~al.}(2019)\citenamefont {Martínez}, \citenamefont {Bisker}, \citenamefont {Horowitz},\ and\ \citenamefont {Parrondo}}]{martinez2019inferring}%
  \BibitemOpen
  \bibfield  {author} {\bibinfo {author} {\bibfnamefont {I.~A.}\ \bibnamefont {Martínez}}, \bibinfo {author} {\bibfnamefont {G.}~\bibnamefont {Bisker}}, \bibinfo {author} {\bibfnamefont {J.~M.}\ \bibnamefont {Horowitz}},\ and\ \bibinfo {author} {\bibfnamefont {J.~M.~R.}\ \bibnamefont {Parrondo}},\ }\bibfield  {title} {\bibinfo {title} {Inferring broken detailed balance in the absence of observable currents},\ }\href@noop {} {\bibfield  {journal} {\bibinfo  {journal} {Nature Communications}\ }\textbf {\bibinfo {volume} {10}} (\bibinfo {year} {2019})}\BibitemShut {NoStop}%
\bibitem [{\citenamefont {Lucente}\ \emph {et~al.}(2022)\citenamefont {Lucente}, \citenamefont {Baldassarri}, \citenamefont {Puglisi}, \citenamefont {Vulpiani},\ and\ \citenamefont {Viale}}]{lucente2022incomplete}%
  \BibitemOpen
  \bibfield  {author} {\bibinfo {author} {\bibfnamefont {D.}~\bibnamefont {Lucente}}, \bibinfo {author} {\bibfnamefont {A.}~\bibnamefont {Baldassarri}}, \bibinfo {author} {\bibfnamefont {A.}~\bibnamefont {Puglisi}}, \bibinfo {author} {\bibfnamefont {A.}~\bibnamefont {Vulpiani}},\ and\ \bibinfo {author} {\bibfnamefont {M.}~\bibnamefont {Viale}},\ }\bibfield  {title} {\bibinfo {title} {Inference of time irreversibility from incomplete information: Linear systems and its pitfalls},\ }\href@noop {} {\bibfield  {journal} {\bibinfo  {journal} {Physical Review Research}\ }\textbf {\bibinfo {volume} {4}} (\bibinfo {year} {2022})}\BibitemShut {NoStop}%
\bibitem [{\citenamefont {Gilson}\ \emph {et~al.}(2023)\citenamefont {Gilson}, \citenamefont {Tagliazucchi},\ and\ \citenamefont {Cofré}}]{gilson2023OU}%
  \BibitemOpen
  \bibfield  {author} {\bibinfo {author} {\bibfnamefont {M.}~\bibnamefont {Gilson}}, \bibinfo {author} {\bibfnamefont {E.}~\bibnamefont {Tagliazucchi}},\ and\ \bibinfo {author} {\bibfnamefont {R.}~\bibnamefont {Cofré}},\ }\bibfield  {title} {\bibinfo {title} {Entropy production of multivariate {Ornstein-Uhlenbeck} processes correlates with consciousness levels in the human brain},\ }\href@noop {} {\bibfield  {journal} {\bibinfo  {journal} {Physical Review E}\ }\textbf {\bibinfo {volume} {107}} (\bibinfo {year} {2023})}\BibitemShut {NoStop}%
\bibitem [{\citenamefont {Benozzo}\ \emph {et~al.}(2023)\citenamefont {Benozzo}, \citenamefont {Baggio}, \citenamefont {Baron}, \citenamefont {Chiuso}, \citenamefont {Zampieri},\ and\ \citenamefont {Bertoldo}}]{benozzo2023linearstatespace}%
  \BibitemOpen
  \bibfield  {author} {\bibinfo {author} {\bibfnamefont {D.}~\bibnamefont {Benozzo}}, \bibinfo {author} {\bibfnamefont {G.}~\bibnamefont {Baggio}}, \bibinfo {author} {\bibfnamefont {G.}~\bibnamefont {Baron}}, \bibinfo {author} {\bibfnamefont {A.}~\bibnamefont {Chiuso}}, \bibinfo {author} {\bibfnamefont {S.}~\bibnamefont {Zampieri}},\ and\ \bibinfo {author} {\bibfnamefont {A.}~\bibnamefont {Bertoldo}},\ }\bibfield  {title} {\bibinfo {title} {Analyzing asymmetry in brain hierarchies with a linear state-space model of resting-state f{MRI} data},\ }\href@noop {} {\bibfield  {journal} {\bibinfo  {journal} {bioRxiv}\ } (\bibinfo {year} {2023})}\BibitemShut {NoStop}%
\bibitem [{\citenamefont {Timme}\ and\ \citenamefont {Casadiego}(2014)}]{timme2014inference}%
  \BibitemOpen
  \bibfield  {author} {\bibinfo {author} {\bibfnamefont {M.}~\bibnamefont {Timme}}\ and\ \bibinfo {author} {\bibfnamefont {J.}~\bibnamefont {Casadiego}},\ }\bibfield  {title} {\bibinfo {title} {Revealing networks from dynamics: an introduction},\ }\href@noop {} {\bibfield  {journal} {\bibinfo  {journal} {Journal of Physics A: Mathematical and Theoretical}\ }\textbf {\bibinfo {volume} {47}} (\bibinfo {year} {2014})}\BibitemShut {NoStop}%
\bibitem [{\citenamefont {Cliff}\ \emph {et~al.}(2023)\citenamefont {Cliff}, \citenamefont {Bryant}, \citenamefont {Lizier}, \citenamefont {Tsuchiya},\ and\ \citenamefont {Fulcher}}]{cliff2023pairwise}%
  \BibitemOpen
  \bibfield  {author} {\bibinfo {author} {\bibfnamefont {O.~M.}\ \bibnamefont {Cliff}}, \bibinfo {author} {\bibfnamefont {A.~G.}\ \bibnamefont {Bryant}}, \bibinfo {author} {\bibfnamefont {J.~T.}\ \bibnamefont {Lizier}}, \bibinfo {author} {\bibfnamefont {N.}~\bibnamefont {Tsuchiya}},\ and\ \bibinfo {author} {\bibfnamefont {B.~D.}\ \bibnamefont {Fulcher}},\ }\bibfield  {title} {\bibinfo {title} {Unifying pairwise interactions in complex dynamics},\ }\href@noop {} {\bibfield  {journal} {\bibinfo  {journal} {Nature Computational Science}\ }\textbf {\bibinfo {volume} {3}},\ \bibinfo {pages} {883} (\bibinfo {year} {2023})}\BibitemShut {NoStop}%
\bibitem [{\citenamefont {Novelli}\ \emph {et~al.}(2019)\citenamefont {Novelli}, \citenamefont {Wollstadt}, \citenamefont {Mediano}, \citenamefont {Wibral},\ and\ \citenamefont {Lizier}}]{novelli2019multivariatetransferentropy}%
  \BibitemOpen
  \bibfield  {author} {\bibinfo {author} {\bibfnamefont {L.}~\bibnamefont {Novelli}}, \bibinfo {author} {\bibfnamefont {P.}~\bibnamefont {Wollstadt}}, \bibinfo {author} {\bibfnamefont {P.}~\bibnamefont {Mediano}}, \bibinfo {author} {\bibfnamefont {M.}~\bibnamefont {Wibral}},\ and\ \bibinfo {author} {\bibfnamefont {J.~T.}\ \bibnamefont {Lizier}},\ }\bibfield  {title} {\bibinfo {title} {Large-scale directed network inference with multivariate transfer entropy and hierarchical statistical testing},\ }\href@noop {} {\bibfield  {journal} {\bibinfo  {journal} {Network Neuroscience}\ }\textbf {\bibinfo {volume} {3}},\ \bibinfo {pages} {827} (\bibinfo {year} {2019})}\BibitemShut {NoStop}%
\bibitem [{\citenamefont {Zou}\ and\ \citenamefont {Feng}(2009)}]{zou2009grangercausality}%
  \BibitemOpen
  \bibfield  {author} {\bibinfo {author} {\bibfnamefont {C.}~\bibnamefont {Zou}}\ and\ \bibinfo {author} {\bibfnamefont {J.}~\bibnamefont {Feng}},\ }\bibfield  {title} {\bibinfo {title} {Granger causality vs. dynamic {B}ayesian network inference: a comparative study},\ }\href@noop {} {\bibfield  {journal} {\bibinfo  {journal} {BMC Bioinformatics}\ }\textbf {\bibinfo {volume} {10}} (\bibinfo {year} {2009})}\BibitemShut {NoStop}%
\bibitem [{\citenamefont {Gilson}\ \emph {et~al.}(2016)\citenamefont {Gilson}, \citenamefont {Moreno-Bote}, \citenamefont {Ponce-Alvarez}, \citenamefont {Ritter},\ and\ \citenamefont {Deco}}]{Gilson2016EC}%
  \BibitemOpen
  \bibfield  {author} {\bibinfo {author} {\bibfnamefont {M.}~\bibnamefont {Gilson}}, \bibinfo {author} {\bibfnamefont {R.}~\bibnamefont {Moreno-Bote}}, \bibinfo {author} {\bibfnamefont {A.}~\bibnamefont {Ponce-Alvarez}}, \bibinfo {author} {\bibfnamefont {P.}~\bibnamefont {Ritter}},\ and\ \bibinfo {author} {\bibfnamefont {G.}~\bibnamefont {Deco}},\ }\bibfield  {title} {\bibinfo {title} {Estimation of directed effective connectivity from {fMRI} functional connectivity hints at asymmetries of cortical connectome},\ }\href@noop {} {\bibfield  {journal} {\bibinfo  {journal} {PLOS Computational Biology}\ }\textbf {\bibinfo {volume} {12}} (\bibinfo {year} {2016})}\BibitemShut {NoStop}%
\bibitem [{\citenamefont {Gilson}\ \emph {et~al.}(2020)\citenamefont {Gilson}, \citenamefont {Zamora-López}, \citenamefont {Pallarés}, \citenamefont {Adhikari}, \citenamefont {Senden}, \citenamefont {Campo}, \citenamefont {Mantini}, \citenamefont {Corbetta}, \citenamefont {Deco},\ and\ \citenamefont {Insabato}}]{Gilson2020EC}%
  \BibitemOpen
  \bibfield  {author} {\bibinfo {author} {\bibfnamefont {M.}~\bibnamefont {Gilson}}, \bibinfo {author} {\bibfnamefont {G.}~\bibnamefont {Zamora-López}}, \bibinfo {author} {\bibfnamefont {V.}~\bibnamefont {Pallarés}}, \bibinfo {author} {\bibfnamefont {M.~H.}\ \bibnamefont {Adhikari}}, \bibinfo {author} {\bibfnamefont {M.}~\bibnamefont {Senden}}, \bibinfo {author} {\bibfnamefont {A.~T.}\ \bibnamefont {Campo}}, \bibinfo {author} {\bibfnamefont {D.}~\bibnamefont {Mantini}}, \bibinfo {author} {\bibfnamefont {M.}~\bibnamefont {Corbetta}}, \bibinfo {author} {\bibfnamefont {G.}~\bibnamefont {Deco}},\ and\ \bibinfo {author} {\bibfnamefont {A.}~\bibnamefont {Insabato}},\ }\bibfield  {title} {\bibinfo {title} {Model-based whole-brain effective connectivity to study distributed cognition in health and disease},\ }\href@noop {} {\bibfield  {journal} {\bibinfo  {journal} {Network Neuroscience}\ }\textbf {\bibinfo {volume} {4}},\ \bibinfo {pages} {338} (\bibinfo {year} {2020})}\BibitemShut {NoStop}%
\bibitem [{\citenamefont {Villaverde}\ \emph {et~al.}(2014)\citenamefont {Villaverde}, \citenamefont {Ross}, \citenamefont {Morán},\ and\ \citenamefont {Banga}}]{villaverde2014mider}%
  \BibitemOpen
  \bibfield  {author} {\bibinfo {author} {\bibfnamefont {A.~F.}\ \bibnamefont {Villaverde}}, \bibinfo {author} {\bibfnamefont {J.}~\bibnamefont {Ross}}, \bibinfo {author} {\bibfnamefont {F.}~\bibnamefont {Morán}},\ and\ \bibinfo {author} {\bibfnamefont {J.~R.}\ \bibnamefont {Banga}},\ }\bibfield  {title} {\bibinfo {title} {Mider: Network inference with mutual information distance and entropy reduction},\ }\href@noop {} {\bibfield  {journal} {\bibinfo  {journal} {PLOS ONE}\ }\textbf {\bibinfo {volume} {9}} (\bibinfo {year} {2014})}\BibitemShut {NoStop}%
\bibitem [{\citenamefont {Friston}(2011)}]{friston2011FCEC}%
  \BibitemOpen
  \bibfield  {author} {\bibinfo {author} {\bibfnamefont {K.}~\bibnamefont {Friston}},\ }\bibfield  {title} {\bibinfo {title} {Functional and effective connectivity: A review},\ }\href@noop {} {\bibfield  {journal} {\bibinfo  {journal} {Brain Connectivity}\ }\textbf {\bibinfo {volume} {1}} (\bibinfo {year} {2011})}\BibitemShut {NoStop}%
\bibitem [{\citenamefont {Nozari}\ \emph {et~al.}(2023)\citenamefont {Nozari}, \citenamefont {Bertolero}, \citenamefont {Stiso}, \citenamefont {Caciagli}, \citenamefont {Cornblath}, \citenamefont {He}, \citenamefont {Mahadevan}, \citenamefont {Pappas},\ and\ \citenamefont {Bassett}}]{nozari2023linearbrain}%
  \BibitemOpen
  \bibfield  {author} {\bibinfo {author} {\bibfnamefont {E.}~\bibnamefont {Nozari}}, \bibinfo {author} {\bibfnamefont {M.~A.}\ \bibnamefont {Bertolero}}, \bibinfo {author} {\bibfnamefont {J.}~\bibnamefont {Stiso}}, \bibinfo {author} {\bibfnamefont {L.}~\bibnamefont {Caciagli}}, \bibinfo {author} {\bibfnamefont {E.~J.}\ \bibnamefont {Cornblath}}, \bibinfo {author} {\bibfnamefont {X.}~\bibnamefont {He}}, \bibinfo {author} {\bibfnamefont {A.~S.}\ \bibnamefont {Mahadevan}}, \bibinfo {author} {\bibfnamefont {G.~J.}\ \bibnamefont {Pappas}},\ and\ \bibinfo {author} {\bibfnamefont {D.~S.}\ \bibnamefont {Bassett}},\ }\bibfield  {title} {\bibinfo {title} {Macroscopic resting-state brain dynamics are best described by linear models},\ }\href@noop {} {\bibfield  {journal} {\bibinfo  {journal} {Nature Biomedical Engineering}\ } (\bibinfo {year} {2023})}\BibitemShut {NoStop}%
\bibitem [{\citenamefont {Brunton}\ \emph {et~al.}(2017)\citenamefont {Brunton}, \citenamefont {Brunton}, \citenamefont {Proctor}, \citenamefont {Kaiser},\ and\ \citenamefont {Kutz}}]{brunton2017chaos}%
  \BibitemOpen
  \bibfield  {author} {\bibinfo {author} {\bibfnamefont {S.~L.}\ \bibnamefont {Brunton}}, \bibinfo {author} {\bibfnamefont {B.~W.}\ \bibnamefont {Brunton}}, \bibinfo {author} {\bibfnamefont {J.~L.}\ \bibnamefont {Proctor}}, \bibinfo {author} {\bibfnamefont {E.}~\bibnamefont {Kaiser}},\ and\ \bibinfo {author} {\bibfnamefont {J.~N.}\ \bibnamefont {Kutz}},\ }\bibfield  {title} {\bibinfo {title} {Chaos as an intermittently forced linear system},\ }\href@noop {} {\bibfield  {journal} {\bibinfo  {journal} {Nature Communications}\ }\textbf {\bibinfo {volume} {8}} (\bibinfo {year} {2017})}\BibitemShut {NoStop}%
\bibitem [{\citenamefont {Harrison}\ \emph {et~al.}(2003)\citenamefont {Harrison}, \citenamefont {Penny},\ and\ \citenamefont {Friston}}]{harrison2003autoregressionfmri}%
  \BibitemOpen
  \bibfield  {author} {\bibinfo {author} {\bibfnamefont {L.}~\bibnamefont {Harrison}}, \bibinfo {author} {\bibfnamefont {W.}~\bibnamefont {Penny}},\ and\ \bibinfo {author} {\bibfnamefont {K.}~\bibnamefont {Friston}},\ }\bibfield  {title} {\bibinfo {title} {Multivariate autoregressive modeling of {fMRI} time series},\ }\href@noop {} {\bibfield  {journal} {\bibinfo  {journal} {NeuroImage}\ }\textbf {\bibinfo {volume} {19}},\ \bibinfo {pages} {1477} (\bibinfo {year} {2003})}\BibitemShut {NoStop}%
\bibitem [{\citenamefont {Kloeden}\ and\ \citenamefont {Platen}(1992)}]{kloeden1992sdes}%
  \BibitemOpen
  \bibfield  {author} {\bibinfo {author} {\bibfnamefont {P.}~\bibnamefont {Kloeden}}\ and\ \bibinfo {author} {\bibfnamefont {E.}~\bibnamefont {Platen}},\ }\href@noop {} {\emph {\bibinfo {title} {Numerical Solution of Stochastic Differential Equations}}}\ (\bibinfo  {publisher} {Springer},\ \bibinfo {year} {1992})\BibitemShut {NoStop}%
\bibitem [{\citenamefont {Chen}\ and\ \citenamefont {Plemmons}(2009)}]{Chen2009nnls}%
  \BibitemOpen
  \bibfield  {author} {\bibinfo {author} {\bibfnamefont {D.}~\bibnamefont {Chen}}\ and\ \bibinfo {author} {\bibfnamefont {R.}~\bibnamefont {Plemmons}},\ }\bibfield  {title} {\bibinfo {title} {Nonnegativity constraints in numerical analysis},\ }in\ \href@noop {} {\emph {\bibinfo {booktitle} {The Birth of Numerical Analysis}}}\ (\bibinfo  {publisher} {World Scientific},\ \bibinfo {year} {2009})\ pp.\ \bibinfo {pages} {109--139}\BibitemShut {NoStop}%
\bibitem [{\citenamefont {Van~Essen}\ \emph {et~al.}(2013)\citenamefont {Van~Essen}, \citenamefont {Smith}, \citenamefont {Barch}, \citenamefont {Behrens}, \citenamefont {Yacoub}, \citenamefont {Ugurbil},\ and\ \citenamefont {for~the WU-Minn HCP~Consortium}}]{vanessen2013hcp}%
  \BibitemOpen
  \bibfield  {author} {\bibinfo {author} {\bibfnamefont {D.}~\bibnamefont {Van~Essen}}, \bibinfo {author} {\bibfnamefont {S.}~\bibnamefont {Smith}}, \bibinfo {author} {\bibfnamefont {D.}~\bibnamefont {Barch}}, \bibinfo {author} {\bibfnamefont {T.}~\bibnamefont {Behrens}}, \bibinfo {author} {\bibfnamefont {E.}~\bibnamefont {Yacoub}}, \bibinfo {author} {\bibfnamefont {K.}~\bibnamefont {Ugurbil}},\ and\ \bibinfo {author} {\bibnamefont {for~the WU-Minn HCP~Consortium}},\ }\bibfield  {title} {\bibinfo {title} {The {WU-Minn} {H}uman {C}onnectome {P}roject: An overview},\ }\href@noop {} {\bibfield  {journal} {\bibinfo  {journal} {NeuroImage}\ }\textbf {\bibinfo {volume} {80}},\ \bibinfo {pages} {62} (\bibinfo {year} {2013})}\BibitemShut {NoStop}%
\bibitem [{\citenamefont {Desikan}\ \emph {et~al.}(2006)\citenamefont {Desikan}, \citenamefont {Ségonne}, \citenamefont {Fischl}, \citenamefont {Quinn}, \citenamefont {Dickerson}, \citenamefont {Blacker}, \citenamefont {Buckner}, \citenamefont {Dale}, \citenamefont {Maguire}, \citenamefont {Hyman}, \citenamefont {Albert},\ and\ \citenamefont {Killiany}}]{desikan2006dk80}%
  \BibitemOpen
  \bibfield  {author} {\bibinfo {author} {\bibfnamefont {R.}~\bibnamefont {Desikan}}, \bibinfo {author} {\bibfnamefont {F.}~\bibnamefont {Ségonne}}, \bibinfo {author} {\bibfnamefont {B.}~\bibnamefont {Fischl}}, \bibinfo {author} {\bibfnamefont {B.}~\bibnamefont {Quinn}}, \bibinfo {author} {\bibfnamefont {B.}~\bibnamefont {Dickerson}}, \bibinfo {author} {\bibfnamefont {D.}~\bibnamefont {Blacker}}, \bibinfo {author} {\bibfnamefont {R.}~\bibnamefont {Buckner}}, \bibinfo {author} {\bibfnamefont {A.}~\bibnamefont {Dale}}, \bibinfo {author} {\bibfnamefont {R.}~\bibnamefont {Maguire}}, \bibinfo {author} {\bibfnamefont {B.}~\bibnamefont {Hyman}}, \bibinfo {author} {\bibfnamefont {M.}~\bibnamefont {Albert}},\ and\ \bibinfo {author} {\bibfnamefont {R.}~\bibnamefont {Killiany}},\ }\bibfield  {title} {\bibinfo {title} {An automated labeling system for subdividing the human cerebral cortex on {MRI} scans into gyral based regions of interest},\ }\href@noop {} {\bibfield  {journal} {\bibinfo  {journal} {Neuroimage}\
  }\textbf {\bibinfo {volume} {31}},\ \bibinfo {pages} {968} (\bibinfo {year} {2006})}\BibitemShut {NoStop}%
\bibitem [{\citenamefont {Deco}\ \emph {et~al.}(2021)\citenamefont {Deco}, \citenamefont {Vidaurre},\ and\ \citenamefont {Kringelbach}}]{deco2021workspace}%
  \BibitemOpen
  \bibfield  {author} {\bibinfo {author} {\bibfnamefont {G.}~\bibnamefont {Deco}}, \bibinfo {author} {\bibfnamefont {D.}~\bibnamefont {Vidaurre}},\ and\ \bibinfo {author} {\bibfnamefont {M.~L.}\ \bibnamefont {Kringelbach}},\ }\bibfield  {title} {\bibinfo {title} {Revisiting the global workspace orchestrating the hierarchical organization of the human brain},\ }\href@noop {} {\bibfield  {journal} {\bibinfo  {journal} {Nature Human Behaviour}\ }\textbf {\bibinfo {volume} {5}},\ \bibinfo {pages} {497} (\bibinfo {year} {2021})}\BibitemShut {NoStop}%
\bibitem [{\citenamefont {Yeo}\ \emph {et~al.}(2011)\citenamefont {Yeo}, \citenamefont {Krienen}, \citenamefont {Sepulcre}, \citenamefont {Sabuncu}, \citenamefont {Lashkari}, \citenamefont {Hollinshead}, \citenamefont {Roffman}, \citenamefont {Smoller}, \citenamefont {Zöllei}, \citenamefont {Polimeni}, \citenamefont {Fischl}, \citenamefont {Liu},\ and\ \citenamefont {Buckner}}]{yeo2011restingnetworks}%
  \BibitemOpen
  \bibfield  {author} {\bibinfo {author} {\bibfnamefont {B.~T.~T.}\ \bibnamefont {Yeo}}, \bibinfo {author} {\bibfnamefont {F.~M.}\ \bibnamefont {Krienen}}, \bibinfo {author} {\bibfnamefont {J.}~\bibnamefont {Sepulcre}}, \bibinfo {author} {\bibfnamefont {M.~R.}\ \bibnamefont {Sabuncu}}, \bibinfo {author} {\bibfnamefont {D.}~\bibnamefont {Lashkari}}, \bibinfo {author} {\bibfnamefont {M.}~\bibnamefont {Hollinshead}}, \bibinfo {author} {\bibfnamefont {J.~L.}\ \bibnamefont {Roffman}}, \bibinfo {author} {\bibfnamefont {J.~W.}\ \bibnamefont {Smoller}}, \bibinfo {author} {\bibfnamefont {L.}~\bibnamefont {Zöllei}}, \bibinfo {author} {\bibfnamefont {J.~R.}\ \bibnamefont {Polimeni}}, \bibinfo {author} {\bibfnamefont {B.}~\bibnamefont {Fischl}}, \bibinfo {author} {\bibfnamefont {H.}~\bibnamefont {Liu}},\ and\ \bibinfo {author} {\bibfnamefont {R.~L.}\ \bibnamefont {Buckner}},\ }\bibfield  {title} {\bibinfo {title} {The organization of the human cerebral cortex estimated by intrinsic functional connectivity},\ }\href@noop
  {} {\bibfield  {journal} {\bibinfo  {journal} {Journal of Neurophysiology}\ }\textbf {\bibinfo {volume} {106}},\ \bibinfo {pages} {1125} (\bibinfo {year} {2011})}\BibitemShut {NoStop}%
\bibitem [{\citenamefont {Kringelbach}\ \emph {et~al.}(2023)\citenamefont {Kringelbach}, \citenamefont {Sanz-Perl}, \citenamefont {Tagliazucchi},\ and\ \citenamefont {Deco}}]{kringelbach2023movie}%
  \BibitemOpen
  \bibfield  {author} {\bibinfo {author} {\bibfnamefont {M.~L.}\ \bibnamefont {Kringelbach}}, \bibinfo {author} {\bibfnamefont {Y.}~\bibnamefont {Sanz-Perl}}, \bibinfo {author} {\bibfnamefont {E.}~\bibnamefont {Tagliazucchi}},\ and\ \bibinfo {author} {\bibfnamefont {G.}~\bibnamefont {Deco}},\ }\bibfield  {title} {\bibinfo {title} {Toward naturalistic neuroscience: Mechanisms underlying the flattening of brain hierarchy in movie-watching compared to rest and task},\ }\href@noop {} {\bibfield  {journal} {\bibinfo  {journal} {Science Advances}\ }\textbf {\bibinfo {volume} {9}} (\bibinfo {year} {2023})}\BibitemShut {NoStop}%
\bibitem [{\citenamefont {Dehaene}\ \emph {et~al.}(1998)\citenamefont {Dehaene}, \citenamefont {Kerszberg},\ and\ \citenamefont {Changeux}}]{dehane1998globalworkspace}%
  \BibitemOpen
  \bibfield  {author} {\bibinfo {author} {\bibfnamefont {S.}~\bibnamefont {Dehaene}}, \bibinfo {author} {\bibfnamefont {M.}~\bibnamefont {Kerszberg}},\ and\ \bibinfo {author} {\bibfnamefont {J.-P.}\ \bibnamefont {Changeux}},\ }\bibfield  {title} {\bibinfo {title} {A neuronal model of a global workspace in effortful cognitive tasks},\ }\href@noop {} {\bibfield  {journal} {\bibinfo  {journal} {Proceedings of the National Academy of Sciences of the United States of America}\ }\textbf {\bibinfo {volume} {95}},\ \bibinfo {pages} {14529} (\bibinfo {year} {1998})}\BibitemShut {NoStop}%
\bibitem [{\citenamefont {Santoro}\ \emph {et~al.}(2023)\citenamefont {Santoro}, \citenamefont {Battiston}, \citenamefont {Petri},\ and\ \citenamefont {Amico}}]{santoro2023higherorder}%
  \BibitemOpen
  \bibfield  {author} {\bibinfo {author} {\bibfnamefont {A.}~\bibnamefont {Santoro}}, \bibinfo {author} {\bibfnamefont {F.}~\bibnamefont {Battiston}}, \bibinfo {author} {\bibfnamefont {G.}~\bibnamefont {Petri}},\ and\ \bibinfo {author} {\bibfnamefont {E.}~\bibnamefont {Amico}},\ }\bibfield  {title} {\bibinfo {title} {Higher-order organization of multivariate time series},\ }\href@noop {} {\bibfield  {journal} {\bibinfo  {journal} {Nature Physics}\ }\textbf {\bibinfo {volume} {19}},\ \bibinfo {pages} {221–229} (\bibinfo {year} {2023})}\BibitemShut {NoStop}%
\bibitem [{\citenamefont {Sieniutycz}\ and\ \citenamefont {Salamon}(1990)}]{Sieniutycz1999thermoeconomics}%
  \BibitemOpen
  \bibfield  {author} {\bibinfo {author} {\bibfnamefont {S.}~\bibnamefont {Sieniutycz}}\ and\ \bibinfo {author} {\bibfnamefont {P.}~\bibnamefont {Salamon}},\ }\href@noop {} {\emph {\bibinfo {title} {Finite-time thermodynamics and thermoeconomics}}}\ (\bibinfo  {publisher} {Taylor \& Francis},\ \bibinfo {year} {1990})\BibitemShut {NoStop}%
\bibitem [{\citenamefont {Pokrovskii}(2020)}]{Pokrovskii2020thermodynamics}%
  \BibitemOpen
  \bibfield  {author} {\bibinfo {author} {\bibfnamefont {V.~N.}\ \bibnamefont {Pokrovskii}},\ }\href@noop {} {\emph {\bibinfo {title} {Thermodynamics of Complex Systems: Principles and applications}}}\ (\bibinfo  {publisher} {IOP Publishing},\ \bibinfo {year} {2020})\BibitemShut {NoStop}%
\bibitem [{\citenamefont {Chen}\ \emph {et~al.}(1986)\citenamefont {Chen}, \citenamefont {Roll},\ and\ \citenamefont {Ross}}]{chen1986economicsforces}%
  \BibitemOpen
  \bibfield  {author} {\bibinfo {author} {\bibfnamefont {N.-F.}\ \bibnamefont {Chen}}, \bibinfo {author} {\bibfnamefont {R.}~\bibnamefont {Roll}},\ and\ \bibinfo {author} {\bibfnamefont {S.~A.}\ \bibnamefont {Ross}},\ }\bibfield  {title} {\bibinfo {title} {Economic forces and the stock market},\ }\href@noop {} {\bibfield  {journal} {\bibinfo  {journal} {The Journal of Business}\ }\textbf {\bibinfo {volume} {59}},\ \bibinfo {pages} {383} (\bibinfo {year} {1986})}\BibitemShut {NoStop}%
\bibitem [{\citenamefont {Rodgers}\ \emph {et~al.}(2023)\citenamefont {Rodgers}, \citenamefont {Tiňo},\ and\ \citenamefont {Johnson}}]{rodgers2023strongconnectivity}%
  \BibitemOpen
  \bibfield  {author} {\bibinfo {author} {\bibfnamefont {N.}~\bibnamefont {Rodgers}}, \bibinfo {author} {\bibfnamefont {P.}~\bibnamefont {Tiňo}},\ and\ \bibinfo {author} {\bibfnamefont {S.}~\bibnamefont {Johnson}},\ }\bibfield  {title} {\bibinfo {title} {Strong connectivity in real directed networks},\ }\href@noop {} {\bibfield  {journal} {\bibinfo  {journal} {Proceedings of the National Academy of Sciences of the United States of America}\ }\textbf {\bibinfo {volume} {120}} (\bibinfo {year} {2023})}\BibitemShut {NoStop}%
\bibitem [{\citenamefont {Moretti}\ and\ \citenamefont {Muñoz}(2013)}]{Moretti2013griffiths}%
  \BibitemOpen
  \bibfield  {author} {\bibinfo {author} {\bibfnamefont {P.}~\bibnamefont {Moretti}}\ and\ \bibinfo {author} {\bibfnamefont {M.~A.}\ \bibnamefont {Muñoz}},\ }\bibfield  {title} {\bibinfo {title} {Griffiths phases and the stretching of criticality in brain networks},\ }\href@noop {} {\bibfield  {journal} {\bibinfo  {journal} {Nature Communications}\ }\textbf {\bibinfo {volume} {4}} (\bibinfo {year} {2013})}\BibitemShut {NoStop}%
\bibitem [{\citenamefont {Santoro}\ \emph {et~al.}(2022)\citenamefont {Santoro}, \citenamefont {Battiston}, \citenamefont {Petri},\ and\ \citenamefont {Amico}}]{santoro2023higherorderdata}%
  \BibitemOpen
  \bibfield  {author} {\bibinfo {author} {\bibfnamefont {A.}~\bibnamefont {Santoro}}, \bibinfo {author} {\bibfnamefont {F.}~\bibnamefont {Battiston}}, \bibinfo {author} {\bibfnamefont {G.}~\bibnamefont {Petri}},\ and\ \bibinfo {author} {\bibfnamefont {E.}~\bibnamefont {Amico}},\ }\bibfield  {title} {\bibinfo {title} {Higher-order organization of multivariate time series},\ }\bibfield  {journal} {\bibinfo  {journal} {Zenodo}\ }\href {https://doi.org/https://doi.org/10.5281/zenodo.7210076} {https://doi.org/10.5281/zenodo.7210076} (\bibinfo {year} {2022})\BibitemShut {NoStop}%
\bibitem [{\citenamefont {de~Solla~Price}(1965)}]{desollaprice1965networksofscience}%
  \BibitemOpen
  \bibfield  {author} {\bibinfo {author} {\bibfnamefont {D.~J.}\ \bibnamefont {de~Solla~Price}},\ }\bibfield  {title} {\bibinfo {title} {Networks of scientific papers},\ }\href@noop {} {\bibfield  {journal} {\bibinfo  {journal} {Science}\ }\textbf {\bibinfo {volume} {149}},\ \bibinfo {pages} {510} (\bibinfo {year} {1965})}\BibitemShut {NoStop}%
\bibitem [{\citenamefont {de~Solla~Price}(1976)}]{desollaprice1976bibliometric}%
  \BibitemOpen
  \bibfield  {author} {\bibinfo {author} {\bibfnamefont {D.~J.}\ \bibnamefont {de~Solla~Price}},\ }\bibfield  {title} {\bibinfo {title} {A general theory of bibliometric and other cumulative advantage processes},\ }\href@noop {} {\bibfield  {journal} {\bibinfo  {journal} {Journal of the American Society for Information Science}\ ,\ \bibinfo {pages} {292}} (\bibinfo {year} {1976})}\BibitemShut {NoStop}%
\bibitem [{\citenamefont {Barabási}\ and\ \citenamefont {Albert}(1999)}]{barabasi1999pa}%
  \BibitemOpen
  \bibfield  {author} {\bibinfo {author} {\bibfnamefont {A.-L.}\ \bibnamefont {Barabási}}\ and\ \bibinfo {author} {\bibfnamefont {R.}~\bibnamefont {Albert}},\ }\bibfield  {title} {\bibinfo {title} {Emergence of scaling in random networks},\ }\href@noop {} {\bibfield  {journal} {\bibinfo  {journal} {Science}\ }\textbf {\bibinfo {volume} {286}},\ \bibinfo {pages} {509} (\bibinfo {year} {1999})}\BibitemShut {NoStop}%
\bibitem [{\citenamefont {Lamrous}\ and\ \citenamefont {Taileb}(2006)}]{lamrous2006hierarchicalkmeans}%
  \BibitemOpen
  \bibfield  {author} {\bibinfo {author} {\bibfnamefont {S.}~\bibnamefont {Lamrous}}\ and\ \bibinfo {author} {\bibfnamefont {M.}~\bibnamefont {Taileb}},\ }\bibfield  {title} {\bibinfo {title} {Divisive hierarchical $k$-means},\ }\href@noop {} {\bibfield  {journal} {\bibinfo  {journal} {IEEE International Conference on Computational Inteligence for Modelling Control and Automation}\ } (\bibinfo {year} {2006})}\BibitemShut {NoStop}%
\bibitem [{\citenamefont {Kappen}\ and\ \citenamefont {Spanjers}(2000)}]{kappen2000meanfield}%
  \BibitemOpen
  \bibfield  {author} {\bibinfo {author} {\bibfnamefont {H.~J.}\ \bibnamefont {Kappen}}\ and\ \bibinfo {author} {\bibfnamefont {J.~J.}\ \bibnamefont {Spanjers}},\ }\bibfield  {title} {\bibinfo {title} {Mean field theory for asymmetric neural networks},\ }\href@noop {} {\bibfield  {journal} {\bibinfo  {journal} {Physical Review E}\ }\textbf {\bibinfo {volume} {61}} (\bibinfo {year} {2000})}\BibitemShut {NoStop}%
\bibitem [{\citenamefont {Tanaka}(2001)}]{tanaka2001InformationGeometry}%
  \BibitemOpen
  \bibfield  {author} {\bibinfo {author} {\bibfnamefont {T.}~\bibnamefont {Tanaka}},\ }\bibfield  {title} {\bibinfo {title} {Information geometry of mean-field approximation},\ }in\ \href@noop {} {\emph {\bibinfo {booktitle} {Advanced mean field methods: Theory and practice}}}\ (\bibinfo  {publisher} {MIT Press},\ \bibinfo {year} {2001})\ pp.\ \bibinfo {pages} {351--360}\BibitemShut {NoStop}%
\bibitem [{\citenamefont {Amari}\ \emph {et~al.}(2001)\citenamefont {Amari}, \citenamefont {Ikeda},\ and\ \citenamefont {Shimokawa}}]{amari2001alphaprojection}%
  \BibitemOpen
  \bibfield  {author} {\bibinfo {author} {\bibfnamefont {S.}~\bibnamefont {Amari}}, \bibinfo {author} {\bibfnamefont {S.}~\bibnamefont {Ikeda}},\ and\ \bibinfo {author} {\bibfnamefont {H.}~\bibnamefont {Shimokawa}},\ }\bibfield  {title} {\bibinfo {title} {Information geometry of alpha-projection in mean field approximation},\ }in\ \href@noop {} {\emph {\bibinfo {booktitle} {Advanced mean field methods: Theory and practice}}}\ (\bibinfo  {publisher} {MIT Press},\ \bibinfo {year} {2001})\ pp.\ \bibinfo {pages} {241--257}\BibitemShut {NoStop}%
\bibitem [{\citenamefont {Dunne}\ \emph {et~al.}(2013)\citenamefont {Dunne}, \citenamefont {Lafferty}, \citenamefont {Dobson}, \citenamefont {Hechinger}, \citenamefont {Kuris}, \citenamefont {Martinez}, \citenamefont {McLaughlin}, \citenamefont {Mouritsen}, \citenamefont {Poulin}, \citenamefont {Reise}, \citenamefont {Stouffer}, \citenamefont {Thieltges}, \citenamefont {Williams},\ and\ \citenamefont {Zander}}]{dunne2013fooweb}%
  \BibitemOpen
  \bibfield  {author} {\bibinfo {author} {\bibfnamefont {J.~A.}\ \bibnamefont {Dunne}}, \bibinfo {author} {\bibfnamefont {K.~D.}\ \bibnamefont {Lafferty}}, \bibinfo {author} {\bibfnamefont {A.~P.}\ \bibnamefont {Dobson}}, \bibinfo {author} {\bibfnamefont {R.~F.}\ \bibnamefont {Hechinger}}, \bibinfo {author} {\bibfnamefont {A.~M.}\ \bibnamefont {Kuris}}, \bibinfo {author} {\bibfnamefont {N.~D.}\ \bibnamefont {Martinez}}, \bibinfo {author} {\bibfnamefont {J.~P.}\ \bibnamefont {McLaughlin}}, \bibinfo {author} {\bibfnamefont {K.~N.}\ \bibnamefont {Mouritsen}}, \bibinfo {author} {\bibfnamefont {R.}~\bibnamefont {Poulin}}, \bibinfo {author} {\bibfnamefont {K.}~\bibnamefont {Reise}}, \bibinfo {author} {\bibfnamefont {D.~B.}\ \bibnamefont {Stouffer}}, \bibinfo {author} {\bibfnamefont {D.~W.}\ \bibnamefont {Thieltges}}, \bibinfo {author} {\bibfnamefont {R.~J.}\ \bibnamefont {Williams}},\ and\ \bibinfo {author} {\bibfnamefont {C.~D.}\ \bibnamefont {Zander}},\ }\bibfield  {title} {\bibinfo {title} {Parasites affect food
  web structure primarily through increased diversity and complexity},\ }\href@noop {} {\bibfield  {journal} {\bibinfo  {journal} {PLOS Biology}\ }\textbf {\bibinfo {volume} {11}} (\bibinfo {year} {2013})}\BibitemShut {NoStop}%
\bibitem [{\citenamefont {Thompson}\ and\ \citenamefont {Townsend}(2003)}]{thompson2003foodweb}%
  \BibitemOpen
  \bibfield  {author} {\bibinfo {author} {\bibfnamefont {R.}~\bibnamefont {Thompson}}\ and\ \bibinfo {author} {\bibfnamefont {C.}~\bibnamefont {Townsend}},\ }\bibfield  {title} {\bibinfo {title} {Impacts on stream food webs of native and exotic forest: an intercontinental comparison},\ }\href@noop {} {\bibfield  {journal} {\bibinfo  {journal} {Ecology}\ }\textbf {\bibinfo {volume} {84}},\ \bibinfo {pages} {145} (\bibinfo {year} {2003})}\BibitemShut {NoStop}%
\bibitem [{\citenamefont {Townsend}\ \emph {et~al.}(1998)\citenamefont {Townsend}, \citenamefont {Thompson}, \citenamefont {McIntosh}, \citenamefont {Kilroy}, \citenamefont {Edwards},\ and\ \citenamefont {Scarsbrook}}]{thompson2002foodweb}%
  \BibitemOpen
  \bibfield  {author} {\bibinfo {author} {\bibfnamefont {C.~R.}\ \bibnamefont {Townsend}}, \bibinfo {author} {\bibfnamefont {R.}~\bibnamefont {Thompson}}, \bibinfo {author} {\bibfnamefont {A.}~\bibnamefont {McIntosh}}, \bibinfo {author} {\bibfnamefont {C.}~\bibnamefont {Kilroy}}, \bibinfo {author} {\bibfnamefont {E.}~\bibnamefont {Edwards}},\ and\ \bibinfo {author} {\bibfnamefont {M.~R.}\ \bibnamefont {Scarsbrook}},\ }\bibfield  {title} {\bibinfo {title} {Disturbance, resource supply, and food-web architecture in streams},\ }\href@noop {} {\bibfield  {journal} {\bibinfo  {journal} {Ecology Letters}\ }\textbf {\bibinfo {volume} {1}},\ \bibinfo {pages} {200} (\bibinfo {year} {1998})}\BibitemShut {NoStop}%
\bibitem [{\citenamefont {Klaise}\ and\ \citenamefont {Johnson}(2017)}]{klaise2017foodwebs}%
  \BibitemOpen
  \bibfield  {author} {\bibinfo {author} {\bibfnamefont {J.}~\bibnamefont {Klaise}}\ and\ \bibinfo {author} {\bibfnamefont {S.}~\bibnamefont {Johnson}},\ }\bibfield  {title} {\bibinfo {title} {The origin of motif families in food webs},\ }\href@noop {} {\bibfield  {journal} {\bibinfo  {journal} {Scientific Reports}\ }\textbf {\bibinfo {volume} {7}} (\bibinfo {year} {2017})}\BibitemShut {NoStop}%
\bibitem [{\citenamefont {Thompson}\ and\ \citenamefont {Townsend}(2004)}]{thompson2004energy}%
  \BibitemOpen
  \bibfield  {author} {\bibinfo {author} {\bibfnamefont {R.~M.}\ \bibnamefont {Thompson}}\ and\ \bibinfo {author} {\bibfnamefont {C.~R.}\ \bibnamefont {Townsend}},\ }\bibfield  {title} {\bibinfo {title} {Energy availability, spatial heterogeneity and ecosystem size predict food-web structure in streams},\ }\href@noop {} {\bibfield  {journal} {\bibinfo  {journal} {OIKOS}\ }\textbf {\bibinfo {volume} {108}},\ \bibinfo {pages} {137} (\bibinfo {year} {2004})}\BibitemShut {NoStop}%
\bibitem [{\citenamefont {Memmott}\ \emph {et~al.}(2001)\citenamefont {Memmott}, \citenamefont {Martinez},\ and\ \citenamefont {Cohen}}]{memmott2001predators}%
  \BibitemOpen
  \bibfield  {author} {\bibinfo {author} {\bibfnamefont {J.}~\bibnamefont {Memmott}}, \bibinfo {author} {\bibfnamefont {N.}~\bibnamefont {Martinez}},\ and\ \bibinfo {author} {\bibfnamefont {J.}~\bibnamefont {Cohen}},\ }\bibfield  {title} {\bibinfo {title} {Predators, parasitoids and pathogens: species richness, trophic generality and body sizes in a natural food web},\ }\href@noop {} {\bibfield  {journal} {\bibinfo  {journal} {Journal of Animal Ecology}\ }\textbf {\bibinfo {volume} {69}},\ \bibinfo {pages} {1} (\bibinfo {year} {2001})}\BibitemShut {NoStop}%
\bibitem [{\citenamefont {Bascompte}\ \emph {et~al.}(2005)\citenamefont {Bascompte}, \citenamefont {Melián},\ and\ \citenamefont {Sala}}]{bascompte2005marinefoodweb}%
  \BibitemOpen
  \bibfield  {author} {\bibinfo {author} {\bibfnamefont {J.}~\bibnamefont {Bascompte}}, \bibinfo {author} {\bibfnamefont {C.~J.}\ \bibnamefont {Melián}},\ and\ \bibinfo {author} {\bibfnamefont {E.}~\bibnamefont {Sala}},\ }\bibfield  {title} {\bibinfo {title} {Interaction strength combinations and the overfishing of a marine food web},\ }\href@noop {} {\bibfield  {journal} {\bibinfo  {journal} {Proceedings of the National Academy of Sciences of the United States of America}\ }\textbf {\bibinfo {volume} {102}},\ \bibinfo {pages} {5443} (\bibinfo {year} {2005})}\BibitemShut {NoStop}%
\bibitem [{\citenamefont {Ulanowicz}\ and\ \citenamefont {Baird}(1999)}]{ulanowicz1999nutrients}%
  \BibitemOpen
  \bibfield  {author} {\bibinfo {author} {\bibfnamefont {R.~E.}\ \bibnamefont {Ulanowicz}}\ and\ \bibinfo {author} {\bibfnamefont {D.}~\bibnamefont {Baird}},\ }\bibfield  {title} {\bibinfo {title} {Nutrient controls on ecosystem dynamics: the chesapeake mesohaline community},\ }\href@noop {} {\bibfield  {journal} {\bibinfo  {journal} {Journal of Marine Systems}\ }\textbf {\bibinfo {volume} {19}},\ \bibinfo {pages} {159} (\bibinfo {year} {1999})}\BibitemShut {NoStop}%
\bibitem [{\citenamefont {Cole}(1981)}]{cole1981ants}%
  \BibitemOpen
  \bibfield  {author} {\bibinfo {author} {\bibfnamefont {B.~J.}\ \bibnamefont {Cole}},\ }\bibfield  {title} {\bibinfo {title} {Dominance hierarchies in leptothorax ants},\ }\href@noop {} {\bibfield  {journal} {\bibinfo  {journal} {Science}\ }\textbf {\bibinfo {volume} {212}},\ \bibinfo {pages} {83} (\bibinfo {year} {1981})}\BibitemShut {NoStop}%
\bibitem [{\citenamefont {Grant}(1973)}]{grant1973kangaroos}%
  \BibitemOpen
  \bibfield  {author} {\bibinfo {author} {\bibfnamefont {T.}~\bibnamefont {Grant}},\ }\bibfield  {title} {\bibinfo {title} {Dominance and association among members of a captive and a free-ranging group of grey kangaroos (macropus giganteus)},\ }\href@noop {} {\bibfield  {journal} {\bibinfo  {journal} {Animal Behaviour}\ }\textbf {\bibinfo {volume} {21}},\ \bibinfo {pages} {449} (\bibinfo {year} {1973})}\BibitemShut {NoStop}%
\bibitem [{\citenamefont {Christian}\ and\ \citenamefont {Luczkovich}(1999)}]{christian1999foodweb}%
  \BibitemOpen
  \bibfield  {author} {\bibinfo {author} {\bibfnamefont {R.~R.}\ \bibnamefont {Christian}}\ and\ \bibinfo {author} {\bibfnamefont {J.~J.}\ \bibnamefont {Luczkovich}},\ }\bibfield  {title} {\bibinfo {title} {Organizing and understanding a winter’s seagrass foodweb network through effective trophic levels},\ }\href@noop {} {\bibfield  {journal} {\bibinfo  {journal} {Ecological Modelling}\ }\textbf {\bibinfo {volume} {117}},\ \bibinfo {pages} {99} (\bibinfo {year} {1999})}\BibitemShut {NoStop}%
\bibitem [{\citenamefont {Goldwasser}\ and\ \citenamefont {Roughgarden}(1993)}]{goldwasser1993foodweb}%
  \BibitemOpen
  \bibfield  {author} {\bibinfo {author} {\bibfnamefont {L.}~\bibnamefont {Goldwasser}}\ and\ \bibinfo {author} {\bibfnamefont {J.}~\bibnamefont {Roughgarden}},\ }\bibfield  {title} {\bibinfo {title} {Construction and analysis of a large {C}aribbean food web},\ }\href@noop {} {\bibfield  {journal} {\bibinfo  {journal} {Ecology}\ }\textbf {\bibinfo {volume} {74}},\ \bibinfo {pages} {1216} (\bibinfo {year} {1993})}\BibitemShut {NoStop}%
\bibitem [{\citenamefont {Huxham}\ \emph {et~al.}(1996)\citenamefont {Huxham}, \citenamefont {Beaney},\ and\ \citenamefont {Raffaelli}}]{huxham1996parasites}%
  \BibitemOpen
  \bibfield  {author} {\bibinfo {author} {\bibfnamefont {M.}~\bibnamefont {Huxham}}, \bibinfo {author} {\bibfnamefont {S.}~\bibnamefont {Beaney}},\ and\ \bibinfo {author} {\bibfnamefont {D.}~\bibnamefont {Raffaelli}},\ }\bibfield  {title} {\bibinfo {title} {Do parasites reduce the chances of triangulation in a real food web?},\ }\href@noop {} {\bibfield  {journal} {\bibinfo  {journal} {OIKOS}\ }\textbf {\bibinfo {volume} {76}},\ \bibinfo {pages} {284} (\bibinfo {year} {1996})}\BibitemShut {NoStop}%
\bibitem [{\citenamefont {Eklöf}\ \emph {et~al.}(2013)\citenamefont {Eklöf}, \citenamefont {Jacob}, \citenamefont {Kopp}, \citenamefont {Bosch}, \citenamefont {Castro-Urgal}, \citenamefont {Chacoff}, \citenamefont {Dalsgaard}, \citenamefont {de~Sassi}, \citenamefont {Galetti}, \citenamefont {Guimarães}, \citenamefont {Lomáscolo}, \citenamefont {González}, \citenamefont {Pizo}, \citenamefont {Rader}, \citenamefont {Rodrigo}, \citenamefont {Tylianakis}, \citenamefont {Vázquez},\ and\ \citenamefont {Allesina}}]{eklof2013ecology}%
  \BibitemOpen
  \bibfield  {author} {\bibinfo {author} {\bibfnamefont {A.}~\bibnamefont {Eklöf}}, \bibinfo {author} {\bibfnamefont {U.}~\bibnamefont {Jacob}}, \bibinfo {author} {\bibfnamefont {J.}~\bibnamefont {Kopp}}, \bibinfo {author} {\bibfnamefont {J.}~\bibnamefont {Bosch}}, \bibinfo {author} {\bibfnamefont {R.}~\bibnamefont {Castro-Urgal}}, \bibinfo {author} {\bibfnamefont {N.~P.}\ \bibnamefont {Chacoff}}, \bibinfo {author} {\bibfnamefont {B.}~\bibnamefont {Dalsgaard}}, \bibinfo {author} {\bibfnamefont {C.}~\bibnamefont {de~Sassi}}, \bibinfo {author} {\bibfnamefont {M.}~\bibnamefont {Galetti}}, \bibinfo {author} {\bibfnamefont {P.~R.}\ \bibnamefont {Guimarães}}, \bibinfo {author} {\bibfnamefont {S.~B.}\ \bibnamefont {Lomáscolo}}, \bibinfo {author} {\bibfnamefont {A.~M.~M.}\ \bibnamefont {González}}, \bibinfo {author} {\bibfnamefont {M.~A.}\ \bibnamefont {Pizo}}, \bibinfo {author} {\bibfnamefont {R.}~\bibnamefont {Rader}}, \bibinfo {author} {\bibfnamefont {A.}~\bibnamefont {Rodrigo}}, \bibinfo {author}
  {\bibfnamefont {J.~M.}\ \bibnamefont {Tylianakis}}, \bibinfo {author} {\bibfnamefont {D.~P.}\ \bibnamefont {Vázquez}},\ and\ \bibinfo {author} {\bibfnamefont {S.}~\bibnamefont {Allesina}},\ }\bibfield  {title} {\bibinfo {title} {The dimensionality of ecological networks},\ }\href@noop {} {\bibfield  {journal} {\bibinfo  {journal} {Ecology Letters}\ }\textbf {\bibinfo {volume} {16}},\ \bibinfo {pages} {577} (\bibinfo {year} {2013})}\BibitemShut {NoStop}%
\bibitem [{\citenamefont {Dunne}\ \emph {et~al.}(2008)\citenamefont {Dunne}, \citenamefont {Williams}, \citenamefont {Martinez}, \citenamefont {Wood},\ and\ \citenamefont {Erwin}}]{dunne2008cambrianfood}%
  \BibitemOpen
  \bibfield  {author} {\bibinfo {author} {\bibfnamefont {J.~A.}\ \bibnamefont {Dunne}}, \bibinfo {author} {\bibfnamefont {R.~J.}\ \bibnamefont {Williams}}, \bibinfo {author} {\bibfnamefont {N.~D.}\ \bibnamefont {Martinez}}, \bibinfo {author} {\bibfnamefont {R.~A.}\ \bibnamefont {Wood}},\ and\ \bibinfo {author} {\bibfnamefont {D.~H.}\ \bibnamefont {Erwin}},\ }\bibfield  {title} {\bibinfo {title} {Compilation and network analyses of {C}ambrian food webs},\ }\href@noop {} {\bibfield  {journal} {\bibinfo  {journal} {PLOS Biology}\ }\textbf {\bibinfo {volume} {6}} (\bibinfo {year} {2008})}\BibitemShut {NoStop}%
\bibitem [{\citenamefont {Havens}(1992)}]{havens1992foodwebs}%
  \BibitemOpen
  \bibfield  {author} {\bibinfo {author} {\bibfnamefont {K.}~\bibnamefont {Havens}},\ }\bibfield  {title} {\bibinfo {title} {Scale and structure in natural food webs},\ }\href@noop {} {\bibfield  {journal} {\bibinfo  {journal} {Science}\ }\textbf {\bibinfo {volume} {257}},\ \bibinfo {pages} {1107} (\bibinfo {year} {1992})}\BibitemShut {NoStop}%
\bibitem [{\citenamefont {van Hooff}\ and\ \citenamefont {Wensing}(1987)}]{vanhooff1987wolf}%
  \BibitemOpen
  \bibfield  {author} {\bibinfo {author} {\bibfnamefont {J.~A.}\ \bibnamefont {van Hooff}}\ and\ \bibinfo {author} {\bibfnamefont {J.~A.~B.}\ \bibnamefont {Wensing}},\ }in\ \href@noop {} {\emph {\bibinfo {booktitle} {Man and Wolf: Advances, Issues, and Problems in Captive Wolf Research}}}\ (\bibinfo  {publisher} {Springer Science \& Business Media},\ \bibinfo {year} {1987})\BibitemShut {NoStop}%
\bibitem [{\citenamefont {Martinez}(1991)}]{martinez1991littlerock}%
  \BibitemOpen
  \bibfield  {author} {\bibinfo {author} {\bibfnamefont {N.~D.}\ \bibnamefont {Martinez}},\ }\bibfield  {title} {\bibinfo {title} {Artifacts or attributes? {E}ffects of resolution on the {Little Rock Lake} food web},\ }\href@noop {} {\bibfield  {journal} {\bibinfo  {journal} {Ecological Monographs}\ }\textbf {\bibinfo {volume} {61}},\ \bibinfo {pages} {367} (\bibinfo {year} {1991})}\BibitemShut {NoStop}%
\bibitem [{\citenamefont {Johnson}()}]{johnsondata}%
  \BibitemOpen
  \bibfield  {author} {\bibinfo {author} {\bibfnamefont {S.}~\bibnamefont {Johnson}},\ }\href {https://www.samuel-johnson.org/data} {\bibinfo {title} {Network data repository from various sources}}\BibitemShut {NoStop}%
\bibitem [{\citenamefont {Link}(2002)}]{link2002marine}%
  \BibitemOpen
  \bibfield  {author} {\bibinfo {author} {\bibfnamefont {J.}~\bibnamefont {Link}},\ }\bibfield  {title} {\bibinfo {title} {Does food web theory work for marine ecosystems?},\ }\href@noop {} {\bibfield  {journal} {\bibinfo  {journal} {Marine Ecology Progress Series}\ }\textbf {\bibinfo {volume} {230}},\ \bibinfo {pages} {1} (\bibinfo {year} {2002})}\BibitemShut {NoStop}%
\bibitem [{\citenamefont {Warren}(1989)}]{warren1989foodweb}%
  \BibitemOpen
  \bibfield  {author} {\bibinfo {author} {\bibfnamefont {P.~H.}\ \bibnamefont {Warren}},\ }\bibfield  {title} {\bibinfo {title} {Spatial and temporal variation in the structure of a freshwater food web},\ }\href@noop {} {\bibfield  {journal} {\bibinfo  {journal} {OIKOS}\ }\textbf {\bibinfo {volume} {55}},\ \bibinfo {pages} {299} (\bibinfo {year} {1989})}\BibitemShut {NoStop}%
\bibitem [{\citenamefont {Yodzis}(1998)}]{yodzis1998benguela}%
  \BibitemOpen
  \bibfield  {author} {\bibinfo {author} {\bibfnamefont {P.}~\bibnamefont {Yodzis}},\ }\bibfield  {title} {\bibinfo {title} {Local trophodynamics and the interaction of marine mammals and fisheries in the {B}enguela ecosystem},\ }\href@noop {} {\bibfield  {journal} {\bibinfo  {journal} {Journal of Animal Ecology}\ }\textbf {\bibinfo {volume} {67}},\ \bibinfo {pages} {635} (\bibinfo {year} {1998})}\BibitemShut {NoStop}%
\bibitem [{\citenamefont {Ulanowicz}\ and\ \citenamefont {DeAngelis}(1999)}]{ulanowicz1999southflorida}%
  \BibitemOpen
  \bibfield  {author} {\bibinfo {author} {\bibfnamefont {R.}~\bibnamefont {Ulanowicz}}\ and\ \bibinfo {author} {\bibfnamefont {D.~L.}\ \bibnamefont {DeAngelis}},\ }\bibfield  {title} {\bibinfo {title} {Network analysis of trophic dynamics in {South Florida} ecosystems},\ }in\ \href@noop {} {\emph {\bibinfo {booktitle} {U.S. Geological Survey Programon the South Florida Ecosystem; Proceedings of South Florida Restoration ScienceForum}}}\ (\bibinfo  {publisher} {BiblioGov},\ \bibinfo {year} {1999})\BibitemShut {NoStop}%
\bibitem [{\citenamefont {Clutton-Brock}\ \emph {et~al.}(1976)\citenamefont {Clutton-Brock}, \citenamefont {Greenwood},\ and\ \citenamefont {Powell}}]{cluttonbrock1976ponies}%
  \BibitemOpen
  \bibfield  {author} {\bibinfo {author} {\bibfnamefont {T.~H.}\ \bibnamefont {Clutton-Brock}}, \bibinfo {author} {\bibfnamefont {P.~J.}\ \bibnamefont {Greenwood}},\ and\ \bibinfo {author} {\bibfnamefont {R.~P.}\ \bibnamefont {Powell}},\ }\bibfield  {title} {\bibinfo {title} {Ranks and relationships in highland ponies and highland cows},\ }\href@noop {} {\bibfield  {journal} {\bibinfo  {journal} {Zeitschrift für Tierpsychologie}\ }\textbf {\bibinfo {volume} {41}},\ \bibinfo {pages} {202} (\bibinfo {year} {1976})}\BibitemShut {NoStop}%
\bibitem [{\citenamefont {Schein}\ and\ \citenamefont {Fohrman}(1955)}]{schein1955cattle}%
  \BibitemOpen
  \bibfield  {author} {\bibinfo {author} {\bibfnamefont {M.~W.}\ \bibnamefont {Schein}}\ and\ \bibinfo {author} {\bibfnamefont {M.~H.}\ \bibnamefont {Fohrman}},\ }\bibfield  {title} {\bibinfo {title} {Social dominance relationships in a herd of dairy cattle},\ }\href@noop {} {\bibfield  {journal} {\bibinfo  {journal} {The British Journal of Animal Behaviour}\ }\textbf {\bibinfo {volume} {3}},\ \bibinfo {pages} {45} (\bibinfo {year} {1955})}\BibitemShut {NoStop}%
\bibitem [{\citenamefont {Hass}(1991)}]{hass1991sheep}%
  \BibitemOpen
  \bibfield  {author} {\bibinfo {author} {\bibfnamefont {C.~C.}\ \bibnamefont {Hass}},\ }\bibfield  {title} {\bibinfo {title} {Social status in female bighorn sheep ({O}vis canadensis): expression, development and reproductive correlates},\ }\href@noop {} {\bibfield  {journal} {\bibinfo  {journal} {Journal of Zoology}\ }\textbf {\bibinfo {volume} {225}},\ \bibinfo {pages} {509} (\bibinfo {year} {1991})}\BibitemShut {NoStop}%
\bibitem [{\citenamefont {Lott}(1979)}]{lott1979bison}%
  \BibitemOpen
  \bibfield  {author} {\bibinfo {author} {\bibfnamefont {D.~F.}\ \bibnamefont {Lott}},\ }\bibfield  {title} {\bibinfo {title} {Dominance relations and breeding rate in mature male {A}merican bison},\ }\href@noop {} {\bibfield  {journal} {\bibinfo  {journal} {Zeitschrift für Tierpsychologie}\ }\textbf {\bibinfo {volume} {49}},\ \bibinfo {pages} {418} (\bibinfo {year} {1979})}\BibitemShut {NoStop}%
\bibitem [{\citenamefont {Strayer}\ and\ \citenamefont {Cummins}(1980)}]{strayer1980monkeys}%
  \BibitemOpen
  \bibfield  {author} {\bibinfo {author} {\bibfnamefont {F.~F.}\ \bibnamefont {Strayer}}\ and\ \bibinfo {author} {\bibfnamefont {M.~S.}\ \bibnamefont {Cummins}},\ }\href@noop {} {\emph {\bibinfo {title} {Dominance Relations: An Ethological View of Human Conflict and Social Interaction}}}\ (\bibinfo  {publisher} {Livingstone},\ \bibinfo {year} {1980})\BibitemShut {NoStop}%
\bibitem [{\citenamefont {Polis}(1991)}]{polis1991coachella}%
  \BibitemOpen
  \bibfield  {author} {\bibinfo {author} {\bibfnamefont {G.~A.}\ \bibnamefont {Polis}},\ }\bibfield  {title} {\bibinfo {title} {Complex trophic interactions in deserts: An empirical critique of food-web theory},\ }\href@noop {} {\bibfield  {journal} {\bibinfo  {journal} {The American Naturalist}\ }\textbf {\bibinfo {volume} {138}},\ \bibinfo {pages} {123} (\bibinfo {year} {1991})}\BibitemShut {NoStop}%
\bibitem [{\citenamefont {Opitz}(1996)}]{opitz1996caribbean}%
  \BibitemOpen
  \bibfield  {author} {\bibinfo {author} {\bibfnamefont {S.}~\bibnamefont {Opitz}},\ }\bibfield  {title} {\bibinfo {title} {Trophic interactions in {Caribbean} coral reefs},\ }\href {https://pdf.usaid.gov/pdf_docs/PNACB197.pdf} {\bibfield  {journal} {\bibinfo  {journal} {International Center for Living Aquatic Resources Management}\ } (\bibinfo {year} {1996})}\BibitemShut {NoStop}%
\bibitem [{\citenamefont {Adamic}\ and\ \citenamefont {Glance}(2005)}]{adamic2005blogs}%
  \BibitemOpen
  \bibfield  {author} {\bibinfo {author} {\bibfnamefont {L.~A.}\ \bibnamefont {Adamic}}\ and\ \bibinfo {author} {\bibfnamefont {N.}~\bibnamefont {Glance}},\ }\bibfield  {title} {\bibinfo {title} {The political blogosphere and the 2004 {U.S.} election: divided they blog},\ }\href@noop {} {\bibfield  {journal} {\bibinfo  {journal} {LinkKDD '05: Proceedings of the 3rd international workshop on Link discovery}\ ,\ \bibinfo {pages} {36}} (\bibinfo {year} {2005})}\BibitemShut {NoStop}%
\bibitem [{\citenamefont {Milo}\ \emph {et~al.}(2004)\citenamefont {Milo}, \citenamefont {Itzkovitz}, \citenamefont {Kashtan}, \citenamefont {Levitt}, \citenamefont {Shen-Orr}, \citenamefont {Ayzenshtat}, \citenamefont {Sheffer},\ and\ \citenamefont {Alon}}]{milo2004networks}%
  \BibitemOpen
  \bibfield  {author} {\bibinfo {author} {\bibfnamefont {R.}~\bibnamefont {Milo}}, \bibinfo {author} {\bibfnamefont {S.}~\bibnamefont {Itzkovitz}}, \bibinfo {author} {\bibfnamefont {N.}~\bibnamefont {Kashtan}}, \bibinfo {author} {\bibfnamefont {R.}~\bibnamefont {Levitt}}, \bibinfo {author} {\bibfnamefont {S.}~\bibnamefont {Shen-Orr}}, \bibinfo {author} {\bibfnamefont {I.}~\bibnamefont {Ayzenshtat}}, \bibinfo {author} {\bibfnamefont {M.}~\bibnamefont {Sheffer}},\ and\ \bibinfo {author} {\bibfnamefont {U.}~\bibnamefont {Alon}},\ }\bibfield  {title} {\bibinfo {title} {Superfamilies of evolved and designed networks},\ }\href@noop {} {\bibfield  {journal} {\bibinfo  {journal} {Science}\ }\textbf {\bibinfo {volume} {303}},\ \bibinfo {pages} {1538} (\bibinfo {year} {2004})}\BibitemShut {NoStop}%
\bibitem [{\citenamefont {Clauset}\ \emph {et~al.}(2016)\citenamefont {Clauset}, \citenamefont {Tucker},\ and\ \citenamefont {Sainz}}]{clauset2016icon}%
  \BibitemOpen
  \bibfield  {author} {\bibinfo {author} {\bibfnamefont {A.}~\bibnamefont {Clauset}}, \bibinfo {author} {\bibfnamefont {E.}~\bibnamefont {Tucker}},\ and\ \bibinfo {author} {\bibfnamefont {M.}~\bibnamefont {Sainz}},\ }\bibfield  {title} {\bibinfo {title} {The {C}olorado index of complex networks},\ }\href@noop {} {\  (\bibinfo {year} {2016})}\BibitemShut {NoStop}%
\bibitem [{\citenamefont {Duncan~MacRae}(1960)}]{macrae1960prison}%
  \BibitemOpen
  \bibfield  {author} {\bibinfo {author} {\bibfnamefont {J.}~\bibnamefont {Duncan~MacRae}},\ }\bibfield  {title} {\bibinfo {title} {Direct factor analysis of sociometric data},\ }\href@noop {} {\bibfield  {journal} {\bibinfo  {journal} {Sociometry}\ }\textbf {\bibinfo {volume} {23}},\ \bibinfo {pages} {360} (\bibinfo {year} {1960})}\BibitemShut {NoStop}%
\bibitem [{\citenamefont {Harbison}\ \emph {et~al.}(2004)\citenamefont {Harbison}, \citenamefont {Gordon}, \citenamefont {Lee}, \citenamefont {Rinaldi}, \citenamefont {Macisaac}, \citenamefont {Danford}, \citenamefont {Hannett}, \citenamefont {Tagne}, \citenamefont {Reynolds}, \citenamefont {Yoo}, \citenamefont {Jennings}, \citenamefont {Zeitlinger}, \citenamefont {Pokholok}, \citenamefont {Kellis}, \citenamefont {Rolfe}, \citenamefont {Takusagawa}, \citenamefont {Lander}, \citenamefont {Gifford}, \citenamefont {Fraenkel},\ and\ \citenamefont {Young}}]{harbison2004genome}%
  \BibitemOpen
  \bibfield  {author} {\bibinfo {author} {\bibfnamefont {C.~T.}\ \bibnamefont {Harbison}}, \bibinfo {author} {\bibfnamefont {D.~B.}\ \bibnamefont {Gordon}}, \bibinfo {author} {\bibfnamefont {T.~I.}\ \bibnamefont {Lee}}, \bibinfo {author} {\bibfnamefont {N.~J.}\ \bibnamefont {Rinaldi}}, \bibinfo {author} {\bibfnamefont {K.~D.}\ \bibnamefont {Macisaac}}, \bibinfo {author} {\bibfnamefont {T.~W.}\ \bibnamefont {Danford}}, \bibinfo {author} {\bibfnamefont {N.~M.}\ \bibnamefont {Hannett}}, \bibinfo {author} {\bibfnamefont {J.-B.}\ \bibnamefont {Tagne}}, \bibinfo {author} {\bibfnamefont {D.~B.}\ \bibnamefont {Reynolds}}, \bibinfo {author} {\bibfnamefont {J.}~\bibnamefont {Yoo}}, \bibinfo {author} {\bibfnamefont {E.~G.}\ \bibnamefont {Jennings}}, \bibinfo {author} {\bibfnamefont {J.}~\bibnamefont {Zeitlinger}}, \bibinfo {author} {\bibfnamefont {D.~K.}\ \bibnamefont {Pokholok}}, \bibinfo {author} {\bibfnamefont {M.}~\bibnamefont {Kellis}}, \bibinfo {author} {\bibfnamefont {P.~A.}\ \bibnamefont {Rolfe}}, \bibinfo
  {author} {\bibfnamefont {K.~T.}\ \bibnamefont {Takusagawa}}, \bibinfo {author} {\bibfnamefont {E.~S.}\ \bibnamefont {Lander}}, \bibinfo {author} {\bibfnamefont {D.~K.}\ \bibnamefont {Gifford}}, \bibinfo {author} {\bibfnamefont {E.}~\bibnamefont {Fraenkel}},\ and\ \bibinfo {author} {\bibfnamefont {R.~A.}\ \bibnamefont {Young}},\ }\bibfield  {title} {\bibinfo {title} {Transcriptional regulatory code of a eukaryotic genome},\ }\href@noop {} {\bibfield  {journal} {\bibinfo  {journal} {Nature}\ }\textbf {\bibinfo {volume} {431}},\ \bibinfo {pages} {99–} (\bibinfo {year} {2004})}\BibitemShut {NoStop}%
\bibitem [{\citenamefont {Gerstein}\ \emph {et~al.}(2012)\citenamefont {Gerstein}, \citenamefont {Kundaje}, \citenamefont {Hariharan}, \citenamefont {Landt}, \citenamefont {Yan}, \citenamefont {Cheng}, \citenamefont {Mu}, \citenamefont {Khurana}, \citenamefont {Rozowsky}, \citenamefont {Alexander}, \citenamefont {Min}, \citenamefont {Alves}, \citenamefont {Abyzov}, \citenamefont {Addleman}, \citenamefont {Bhardwaj}, \citenamefont {Boyle}, \citenamefont {Cayting}, \citenamefont {Alexandra~Charos}, \citenamefont {Cheng}, \citenamefont {Clarke}, \citenamefont {Eastman}, \citenamefont {Euskirchen}, \citenamefont {Frietze}, \citenamefont {Fu}, \citenamefont {Gertz}, \citenamefont {Grubert}, \citenamefont {Harmanci}, \citenamefont {Jain}, \citenamefont {Kasowski}, \citenamefont {Lacroute}, \citenamefont {Leng}, \citenamefont {Lian}, \citenamefont {Monahan}, \citenamefont {O’Geen}, \citenamefont {Ouyang}, \citenamefont {Partridge}, \citenamefont {Patacsil}, \citenamefont {Pauli}, \citenamefont {Raha},
  \citenamefont {Ramirez}, \citenamefont {Reddy}, \citenamefont {Reed}, \citenamefont {Shi}, \citenamefont {Slifer}, \citenamefont {Wang}, \citenamefont {Wu}, \citenamefont {Yang}, \citenamefont {Yip}, \citenamefont {Zilberman-Schapira}, \citenamefont {Batzoglou}, \citenamefont {Sidow}, \citenamefont {Farnham}, \citenamefont {Myers}, \citenamefont {Weissman},\ and\ \citenamefont {Snyder}}]{gerstein2012encode}%
  \BibitemOpen
  \bibfield  {author} {\bibinfo {author} {\bibfnamefont {M.~B.}\ \bibnamefont {Gerstein}}, \bibinfo {author} {\bibfnamefont {A.}~\bibnamefont {Kundaje}}, \bibinfo {author} {\bibfnamefont {M.}~\bibnamefont {Hariharan}}, \bibinfo {author} {\bibfnamefont {S.~G.}\ \bibnamefont {Landt}}, \bibinfo {author} {\bibfnamefont {K.-K.}\ \bibnamefont {Yan}}, \bibinfo {author} {\bibfnamefont {C.}~\bibnamefont {Cheng}}, \bibinfo {author} {\bibfnamefont {X.~J.}\ \bibnamefont {Mu}}, \bibinfo {author} {\bibfnamefont {E.}~\bibnamefont {Khurana}}, \bibinfo {author} {\bibfnamefont {J.}~\bibnamefont {Rozowsky}}, \bibinfo {author} {\bibfnamefont {R.}~\bibnamefont {Alexander}}, \bibinfo {author} {\bibfnamefont {R.}~\bibnamefont {Min}}, \bibinfo {author} {\bibfnamefont {P.}~\bibnamefont {Alves}}, \bibinfo {author} {\bibfnamefont {A.}~\bibnamefont {Abyzov}}, \bibinfo {author} {\bibfnamefont {N.}~\bibnamefont {Addleman}}, \bibinfo {author} {\bibfnamefont {N.}~\bibnamefont {Bhardwaj}}, \bibinfo {author} {\bibfnamefont {A.~P.}\
  \bibnamefont {Boyle}}, \bibinfo {author} {\bibfnamefont {P.}~\bibnamefont {Cayting}}, \bibinfo {author} {\bibfnamefont {D.~Z.~C.}\ \bibnamefont {Alexandra~Charos}}, \bibinfo {author} {\bibfnamefont {Y.}~\bibnamefont {Cheng}}, \bibinfo {author} {\bibfnamefont {D.}~\bibnamefont {Clarke}}, \bibinfo {author} {\bibfnamefont {C.}~\bibnamefont {Eastman}}, \bibinfo {author} {\bibfnamefont {G.}~\bibnamefont {Euskirchen}}, \bibinfo {author} {\bibfnamefont {S.}~\bibnamefont {Frietze}}, \bibinfo {author} {\bibfnamefont {Y.}~\bibnamefont {Fu}}, \bibinfo {author} {\bibfnamefont {J.}~\bibnamefont {Gertz}}, \bibinfo {author} {\bibfnamefont {F.}~\bibnamefont {Grubert}}, \bibinfo {author} {\bibfnamefont {A.}~\bibnamefont {Harmanci}}, \bibinfo {author} {\bibfnamefont {P.}~\bibnamefont {Jain}}, \bibinfo {author} {\bibfnamefont {M.}~\bibnamefont {Kasowski}}, \bibinfo {author} {\bibfnamefont {P.}~\bibnamefont {Lacroute}}, \bibinfo {author} {\bibfnamefont {J.}~\bibnamefont {Leng}}, \bibinfo {author} {\bibfnamefont
  {J.}~\bibnamefont {Lian}}, \bibinfo {author} {\bibfnamefont {H.}~\bibnamefont {Monahan}}, \bibinfo {author} {\bibfnamefont {H.}~\bibnamefont {O’Geen}}, \bibinfo {author} {\bibfnamefont {Z.}~\bibnamefont {Ouyang}}, \bibinfo {author} {\bibfnamefont {E.~C.}\ \bibnamefont {Partridge}}, \bibinfo {author} {\bibfnamefont {D.}~\bibnamefont {Patacsil}}, \bibinfo {author} {\bibfnamefont {F.}~\bibnamefont {Pauli}}, \bibinfo {author} {\bibfnamefont {D.}~\bibnamefont {Raha}}, \bibinfo {author} {\bibfnamefont {L.}~\bibnamefont {Ramirez}}, \bibinfo {author} {\bibfnamefont {T.~E.}\ \bibnamefont {Reddy}}, \bibinfo {author} {\bibfnamefont {B.}~\bibnamefont {Reed}}, \bibinfo {author} {\bibfnamefont {M.}~\bibnamefont {Shi}}, \bibinfo {author} {\bibfnamefont {T.}~\bibnamefont {Slifer}}, \bibinfo {author} {\bibfnamefont {J.}~\bibnamefont {Wang}}, \bibinfo {author} {\bibfnamefont {L.}~\bibnamefont {Wu}}, \bibinfo {author} {\bibfnamefont {X.}~\bibnamefont {Yang}}, \bibinfo {author} {\bibfnamefont {K.~Y.}\ \bibnamefont {Yip}},
  \bibinfo {author} {\bibfnamefont {G.}~\bibnamefont {Zilberman-Schapira}}, \bibinfo {author} {\bibfnamefont {S.}~\bibnamefont {Batzoglou}}, \bibinfo {author} {\bibfnamefont {A.}~\bibnamefont {Sidow}}, \bibinfo {author} {\bibfnamefont {P.~J.}\ \bibnamefont {Farnham}}, \bibinfo {author} {\bibfnamefont {R.~M.}\ \bibnamefont {Myers}}, \bibinfo {author} {\bibfnamefont {S.~M.}\ \bibnamefont {Weissman}},\ and\ \bibinfo {author} {\bibfnamefont {M.}~\bibnamefont {Snyder}},\ }\bibfield  {title} {\bibinfo {title} {Architecture of the human regulatory network derived from {ENCODE} data},\ }\href@noop {} {\bibfield  {journal} {\bibinfo  {journal} {Nature}\ }\textbf {\bibinfo {volume} {489}},\ \bibinfo {pages} {91} (\bibinfo {year} {2012})}\BibitemShut {NoStop}%
\bibitem [{\citenamefont {Galán-Vásquez}\ \emph {et~al.}(2011)\citenamefont {Galán-Vásquez}, \citenamefont {Luna},\ and\ \citenamefont {Martínez-Antonio}}]{galan2011regnetwork}%
  \BibitemOpen
  \bibfield  {author} {\bibinfo {author} {\bibfnamefont {E.}~\bibnamefont {Galán-Vásquez}}, \bibinfo {author} {\bibfnamefont {B.}~\bibnamefont {Luna}},\ and\ \bibinfo {author} {\bibfnamefont {A.}~\bibnamefont {Martínez-Antonio}},\ }\bibfield  {title} {\bibinfo {title} {The regulatory network of {Pseudomonas aeruginosa}},\ }\href@noop {} {\bibfield  {journal} {\bibinfo  {journal} {Microbial Informatics and Experimentation}\ }\textbf {\bibinfo {volume} {1}} (\bibinfo {year} {2011})}\BibitemShut {NoStop}%
\bibitem [{\citenamefont {Sanz}\ \emph {et~al.}(2011)\citenamefont {Sanz}, \citenamefont {Navarro}, \citenamefont {Arbués}, \citenamefont {Martín}, \citenamefont {Marijuán},\ and\ \citenamefont {Moreno}}]{sans2011regnetwork}%
  \BibitemOpen
  \bibfield  {author} {\bibinfo {author} {\bibfnamefont {J.}~\bibnamefont {Sanz}}, \bibinfo {author} {\bibfnamefont {J.}~\bibnamefont {Navarro}}, \bibinfo {author} {\bibfnamefont {A.}~\bibnamefont {Arbués}}, \bibinfo {author} {\bibfnamefont {C.}~\bibnamefont {Martín}}, \bibinfo {author} {\bibfnamefont {P.~C.}\ \bibnamefont {Marijuán}},\ and\ \bibinfo {author} {\bibfnamefont {Y.}~\bibnamefont {Moreno}},\ }\bibfield  {title} {\bibinfo {title} {The transcriptional regulatory network of {Mycobacterium tuberculosis}},\ }\href@noop {} {\bibfield  {journal} {\bibinfo  {journal} {PLOS One}\ }\textbf {\bibinfo {volume} {6}} (\bibinfo {year} {2011})}\BibitemShut {NoStop}%
\bibitem [{\citenamefont {Harriger}\ \emph {et~al.}(2012)\citenamefont {Harriger}, \citenamefont {van~den Heuvel},\ and\ \citenamefont {Sporns}}]{harriger2012rhesus}%
  \BibitemOpen
  \bibfield  {author} {\bibinfo {author} {\bibfnamefont {L.}~\bibnamefont {Harriger}}, \bibinfo {author} {\bibfnamefont {M.~P.}\ \bibnamefont {van~den Heuvel}},\ and\ \bibinfo {author} {\bibfnamefont {O.}~\bibnamefont {Sporns}},\ }\bibfield  {title} {\bibinfo {title} {Rich club organization of macaque cerebral cortex and its role in network communication},\ }\href@noop {} {\bibfield  {journal} {\bibinfo  {journal} {PLOS One}\ }\textbf {\bibinfo {volume} {7}} (\bibinfo {year} {2012})}\BibitemShut {NoStop}%
\bibitem [{\citenamefont {Markov}\ \emph {et~al.}(2013)\citenamefont {Markov}, \citenamefont {Ercsey-Ravasz}, \citenamefont {Lamy}, \citenamefont {Gomes}, \citenamefont {Magrou}, \citenamefont {Misery}, \citenamefont {Giroud}, \citenamefont {Barone}, \citenamefont {Dehay}, \citenamefont {Toroczkai}, \citenamefont {Knoblauch}, \citenamefont {Essen},\ and\ \citenamefont {Kennedy}}]{markov2013macaque}%
  \BibitemOpen
  \bibfield  {author} {\bibinfo {author} {\bibfnamefont {N.~T.}\ \bibnamefont {Markov}}, \bibinfo {author} {\bibfnamefont {M.}~\bibnamefont {Ercsey-Ravasz}}, \bibinfo {author} {\bibfnamefont {C.}~\bibnamefont {Lamy}}, \bibinfo {author} {\bibfnamefont {A.~R.~R.}\ \bibnamefont {Gomes}}, \bibinfo {author} {\bibfnamefont {L.}~\bibnamefont {Magrou}}, \bibinfo {author} {\bibfnamefont {P.}~\bibnamefont {Misery}}, \bibinfo {author} {\bibfnamefont {P.}~\bibnamefont {Giroud}}, \bibinfo {author} {\bibfnamefont {P.}~\bibnamefont {Barone}}, \bibinfo {author} {\bibfnamefont {C.}~\bibnamefont {Dehay}}, \bibinfo {author} {\bibfnamefont {Z.}~\bibnamefont {Toroczkai}}, \bibinfo {author} {\bibfnamefont {K.}~\bibnamefont {Knoblauch}}, \bibinfo {author} {\bibfnamefont {D.~C.~V.}\ \bibnamefont {Essen}},\ and\ \bibinfo {author} {\bibfnamefont {H.}~\bibnamefont {Kennedy}},\ }\bibfield  {title} {\bibinfo {title} {The role of long-range connections on the specificity of the macaque interareal cortical network},\ }\href@noop {}
  {\bibfield  {journal} {\bibinfo  {journal} {Proceedings of the National Academy of Sciences of the United States of America}\ }\textbf {\bibinfo {volume} {110}},\ \bibinfo {pages} {5187} (\bibinfo {year} {2013})}\BibitemShut {NoStop}%
\bibitem [{\citenamefont {Watts}\ and\ \citenamefont {Strogatz}(1998)}]{watts1998smallworld}%
  \BibitemOpen
  \bibfield  {author} {\bibinfo {author} {\bibfnamefont {D.~J.}\ \bibnamefont {Watts}}\ and\ \bibinfo {author} {\bibfnamefont {S.~H.}\ \bibnamefont {Strogatz}},\ }\bibfield  {title} {\bibinfo {title} {Collective dynamics of ‘small-world’ networks},\ }\href@noop {} {\bibfield  {journal} {\bibinfo  {journal} {Nature}\ }\textbf {\bibinfo {volume} {393}},\ \bibinfo {pages} {440} (\bibinfo {year} {1998})}\BibitemShut {NoStop}%
\bibitem [{\citenamefont {de~Reus}\ and\ \citenamefont {van~den Heuvel}(2013)}]{dereus2013cat}%
  \BibitemOpen
  \bibfield  {author} {\bibinfo {author} {\bibfnamefont {M.~A.}\ \bibnamefont {de~Reus}}\ and\ \bibinfo {author} {\bibfnamefont {M.~P.}\ \bibnamefont {van~den Heuvel}},\ }\bibfield  {title} {\bibinfo {title} {Rich club organization and intermodule communication in the cat connectome},\ }\href@noop {} {\bibfield  {journal} {\bibinfo  {journal} {Journal of Neuroscience}\ }\textbf {\bibinfo {volume} {33}},\ \bibinfo {pages} {12929} (\bibinfo {year} {2013})}\BibitemShut {NoStop}%
\bibitem [{\citenamefont {Bota}\ and\ \citenamefont {Swanson}(2006)}]{bota2006neuralnetwork}%
  \BibitemOpen
  \bibfield  {author} {\bibinfo {author} {\bibfnamefont {M.}~\bibnamefont {Bota}}\ and\ \bibinfo {author} {\bibfnamefont {L.~W.}\ \bibnamefont {Swanson}},\ }\bibfield  {title} {\bibinfo {title} {Online workbenches for neural network connections},\ }\href@noop {} {\bibfield  {journal} {\bibinfo  {journal} {Journal of Comparative Neurology}\ }\textbf {\bibinfo {volume} {500}},\ \bibinfo {pages} {807} (\bibinfo {year} {2006})}\BibitemShut {NoStop}%
\bibitem [{\citenamefont {Jeong}\ \emph {et~al.}(2000)\citenamefont {Jeong}, \citenamefont {Tombor}, \citenamefont {Albert}, \citenamefont {Oltvai},\ and\ \citenamefont {Barabási}}]{jeong2000metabolic}%
  \BibitemOpen
  \bibfield  {author} {\bibinfo {author} {\bibfnamefont {H.}~\bibnamefont {Jeong}}, \bibinfo {author} {\bibfnamefont {B.}~\bibnamefont {Tombor}}, \bibinfo {author} {\bibfnamefont {R.}~\bibnamefont {Albert}}, \bibinfo {author} {\bibfnamefont {Z.~N.}\ \bibnamefont {Oltvai}},\ and\ \bibinfo {author} {\bibfnamefont {A.}~\bibnamefont {Barabási}},\ }\bibfield  {title} {\bibinfo {title} {The large-scale organization of metabolic networks},\ }\href@noop {} {\bibfield  {journal} {\bibinfo  {journal} {Nature}\ }\textbf {\bibinfo {volume} {407}},\ \bibinfo {pages} {651} (\bibinfo {year} {2000})}\BibitemShut {NoStop}%
\bibitem [{\citenamefont {Garfield}()}]{garfieldhistcite}%
  \BibitemOpen
  \bibfield  {author} {\bibinfo {author} {\bibfnamefont {E.}~\bibnamefont {Garfield}},\ }\href {http://www.garfield.library.upenn.edu/histcomp/index.html} {\bibinfo {title} {Index of citation networks produced by analyses from the software {HistCite}}}\BibitemShut {NoStop}%
\bibitem [{\citenamefont {Nooy}\ \emph {et~al.}(2018)\citenamefont {Nooy}, \citenamefont {Mrvar},\ and\ \citenamefont {Batagelj}}]{denooy2018socialnetworks}%
  \BibitemOpen
  \bibfield  {author} {\bibinfo {author} {\bibfnamefont {W.~D.}\ \bibnamefont {Nooy}}, \bibinfo {author} {\bibfnamefont {A.}~\bibnamefont {Mrvar}},\ and\ \bibinfo {author} {\bibfnamefont {V.}~\bibnamefont {Batagelj}},\ }\href@noop {} {\emph {\bibinfo {title} {Exploratory Social Network Analysis with {Pajek}}}}\ (\bibinfo  {publisher} {Cambridge University Press},\ \bibinfo {year} {2018})\BibitemShut {NoStop}%
\bibitem [{\citenamefont {Williams}\ and\ \citenamefont {Musolesi}(2016)}]{william2016spatiotempnetworks}%
  \BibitemOpen
  \bibfield  {author} {\bibinfo {author} {\bibfnamefont {M.~J.}\ \bibnamefont {Williams}}\ and\ \bibinfo {author} {\bibfnamefont {M.}~\bibnamefont {Musolesi}},\ }\bibfield  {title} {\bibinfo {title} {Spatio-temporal networks: reachability, centrality and robustness},\ }\href@noop {} {\bibfield  {journal} {\bibinfo  {journal} {Royal Society Open Science}\ }\textbf {\bibinfo {volume} {3}} (\bibinfo {year} {2016})}\BibitemShut {NoStop}%
\bibitem [{\citenamefont {Glasser}\ \emph {et~al.}(2013)\citenamefont {Glasser}, \citenamefont {Sotiropoulos}, \citenamefont {Wilson}, \citenamefont {Coalson}, \citenamefont {Fischl}, \citenamefont {Andersson}, \citenamefont {Xu}, \citenamefont {Jbabdi}, \citenamefont {Webster}, \citenamefont {Polimeni}, \citenamefont {Essen}, \citenamefont {Jenkinson},\ and\ \citenamefont {for~the WU-Minn HCP~Consortium}}]{glasser2013HCP}%
  \BibitemOpen
  \bibfield  {author} {\bibinfo {author} {\bibfnamefont {M.~F.}\ \bibnamefont {Glasser}}, \bibinfo {author} {\bibfnamefont {S.~N.}\ \bibnamefont {Sotiropoulos}}, \bibinfo {author} {\bibfnamefont {J.~A.}\ \bibnamefont {Wilson}}, \bibinfo {author} {\bibfnamefont {T.~S.}\ \bibnamefont {Coalson}}, \bibinfo {author} {\bibfnamefont {B.}~\bibnamefont {Fischl}}, \bibinfo {author} {\bibfnamefont {J.~L.}\ \bibnamefont {Andersson}}, \bibinfo {author} {\bibfnamefont {J.}~\bibnamefont {Xu}}, \bibinfo {author} {\bibfnamefont {S.}~\bibnamefont {Jbabdi}}, \bibinfo {author} {\bibfnamefont {M.}~\bibnamefont {Webster}}, \bibinfo {author} {\bibfnamefont {J.~R.}\ \bibnamefont {Polimeni}}, \bibinfo {author} {\bibfnamefont {D.~C.~V.}\ \bibnamefont {Essen}}, \bibinfo {author} {\bibfnamefont {M.}~\bibnamefont {Jenkinson}},\ and\ \bibinfo {author} {\bibnamefont {for~the WU-Minn HCP~Consortium}},\ }\bibfield  {title} {\bibinfo {title} {The minimal preprocessing pipelines for the {Human Connectome Project}},\ }\href@noop {} {\bibfield
  {journal} {\bibinfo  {journal} {NeuroImage}\ }\textbf {\bibinfo {volume} {18}},\ \bibinfo {pages} {105} (\bibinfo {year} {2013})}\BibitemShut {NoStop}%
\bibitem [{\citenamefont {Smith}\ \emph {et~al.}(2013)\citenamefont {Smith}, \citenamefont {Beckmann}, \citenamefont {Andersson}, \citenamefont {Auerbach}, \citenamefont {Bijsterbosch}, \citenamefont {Douaud}, \citenamefont {Duff}, \citenamefont {Feinberg}, \citenamefont {Griffanti}, \citenamefont {Harms}, \citenamefont {Kelly}, \citenamefont {Laumann}, \citenamefont {Miller}, \citenamefont {Moeller}, \citenamefont {Petersen}, \citenamefont {Power}, \citenamefont {Salimi-Khorshidi}, \citenamefont {Snyder}, \citenamefont {Vu}, \citenamefont {Woolrich}, \citenamefont {Xu}, \citenamefont {Yacoub}, \citenamefont {Uğurbil}, \citenamefont {Essen}, \citenamefont {Glasser},\ and\ \citenamefont {for~the WU-Minn HCP~Consortium}}]{smith2013restinghcp}%
  \BibitemOpen
  \bibfield  {author} {\bibinfo {author} {\bibfnamefont {S.~M.}\ \bibnamefont {Smith}}, \bibinfo {author} {\bibfnamefont {C.~F.}\ \bibnamefont {Beckmann}}, \bibinfo {author} {\bibfnamefont {J.}~\bibnamefont {Andersson}}, \bibinfo {author} {\bibfnamefont {E.~J.}\ \bibnamefont {Auerbach}}, \bibinfo {author} {\bibfnamefont {J.}~\bibnamefont {Bijsterbosch}}, \bibinfo {author} {\bibfnamefont {G.}~\bibnamefont {Douaud}}, \bibinfo {author} {\bibfnamefont {E.}~\bibnamefont {Duff}}, \bibinfo {author} {\bibfnamefont {D.~A.}\ \bibnamefont {Feinberg}}, \bibinfo {author} {\bibfnamefont {L.}~\bibnamefont {Griffanti}}, \bibinfo {author} {\bibfnamefont {M.~P.}\ \bibnamefont {Harms}}, \bibinfo {author} {\bibfnamefont {M.}~\bibnamefont {Kelly}}, \bibinfo {author} {\bibfnamefont {T.}~\bibnamefont {Laumann}}, \bibinfo {author} {\bibfnamefont {K.~L.}\ \bibnamefont {Miller}}, \bibinfo {author} {\bibfnamefont {S.}~\bibnamefont {Moeller}}, \bibinfo {author} {\bibfnamefont {S.}~\bibnamefont {Petersen}}, \bibinfo {author}
  {\bibfnamefont {J.}~\bibnamefont {Power}}, \bibinfo {author} {\bibfnamefont {G.}~\bibnamefont {Salimi-Khorshidi}}, \bibinfo {author} {\bibfnamefont {A.~Z.}\ \bibnamefont {Snyder}}, \bibinfo {author} {\bibfnamefont {A.~T.}\ \bibnamefont {Vu}}, \bibinfo {author} {\bibfnamefont {M.~W.}\ \bibnamefont {Woolrich}}, \bibinfo {author} {\bibfnamefont {J.}~\bibnamefont {Xu}}, \bibinfo {author} {\bibfnamefont {E.}~\bibnamefont {Yacoub}}, \bibinfo {author} {\bibfnamefont {K.}~\bibnamefont {Uğurbil}}, \bibinfo {author} {\bibfnamefont {D.~C.~V.}\ \bibnamefont {Essen}}, \bibinfo {author} {\bibfnamefont {M.~F.}\ \bibnamefont {Glasser}},\ and\ \bibinfo {author} {\bibnamefont {for~the WU-Minn HCP~Consortium}},\ }\bibfield  {title} {\bibinfo {title} {Resting-state {fMRI} in the {Human Connectome Project}},\ }\href@noop {} {\bibfield  {journal} {\bibinfo  {journal} {NeuroImage}\ }\textbf {\bibinfo {volume} {80}},\ \bibinfo {pages} {144} (\bibinfo {year} {2013})}\BibitemShut {NoStop}%
\bibitem [{\citenamefont {Schröder}\ \emph {et~al.}(2015)\citenamefont {Schröder}, \citenamefont {Haak}, \citenamefont {Jimenez}, \citenamefont {Beckmann},\ and\ \citenamefont {Doeller}}]{schroder2015topography}%
  \BibitemOpen
  \bibfield  {author} {\bibinfo {author} {\bibfnamefont {T.~N.}\ \bibnamefont {Schröder}}, \bibinfo {author} {\bibfnamefont {K.~V.}\ \bibnamefont {Haak}}, \bibinfo {author} {\bibfnamefont {N.~I.~Z.}\ \bibnamefont {Jimenez}}, \bibinfo {author} {\bibfnamefont {C.~F.}\ \bibnamefont {Beckmann}},\ and\ \bibinfo {author} {\bibfnamefont {C.~F.}\ \bibnamefont {Doeller}},\ }\bibfield  {title} {\bibinfo {title} {Functional topography of the human entorhinal cortex},\ }\href@noop {} {\bibfield  {journal} {\bibinfo  {journal} {eLife}\ }\textbf {\bibinfo {volume} {4}} (\bibinfo {year} {2015})}\BibitemShut {NoStop}%
\bibitem [{\citenamefont {Salimi-Khorshidi}\ \emph {et~al.}(2014)\citenamefont {Salimi-Khorshidi}, \citenamefont {Douaud}, \citenamefont {Beckmann}, \citenamefont {Glasser}, \citenamefont {Griffanti},\ and\ \citenamefont {Smith}}]{salimi2014autodenoise}%
  \BibitemOpen
  \bibfield  {author} {\bibinfo {author} {\bibfnamefont {G.}~\bibnamefont {Salimi-Khorshidi}}, \bibinfo {author} {\bibfnamefont {G.}~\bibnamefont {Douaud}}, \bibinfo {author} {\bibfnamefont {C.~F.}\ \bibnamefont {Beckmann}}, \bibinfo {author} {\bibfnamefont {M.~F.}\ \bibnamefont {Glasser}}, \bibinfo {author} {\bibfnamefont {L.}~\bibnamefont {Griffanti}},\ and\ \bibinfo {author} {\bibfnamefont {S.~M.}\ \bibnamefont {Smith}},\ }\bibfield  {title} {\bibinfo {title} {Automatic denoising of functional {MRI} data: combining independent component analysis and hierarchical fusion of classifiers},\ }\href@noop {} {\bibfield  {journal} {\bibinfo  {journal} {NeuroImage}\ }\textbf {\bibinfo {volume} {90}},\ \bibinfo {pages} {449} (\bibinfo {year} {2014})}\BibitemShut {NoStop}%
\bibitem [{\citenamefont {Griffanti}\ \emph {et~al.}(2014)\citenamefont {Griffanti}, \citenamefont {Salimi-Khorshidi}, \citenamefont {Beckmann}, \citenamefont {Auerbach}, \citenamefont {Douaud}, \citenamefont {Sexton}, \citenamefont {Zsoldos}, \citenamefont {Ebmeier}, \citenamefont {Filippini}, \citenamefont {Mackay}, \citenamefont {Moeller}, \citenamefont {Xu}, \citenamefont {Yacoub}, \citenamefont {Baselli}, \citenamefont {Ugurbil}, \citenamefont {Miller},\ and\ \citenamefont {Smith}}]{griffanti2014ica}%
  \BibitemOpen
  \bibfield  {author} {\bibinfo {author} {\bibfnamefont {L.}~\bibnamefont {Griffanti}}, \bibinfo {author} {\bibfnamefont {G.}~\bibnamefont {Salimi-Khorshidi}}, \bibinfo {author} {\bibfnamefont {C.~F.}\ \bibnamefont {Beckmann}}, \bibinfo {author} {\bibfnamefont {E.~J.}\ \bibnamefont {Auerbach}}, \bibinfo {author} {\bibfnamefont {G.}~\bibnamefont {Douaud}}, \bibinfo {author} {\bibfnamefont {C.~E.}\ \bibnamefont {Sexton}}, \bibinfo {author} {\bibfnamefont {E.}~\bibnamefont {Zsoldos}}, \bibinfo {author} {\bibfnamefont {K.~P.}\ \bibnamefont {Ebmeier}}, \bibinfo {author} {\bibfnamefont {N.}~\bibnamefont {Filippini}}, \bibinfo {author} {\bibfnamefont {C.~E.}\ \bibnamefont {Mackay}}, \bibinfo {author} {\bibfnamefont {S.}~\bibnamefont {Moeller}}, \bibinfo {author} {\bibfnamefont {J.}~\bibnamefont {Xu}}, \bibinfo {author} {\bibfnamefont {E.}~\bibnamefont {Yacoub}}, \bibinfo {author} {\bibfnamefont {G.}~\bibnamefont {Baselli}}, \bibinfo {author} {\bibfnamefont {K.}~\bibnamefont {Ugurbil}}, \bibinfo {author}
  {\bibfnamefont {K.~L.}\ \bibnamefont {Miller}},\ and\ \bibinfo {author} {\bibfnamefont {S.~M.}\ \bibnamefont {Smith}},\ }\bibfield  {title} {\bibinfo {title} {{ICA}-based artefact removal and accelerated {fMRI} acquisition for improved resting state network imaging},\ }\href@noop {} {\bibfield  {journal} {\bibinfo  {journal} {NeuroImage}\ }\textbf {\bibinfo {volume} {95}},\ \bibinfo {pages} {232} (\bibinfo {year} {2014})}\BibitemShut {NoStop}%
\bibitem [{\citenamefont {Oostenveld}\ \emph {et~al.}(2011)\citenamefont {Oostenveld}, \citenamefont {Fries}, \citenamefont {Maris},\ and\ \citenamefont {Schoffelen}}]{oostenveld2011FT}%
  \BibitemOpen
  \bibfield  {author} {\bibinfo {author} {\bibfnamefont {R.}~\bibnamefont {Oostenveld}}, \bibinfo {author} {\bibfnamefont {P.}~\bibnamefont {Fries}}, \bibinfo {author} {\bibfnamefont {E.}~\bibnamefont {Maris}},\ and\ \bibinfo {author} {\bibfnamefont {J.-M.}\ \bibnamefont {Schoffelen}},\ }\bibfield  {title} {\bibinfo {title} {{FieldTrip}: Open source software for advanced analysis of {MEG, EEG}, and invasive electrophysiological data},\ }\href@noop {} {\bibfield  {journal} {\bibinfo  {journal} {Computational Intelligence and Neuroscience}\ }\textbf {\bibinfo {volume} {2011}} (\bibinfo {year} {2011})}\BibitemShut {NoStop}%
\bibitem [{\citenamefont {Bacco}\ \emph {et~al.}(2018)\citenamefont {Bacco}, \citenamefont {Larremore},\ and\ \citenamefont {Moore}}]{debacco2018springrank}%
  \BibitemOpen
  \bibfield  {author} {\bibinfo {author} {\bibfnamefont {C.~D.}\ \bibnamefont {Bacco}}, \bibinfo {author} {\bibfnamefont {D.~B.}\ \bibnamefont {Larremore}},\ and\ \bibinfo {author} {\bibfnamefont {C.}~\bibnamefont {Moore}},\ }\bibfield  {title} {\bibinfo {title} {A physical model for efficient ranking in networks},\ }\href@noop {} {\bibfield  {journal} {\bibinfo  {journal} {Science Advances}\ }\textbf {\bibinfo {volume} {4}} (\bibinfo {year} {2018})}\BibitemShut {NoStop}%
\end{thebibliography}%

\end{document}